\documentclass{appolb}

\usepackage[T1]{fontenc}

\usepackage{amsmath,amsbsy,amsfonts}
\usepackage{url}
\usepackage{slashed}

\usepackage{graphicx}
\usepackage{hyperref}

\def\fhc#1{{#1}^{\vphantom{\dag}}}
\def\tsqrt#1{{\textstyle \sqrt{#1}}}
\def\eps{\ensuremath{\varepsilon}}

\def\A{\ensuremath{\mathcal{A}}}

\graphicspath{{.}{programy/decays/}}

\def\spinfigscale{0.38}
\def\minscale{0.55}
\def\scanscale{0.45}

\def\c#1{#1}
\def\uev{\c{\ensuremath{\bar{u} \slashed{\varepsilon} v}}}
\def\uGPv{\c{\ensuremath{\bar{u} G P v}}}
\def\ugPv{\c{\ensuremath{\bar{u} \tilde{G} P v}}}
\def\uGpv{\c{\ensuremath{\bar{u} G \tilde{P} v}}}
\def\ugpv{\c{\ensuremath{\bar{u} \tilde{G} \tilde{P} v}}}
\def\fxx{\c{$N \neq 1$}}
\def\nofxx{\c{$N = 1$}}

\def\m#1{\ensuremath{\mathbf{#1}}}
\def\M#1{\ensuremath{\mathbf{#1}}}

\begin{document}
\preprint{ IFT/24/2005 }
\title{ Special relativity in decays of hybrids }
\author{
  Stanis{\l}aw D. G{\l}azek
  \and
  Jakub Nar\k{e}bski
  \address{Institute of Theoretical Physics, Warsaw University, 
           ul. Ho{\.z}a 69, 00-681 Warsaw, Poland}}
\date{\today}

\maketitle
\begin{abstract}
A decay of a heavy hybrid is expected to produce light mesons 
flying out with speeds comparable to the speed of light and 
phenomenological models of the decay must respect symmetries
of special relativity. We study consequences of this requirement 
in a class of simple constituent models with spin. Our models 
respect boost symmetry because they conform to the rules of a 
boost-invariant renormalization group procedure for effective 
particles in light-front QCD. But rotational symmetry of the 
decay amplitude is not guaranteed and the parameters in the 
model wave functions must take special values in order to obtain 
the symmetry. When the effective interaction Hamiltonian responsible 
for a hybrid decay has the same structure as the gluon-quark-antiquark 
interaction term obtained by solving the renormalization group 
equations for Hamiltonians in first order perturbation theory, 
the non-relativistic image of a hybrid as built from a quark 
and an antiquark and a heavy gluon that typically resides between 
the quarks, cannot produce rotationally symmetric amplitude. However, 
there exists an alternative generic picture in the model that does 
satisfy the requirements of special relativity. Namely, the distance 
between the quark and antiquark must be much smaller than the distance 
between the gluon and the pair of quarks, as if a hybrid were similar 
to a gluonium in which one gluon is replaced by a quark-antiquark 
pair.
\end{abstract}

\section{Introduction}

Special relativity symmetry imposes severe
constraints on the constituent picture of
decays of hybrids. When one attempts to
construct a constituent model of hybrids 
based on the weak-coupling expansion in QCD 
\cite{Wilsonetal}, treated as a potential 
candidate to complement the strong-coupling 
lattice picture~\cite{Wilsonlattice, Lattice2004}, 
the parameters of the model cannot be treated 
as independent. In principle, QCD contains a 
small number of free parameters: quark masses, 
a coupling constant, and a scale parameter at 
which the masses and coupling are specified. In
perturbation theory, the coupling depends on
the scale through the ratio of that scale to
$\Lambda_{QCD}$, the latter being adjusted in
the scheme in which the constituent picture
is being considered. However, not every
scheme can clearly define the concept of
constituents. It is particularly hard to describe
constituent particles in the region of small
virtualities where binding mechanism is at
work. In fact, the constituent dynamics is so
difficult to derive in QCD and solve precisely, 
especially for light quarks and gluons, that
no clear quantum-mechanical picture in
Minkowski space has been derived yet despite
extensive studies. Thus, phenomenological
images are based on models and such models
contain additional parameters due to
arbitrary simplifying assumptions that are
not under control by a precise theory.

Nevertheless, a phenomenological model
constructed within a well-defined theoretical
framework may help to finesse the leading
approximation through agreement with data.
Namely, if QCD is correct and does not
require changes to precisely describe data, a
model may hint at the structure that one
should look for when attempting to solve the
theory in a sequence of successive approximations. 
More importantly, data can provide
constraints on models not only through
discrete sets of numbers that correspond to
the magnitudes of the considered quantities,
but also through continuous symmetries. The
chief example is the Lorentz symmetry that
includes boosts and rotations. The symmetry
dictates the shape of functional dependence
of a decay amplitude on the coordinates used 
to describe the outgoing particles. We show 
how this dictum works in a simple model. 

In the standard dynamical approach,
rotational symmetry is kinematic (independent
of interactions) and boost symmetry is
dynamical (a change of frame of reference
involves effects caused by interactions, such
as a change in the number of constituents).
Thus, in the standard approach, a model 
of a decay of a hybrid (or any other hadron)
constructed in the center-of-mass frame of 
reference (CMF) of the hadron can easily 
respect the kinematical rotational symmetry
in that frame of reference. However, it is not clear
in the models that are based on the standard
approach how well they respect the dynamical
boost symmetry~\cite{RQM}. The boost symmetry
would be very useful for checking the
validity of all kinds of constituent models
since QCD is supposed to yield only fully
relativistic answers~\cite{GlazekSzczepaniak,
PoplawskiSzczepaniak}. Also, when one models
a light meson in its own CMF as built from a
fixed number of constituents and then uses
the same constituent picture when the meson
is moving with a speed close to the speed of
light, which happens when a light meson is a 
product in a decay of a hybrid, one has to verify 
if the model satisfies constraints that the
boost symmetry imposes. But the constraints 
imposed by boost symmetry are hard to satisfy 
in constituent models in the standard dynamics 
because boost generators in the standard dynamics 
depend on interactions and change the number of 
constituents. 

In the light-front (LF) approach discussed
here~\cite{Dirac, SGTM}, boost symmetry is
kinematical and it does not involve a change
in the number of constituents no matter how
fast a meson or a hybrid are moving. Instead,
the rotational symmetry is dynamical.
Therefore, if one wants to keep a fixed
number of constituents, or when one seeks a
physical picture in which the contributions
from basis states with different than the
leading number of constituents are small,
it is the rotational symmetry of decay amplitudes
(instead of boost symmetry in the standard
approach) that begins to impose stringent
constraints on the LF wave functions that stand
a chance to approximate solutions to QCD, if
some constituent picture is actually valid
in the theory. Since the group of rotations
is compact (the group of boosts is not) and it 
is already well-understood in non-relativistic 
quantum mechanics, the LF approach provides 
an opportunity for studying dynamical constraints 
of special relativity in an intuitively familiar 
way. The rotational symmetry is a much more 
familiar concept of quantum theory than the boost 
symmetry. This is reflected in the fact that the 
latter symmetry is rarely discussed in terms of 
constituents in the context of QCD. But using the 
LF approach, one can gain some insight concerning 
structures that may emerge from a relativistic 
dynamical quantum theory by checking if a set of 
wave functions chosen in a model can provide a 
rotationally symmetric decay amplitude of a hybrid.
We interpret our findings concerning hybrids
in the LF approach from the point of view of 
models of non-hybrid mesons. 

Section~\ref{sec:intro} discusses theoretical
background for the present study. Our model
assumptions are described in
Section~\ref{sec:key}. Numerical results
obtained from the symmetry constraints on the
decay amplitude of a hybrid are described in
Section~\ref{sec:sym} and discussed in
Section~\ref{sec:discussion}. Section~\ref{sec:concl} 
contains our conclusions regarding
the constituent structure of hybrids.

\section{ Theoretical background }
\label{sec:intro}

A constituent model of a hybrid qualifies as
a theoretically reasonable one when it is
clear, at least in principle, how the
constituents used in the model can be related
to quarks and gluons in QCD within a single
formulation of the theory. Of course, it is
possible that QCD will be eventually solved
and precise comparison with data will demonstrate 
that the theory requires some changes of currently 
unknown nature and implications~\cite{Lattice2004}. 
Before this happens, however, models of hybrid 
decays can be regarded as reasonable from the
theoretical point of view if they are
designed in agreement with some framework for
solving QCD. But since exact solutions are
missing and the constituent picture for
hadrons continues to pose major conceptual
problems (there is no dynamical explanation
in QCD of the phenomenological success of the
constituent quark model, examples of 
studies can be found in~\cite{meson, meson2}),
there exists today a considerable room for
varying parameters and adjusting them to data
even in reasonable models. This status of
theoretically reasonable models of hadrons
will continue until QCD is solved precisely
enough to derive the model pictures and
remove the freedom in choosing their
parameters. One should also keep in mind that
a model may be suitable for representing a
theory for one type of data and not so
reasonable for another type. There is an
exception to this ambiguity: all
theoretically reasonable models of hadrons
must respect symmetries of special
relativity. The alternative is to explain why
the constraints of that symmetry can be
ignored.

The requirement of Lorentz symmetry imposes
constraints on the wave function of the
hybrid as soon as one has a candidate for the
dynamical mechanism of the decay. This
mechanism should also be derived from QCD.
Unfortunately, constituent dynamics is not
understood yet in terms of the theory. In the
LF QCD, one can apply the renormalization
group procedure for effective particles
(RGPEP, see e.g.~\cite{GlazekQQbar} which shows
the method in a considerably simpler case of 
heavy quarkonia, rather than the very difficult
case of hybrids) and derive interactions of 
constituent quarks and gluons order by order 
in perturbation theory assuming an extremely 
small coupling constant, as if $\Lambda_{QCD}$
were much smaller than it actually is. Certainly, 
when $\Lambda_{QCD}$ is set to a realistic value, 
and when the appropriate renormalization group 
parameter is lowered down to the scale of the 
binding mechanism, the asymptotically free coupling 
constant increases to values for which the formal 
perturbative expansion in powers of the coupling 
constant is not expected to work well. But numerical 
tests in asymptotically free models with bound states 
\cite{GlazekMlynik} suggest that perturbation theory 
in RGPEP may be able to identify parts of the structure 
of the interaction terms in effective Hamiltonians that 
dominate in the bound-state dynamics, using as small 
a coupling constant as one wishes in the process. 
When a structure is already found, the main effect 
is the increase of the coupling constant in front of 
the identified structure~\cite{GlazekMlynik}. This is 
true provided that one does not lower the renormalization
group parameter in RGPEP too much and the calculation of 
the effective interaction is not cutting into the 
mechanism of formation of the bound states of interest. 

The first term relevant to hybrid decay that
one derives in perturbation theory is the
term in which an effective gluon turns into 
a pair of an effective quark and an effective 
antiquark. Therefore, we assume here that the 
interaction that leads to the decay of a hybrid 
is that of a constituent gluon decaying into a 
pair of a constituent quark and antiquark. The
constituents are identified with effective particles
in RGPEP. According to RGPEP, the relevant structure 
of the interaction is exactly like in the canonical 
Hamiltonian for LF QCD except for two features. 
One feature is that the annihilation operator for 
the effective gluon, the creation operator for the
effective quark, and the creation operator for the 
effective antiquark, all correspond to a small 
renormalization group scale $\lambda$ just above 
the scale where the binding mechanism is active.
Another feature is that the interaction vertex 
contains a form factor $f_\lambda$, instead of 
being local as in the canonical theory that one 
starts with. The effective interaction term is 
denoted by $H_{I \lambda}$.
 
Once the interaction term $H_{I \lambda}$
responsible for the decay of a hybrid state,
denoted by  $| h \rangle $, into two mesons,
denoted by $ | p \rangle $ and $ | b \rangle$ 
(the letter $p$ is chosen for a light meson, 
like meson $\pi$, and letter $b$ is chosen for 
a much heavier meson, like meson $b_1$ of mass 
1235 MeV), is specified, the decay amplitude, 
denoted by $\cal A$, is evaluated using the 
formula $ {\cal A}= \langle p \, b \, | H_{I \lambda} 
| h \rangle$. The symmetries of the decay
amplitude $\cal A$ depend on the shapes of
the wave functions of the hybrid, meson $p$,
and meson $b$. Our question is: For what wave
functions of the hybrid and two mesons one can 
obtain a spherically symmetric decay amplitude 
$\cal A$ for a $0^{++}$ hybrid using the 
interaction term $H_{I \lambda}$? 

We choose the wave functions to correspond to
the phenomenological images that underlie
constituent models of hadrons. For example,
we choose a Gaussian wave function of the
relative momentum of quarks to model a meson.
We demand that the width of that function is
on the order of masses of the involved
particles. Similar Gaussian wave functions are introduced
for the hybrid state. There are also spin
dependent factors for quarks and gluons in
the states we consider that were not studied
before~\cite{GlazekSzczepaniak}. Our study
includes several choices of these factors.
The factors are built from the spinors and
Dirac matrices which appear in the current
operators that formally can produce meson or
hybrid states with the quantum numbers we
consider. We check the decay amplitudes of
$0^{++}$ hybrids into two types of mesons:
two scalar or two pseudoscalar ones. The 
interaction term we use here differs from the
scalar-gluon term used in the previous study 
\cite{GlazekSzczepaniak} by inclusion of the
gluon spin as dictated by QCD. Typically,
the resulting decay amplitudes are not 
spherically symmetric. However, the degree of 
violation of the spherical symmetry depends 
on the values of the parameters we introduce
in the wave functions and the RGPEP parameter
$\lambda$. But one can vary the parameters and 
check if there exist any choices for which an 
amplitude is spherically symmetric. It turns 
out that such choices do exist and we find 
them here by minimizing the deviation of $\cal A$ 
from spherical symmetry. 

The main assumption that is tested in our
study is that the number of constituents can
be minimal. We know that the effective
particle dynamics derived using RGPEP in LF
QCD includes Hamiltonian terms that change
the number of effective quarks and gluons. In
the standard formulation of particle dynamics
that evolves in time $t=x^0$ instead of the
LF $x^+ = x^0 + x^3$, special relativity
requires that states of hadrons are built
from Fock sectors with different numbers of
virtual quanta. Even the vacuum state, with
no hadrons at all, appears to be a very
complex state whose structure eludes efforts
of physicists to explain it. But the
situation is different in the LF dynamics.
When QCD is regulated in transverse
(perpendicular to the $z$-axis) and
longitudinal (along the front) direction,
there is no creation of quanta from the bare
vacuum and hadrons can be considered using an
expansion into their Fock components.
Moreover, the effective interactions that are
obtained from RGPEP contain the vertex form
factors $f_\lambda$ that may have a small width
$\lambda$ in momentum space. These form
factors prevent the interaction terms from
easily producing additional constituents,
even if the coupling constant is not small in
comparison to 1. But since we do not know if 
one can approximate the solution for hadronic 
states in LF QCD by keeping only the smallest
possible number of constituents with some
$\lambda$, the critical question is if there
exists any reasonable choice of the
parameters in a model with only a minimal
number of constituents for which the model 
renders a spherically symmetric $\cal A$. 
If such choices exist, what do we learn about 
the allowed model parameters from the symmetry
requirement? 

It will be shown that all the sets of parameters
that we obtain share some features. Some of these 
features turn out to be independent of all details 
in our treatment of spin of the effective quarks 
and the gluon. The conclusion obtained earlier in 
Ref.~\cite{GlazekSzczepaniak}, using a model with 
a spinless ``gluon,'' is shown to be also valid 
when one includes spin. We study several options
to do so and we find that the tendency observed in
\cite{GlazekSzczepaniak} for a spinless gluon is a 
generic phenomenon, even though some changes do
occur. But the general conclusion is that the 
probability distribution for the constituent
quarks and a gluon in a scalar hybrid must resemble 
a state built from the gluon and an octet diquark. 
The diquark has a smaller size than the typical 
distance between the gluon and the diquark. The 
structure can be imagined as a gluonium with one 
gluon replaced by a small quark-antiquark pair. 
This is a result of pure fitting of the model 
parameters. Our study does not answer the question 
if or how this picture may arise from the effective
LF dynamics in QCD.

Although it is known how the effective
dynamics can be derived order by order from
QCD, the resulting eigenvalue equations for
mesons or hybrids are too complex and the
number of basis states too large for solving
the equations completely without some guiding
rules for simplifying the mathematics. What
we observe here is that the constraints of
special relativity strongly limit the
acceptable wave functions if one assumes that
physical states are dominated by the Fock
sectors with the smallest possible numbers of
effective constituents. The constraints of
relativity force the wave function parameters
to take values that suggest a dominant role
of gluons in the distribution of matter inside
hybrids. Gluons seem to dictate to quarks what 
the latter must do, rather than vice versa, i.e.,
not as constituent models based on the picture 
for glueless hadrons suggest. Our main point is 
not that our model must be correct, but that the 
constraints of relativity on LF models with a 
minimal number of constituents are quite restrictive, 
can be implemented in practice, and point in new 
directions. 

Let us add that the problems with Lorentz
symmetry in constituent models of bound
states of quarks and gluons occur not only
when an outgoing meson is light and has to move 
fast in the rest frame of a decaying hybrid. They
also occur when one considers a decay of a
hybrid in fast motion, which happens whenever
the decay is a part of a bigger process that
includes production and propagation with a high 
speed of the hybrid itself. Such circumstances
may be of interest, for example, in a photoproduction 
of hybrids in motion.

\section{Assumptions}
\label{sec:key}

The model we discuss is an extension of the scalar model from 
Ref.~\cite{GlazekSzczepaniak} and we adopt notation used there 
without changes. The new element here is the spin of a constituent 
gluon. In Ref.~\cite{GlazekSzczepaniak}, gluons were treated as 
scalar particles. Here, the interaction Hamiltonian $H_{I \lambda}$ 
that is responsible for the decay of the constituent gluon is taken 
directly from LF QCD with the RGPEP width parameter $\lambda$ near
the scale of hadronic masses:
\begin{equation}
  \label{eq:H_int}
  \mathcal{H}_{I \lambda} = 
  g f_\lambda \bar \psi_\lambda \gamma^\mu A^a_{\mu,\lambda} t^a \psi_\lambda \quad .
\end{equation}
We display below details of the term that creates a pair of an effective quark and 
an effective antiquark from an effective gluon. This is the only term that
counts in our calculation of the decay amplitude $\cal A$. The interaction 
term contains the vertex form factor $f_\lambda$. If we denote the invariant
mass of the quark-antiquark pair by ${\cal M}_{q \bar q}$ and the gluon mass by 
${\cal M}_g$, then~\cite{GlazekQQbar}
\begin{equation}
  \label{flambda}
  f_\lambda =  e^{- ({\cal M}_{q \bar q}^2 - {\cal M}_g^2)^2/\lambda^4} \, . 
\end{equation}

\subsection{$q{\bar q}$ Mesons}
\label{subsec:qq-mesons}

The $q\bar{q}$ meson wave functions are of the same type as 
in Ref.~\cite{GlazekSzczepaniak}.
\begin{equation}
  \label{eq:|meson>}
  | p \rangle = \sum_{12} \int [12]
  \: p^+ \tilde \delta(1 + 2 - p) \Psi^p_{J^{PC}}(1,2)
  \: b^\dagger_{\lambda 1} d^\dagger_{\lambda 2} | 0 \rangle \; ,
\end{equation}
where $\Psi^p_{J^{PC}}(1,2)$ is a~product of color, flavor (isospin),
spin, and momentum dependent factors:
\begin{equation}
  \label{eq:Psi_meson}
  \Psi^p_{J^{PC}}(1,2) =
  \chi^\dagger_{c_1} C_p \fhc{\chi}_{c_2} \,\, 
  \chi^\dagger_{i_1} I_p \fhc{\chi}_{i_2} \,\, 
  \chi^\dagger_{s_1} S_p(1,2) \fhc{\chi}_{s_2}\: 
  \psi_p(1,2) \; ,
\end{equation}
with $C_p = 1/\sqrt{3}$ (color singlet), $I_p = 1/\sqrt{2}$ (isospin singlet),
and $S_p(1,2)$ is a $2 \times 2$ spin matrix, sandwiched between two-component
spinors. $\tilde \delta$ denotes $16\pi^3$ times a three-dimensional 
$\delta$-function of plus and transverse momenta of the particles indicated in 
the argument, $\tilde{\delta}(k) = 16 \pi^3 \delta(k^+)\delta^{(2)}(k^\perp)$,
see~\cite{GlazekSzczepaniak}.
The wave function $\psi_p(1,2)$ is chosen to be Gaussian function, 
\begin{equation}
  \label{eq:psi_p}
  \psi_p(1,2) = 
  \mathcal{N}_p \, N_{pm}({\vec k}_{12}) \, N_{ps}({\vec k}_{12}) \,\,
  \exp{\left[\frac{-{\vec k}_{12}^{\, 2}}{2 \beta_p^2}\right]}\, ,
\end{equation}
where $\vec k_{12}$ is the relative three-momentum of the
quarks in their center of mass system, see Appendix~\ref{sec:appendix}.  
The additional functions $N_{pm}$ and $N_{ps}$ \cite{GlazekSzczepaniak}
are introduced entirely ad hoc. One option we investigate is
that these functions are kept equal 1. In this case, the
momentum integrals involve the relativistic momentum-space
measure and full complexity of factors resulting from the
relativistic spin structure.  In particular, the
normalization of a meson state is given by an integral of a
function that is a product of the square of the Gaussian function, a
factor resulting from the relativistic momentum-space
measure, and a complex momentum-dependent spin factor. The
other options we investigate are that the functions $N_{pm}$
or $N_{ps}$ are chosen to cancel the relativistic
momentum-space measure or the spin factor in the
normalization integral, respectively. The normalization 
condition is $\langle p|p'\rangle = p^+ \tilde \delta(p-p')$.
When both the measure
and spin factors are canceled by $N_{pm}$ (measure) and
$N_{ps}$ (spin), the normalization integral is a plain
Gaussian integral as in a non-relativistic quantum
mechanics.  We investigate these options to find out how
strongly the relativity constraints on the amplitude
$\cal A$ depend on different factors.  The same type of factors as
$N_{pm}$ or $N_{ps}$ in the meson $p$ are introduced in the
meson $b$ and denoted by $N_{bm}$ or $N_{bs}$.  All factors
$N$ are listed in the Appendix~\ref{sec:appendix}.  
The cases we discuss here are described as $N=1$ 
(full relativistic complexity of the model wave functions) 
or $N \neq 1$ (non-relativistic appearance of the normalization
integrals of the model wave functions).

For $J^{PC} = 0^{++}$ mesons, we have the following spin factor
\begin{equation}
\label{uv}
  \chi^\dagger_{s_1} S_p(1,2) \fhc{\chi}_{s_2} =   
  {\bar u}_1 v_2 \, ,
\end{equation}
(the notation is the same as in Ref.~\cite{GlazekSzczepaniak}), where 
$u_1$ and $v_2$ are Dirac spinors for quarks.  For $J^{PC} = 0^{-+}$ 
mesons (pseudoscalar mesons) we use
\begin{equation}
\label{u5v}
  \chi^\dagger_{s_1} S_p(1,2) \fhc{\chi}_{s_2} =   
  \bar{u}_1 \gamma^5 v_2 \, .
\end{equation}
The LF spinors we use here are described in the Appendix~\ref{sec:appendix}.

The above model meson states require comments concerning
their parameters, spins, and quark content. First of all,
we use the name $b$-meson here to indicate generically that
the meson is relatively heavy, like mesons $b_1$ or $\eta(1295)$
or others in the same range of masses. The spin of the meson
$b$ is assumed to be zero, and we consider only scalars and 
pseudoscalars to find out the consequences of the special 
relativity constraints in simplest and most transparent cases. 
Thus, we do not describe the spin of real $b_1$ mesons. 
The parameters of the wave function of a $b$-meson in our model
are not constrained to explain properties of any real meson
and they are left free within a considerable range in order to
check what, if any, combination of all parameters can produce 
rotationally symmetric decay amplitudes. The issue here is not 
if we can fit a model to data when we ignore gluons in a
model of ordinary mesons, but if ignoring gluons in a 
potentially valid effective constituent picture in QCD is 
allowed by special relativity symmetry even in principle, 
before a dynamical analysis is attempted.

The same applies in the case of a light meson, called here
$p$-meson, which can have a mass as small as a $\pi$-meson.
However, in the case of a pion, the assumption that such
light meson is dominated by a quark-antiquark component may
be considered merely a mock up when one attempts to
understand the structure of light mesons in terms of
canonical (almost massless) quark degrees of freedom in QCD,
where the problem of breakdown of chiral symmetry requires a
careful statement, or when one tries to create a strong
binding effect in a naive potential model.  We are facing a
problem that on the one hand hadrons can be classified in
terms of constituent quarks and lightest mesons belong to
the same scheme as the heavier ones, and on the other the
phenomenon of chiral symmetry breaking in canonical QCD is
not explained quantitatively. Our model study of symmetry
constraints in the effective constituent picture based on
RGPEP in QCD is not solving this problem. What matters is
that the effective particle picture is not necessarily
wrong. Namely, the effective quarks may have large masses
and chiral symmetry may be already explicitly broken in the
effective Hamiltonian that has the width $\lambda$
comparable with hadronic masses. At the same time, the
binding potentials in $H_\lambda$ in QCD may differ in the
case of $\pi$-mesons and in the case of heavier ones.  The
key question here is not what dynamical mechanism might be
responsible for the success of the constituent
classification of hadrons, whether it is vacuum condensates
in standard dynamics or corresponding special terms in a LF
Hamiltonian, but if such minimal constituent picture can
satisfy constraints of special relativity, assuming a most
plausible Hamiltonian term that can lead to the decay of a
hybrid. Therefore, we take the stance at this stage of the
development that anything goes that produces relativity with
constituents and we ask if this condition can be satisfied
in any, even remotely plausible way from the point of view
of currently popular models. Our major issue is if the LF
QCD constituent picture has a chance to obey rotational
symmetry. We do not solve the dynamics of formation of pions
here and we will call the light meson $p$, not $\pi$.
 
\subsection{Hybrid Meson}
\label{subsec:hybrid}

Our model for $0^{++}$ hybrid state is of the form (cf.~\cite{GlazekSzczepaniak})
\begin{equation}
  \label{eq:|hybrid>}
  | h \rangle = \sum_{123} \int[123] 
  \, h^+ \tilde \delta(1+2+3 - h) \Psi^h_{J^{PC}}(1,2,3) 
  \, b^\dagger_{\lambda 1} d^\dagger_{\lambda 2} a^\dagger_{\lambda 3} | 0 \rangle \, .
\end{equation}
The creation operator for an effective gluon of width $\lambda$, $a_{\lambda 3}^\dagger$, 
carries spin, and the wave function $\Psi^h_{J^{PC}}(1,2,3)$ depends on this 
spin. The wave function is a product of color, flavor (isospin), spin, and 
Gaussian functions of the relative momenta of the three particles:  
\begin{equation}
  \label{eq:Psi_h}
  \begin{split}
  \Psi_{J^{PC}}(1,2,3) &= 
  \,\, \chi^\dagger_{c_1}C_h^{c_3} \chi^{\vphantom{\dagger}}_{c_2}
  \,\, \chi^\dagger_{i_1}I_h       \chi^{\vphantom{\dagger}}_{i_2}
  \,\, \chi^\dagger_{s_1}S_h(1,2,3)\chi^{\vphantom{\dagger}}_{s_2} 
  \psi_h(1,2,3) \; ,
  \end{split}
\end{equation}
with $C^{c_3}_h = t^{c_3}/2$ and $I_h = 1/\sqrt{2}$.
The momentum dependent factor (hybrid wave function)
is assumed to have the $q\bar{q}$--cluster 
form~\cite{GlazekSzczepaniak, HornMandula}
\begin{equation}
  \label{eq:psi_h}
  \begin{split}
    \psi_h(1,2,3) &= 
    \mathcal{N}_h \,\, 
    N_{hm}(\vec{k}_q, \vec{k}_g) \,\,
    N_{hs}(\vec{k}_q, \vec{k}_g) 
    \,\, \exp{\left[\frac{-\vec{k}_q^{\, 2}}{2 \beta_{hq}^2}\right]} 
    \,\, \exp{\left[\frac{-\vec{k}_g^{\, 2}}{2 \beta_{hg}^2}\right]} \, ,
  \end{split}
\end{equation}
with typical Gaussian functions of the relative momenta.
Namely, $\vec k_q$ is the three-momentum of the quark in the
CMF of the quark-antiquark pair, and $\vec k_g$ is the
three-momentum of the gluon in the CMF of the quark,
antiquark, and gluon (see Appendix~\ref{sec:appendix} for details). 
Since we use the LF form of dynamics,
the CMFs are well defined and separation of the relative and
CMF motion for any state is purely kinematical. The optional
additional factors $N_{hm}$ and $N_{hs}$ can be again kept
equal 1 or chosen so that the wave function normalization
condition has a non-relativistic appearance for three
constituents similarly to the wave functions of mesons, 
see Appendix~\ref{sec:appendix}.
For consistency, we will always either introduce all factors 
$N$ for both mesons and hybrid equal 1 (the cases labeled $N=1$ ), 
or insert all factors $N$ such that all our states are normalized 
through the same integrals as in a non-relativistic theory
(the cases labeled $N \neq 1$).

The simplest scalar-hybrid spin factor that may be considered 
in a relativistic theory is 
\begin{equation}
\label{uev}
  \chi_1^\dagger S_h(1,2,3) \chi_2  =   
  {\bar u}_1 \gamma_\mu {\varepsilon}^\mu_3 v_2 \, ,
\end{equation}
where ${\varepsilon}_3$ is the gluon polarization four vector.
This case will be referred to as the case $\uev$. For 
the effective gluon in the light-front gauge $A^+ \equiv 0$, the 
polarization vector has the following components:
\begin{equation}
  \label{eq:eps}
  \varepsilon_{k\sigma}^\mu =
  \left(
    \varepsilon_{k\sigma}^+ = 0,
    \varepsilon_{k\sigma}^- = \frac{2 k^\perp \varepsilon_{\sigma}^\perp}{k^+},
    \varepsilon_{k\sigma}^\perp = \varepsilon_{\sigma}^\perp
  \right)\:.
\end{equation}
The sum over gluon polarizations is
\begin{equation}
  \label{eq:sum:ee}
  \begin{split}
    \sum_{\sigma} \varepsilon_{k_3\sigma}^\mu \varepsilon_{k_3\sigma}^{\ast\,\nu} 
    &=
    - g^{\mu\nu} + \frac{k_3^\mu g^{+\nu} + g^{+\mu} k_3^\nu}{k_3^+} \, .
  \end{split}
\end{equation}
In the above formula, the component $k_3^-$  of the gluon momentum is 
the same as for massless gluons, $k_3^- = k_3^{\perp \, 2}/k_3^+$. But 
in evaluating kinetic energy of the effective gluons, we introduce the 
gluon effective mass parameter $m_g$ that can depend on the RGPEP 
parameter $\lambda$. We do not know the value of $m_g$ and we leave it 
as a free parameter in our Gaussian wave function.

An alternative structure for the spin factor, pertaining to 
non-abelian gauge symmetry but not necessarily better than Eq.~\eqref{uev}
from the dynamical point of view in the effective theory, is  
\begin{equation}
  \label{P123}
    \chi^\dag_{s_1}\, S_h(1,2,3)\, \fhc{\chi}_{s_2} = 
    \bar{u}_1 \gamma_\mu v_2\, G^{\mu\nu} P_\nu \, ,
\end{equation}
where $G^{\mu \nu} = k_3^\mu
\varepsilon_{k\sigma}^\nu - k_3^\nu
\varepsilon_{k\sigma}^\mu$ and $P=k_1 + k_2 + k_3$.
The four-vector $\varepsilon_{k\sigma}^\nu$ is the
polarization vector for massless gauge
bosons. But $k_3^-$ can be calculated as if
the gluon mass were 0, or using the parameter
$m_g$, and we do not know which way is more
realistic in a dynamical theory. Therefore,
we insert the unknown mass $m_g$ in the
formula $k_3^- = (k_3^{\perp \, 2} + m_g^2)/k_3^+$ 
and check for what values of $m_g$ the resulting 
decay amplitude $\cal A$ is spherically symmetric 
in the hybrid CMF. This case is referred to as 
$\uGPv$.

We also consider alternative versions of the spin 
factor, where we calculate $k_3^-$ as if the gluon 
mass should be kept 0 in $G^{\mu \nu}$ and/or in $P^\nu$.
These cases are referred to as $\ugpv$,
$\uGpv$, and $\ugPv$, respectively. The tilde
means that we put $m_g = 0$ in evaluating $k_3^-$
in the factor that is labeled with the tilde.

The alternative spin factors resemble the one
that occurs in the operator structures used
in lattice calculations~\cite{lattice, lattice2}. 
In the hybrid CMF, where $\vec P=0$, this factor 
reduces to $\bar{u}\gamma^i v G_{i0}$, which is
is a combination of the components of the quark 
current and the chromoelectric gluon field.

Note that the gluon momentum component $k_3^-
= (k_3^{\perp \, 2} + m_g^2)/k_3^+$ does not
contribute to the Lorentz product
$k_3\varepsilon_{k\sigma}$ in the chosen
gauge and the entire small group of the
Poincare transformations that preserve the
light-front hyperplane also does not change
the gauge condition $A^+ \equiv 0$. Thus, we
can safely boost the hybrid state using
kinematical relations and our assignment of
mass to the effective gluon does not
interfere with our choice of gauge. 
This is important because our model would not be
reasonable otherwise. Namely, if the small
group and gauge choice would not commute, we
would not be able to construct the states of
mesons and hybrids in motion without changing
the gauge. The latter change would be
associated with altering interaction terms in
the effective Hamiltonian of width $\lambda$.
Fortunately, our choice of the LF dynamics,
instead of the standard one, offers a
possibility of keeping boost invariance using
one and the same choice of gauge in all
frames of reference that can be reached by
the boosts. Note also that our RGPEP procedure
respects this commutativity because it is 
invariant under the small group and our model 
is reasonable in this respect.

As explained earlier, in contrast to the standard 
approaches where rotational symmetry is kinematical 
and boosts are dynamical, in the LF approach the 
rotational symmetry is dynamical. Therefore, we 
now have to check to what extent our models can 
guarantee that the resulting decay amplitude is
spherically symmetric in the rest frame of the 
hybrid. We do this in the next section using both 
Eqs.~\eqref{uev} and~\eqref{P123} in a number of
cases that we have introduced above.

\section{Symmetry constraints}
\label{sec:sym}

The decay amplitude of a scalar hybrid ($J^{PC} 
= 0^{++}$) into two mesons, either two $J^{PC} = 
0^{++}$ mesons or two $0^{-+}$ mesons, should be 
spherically symmetric. But a constituent model 
built in the LF scheme introduces dependence on 
the angle $\theta$ between the $z$-axis and the 
direction of flight of the light meson. This 
effect is the price we pay for boost invariance. 
The effect was discovered and initially studied 
in a scalar model in Ref.~\cite{GlazekSzczepaniak}.

\begin{figure}[!tbp]
\centering
\includegraphics[scale=\spinfigscale,trim=20  0  0  10]{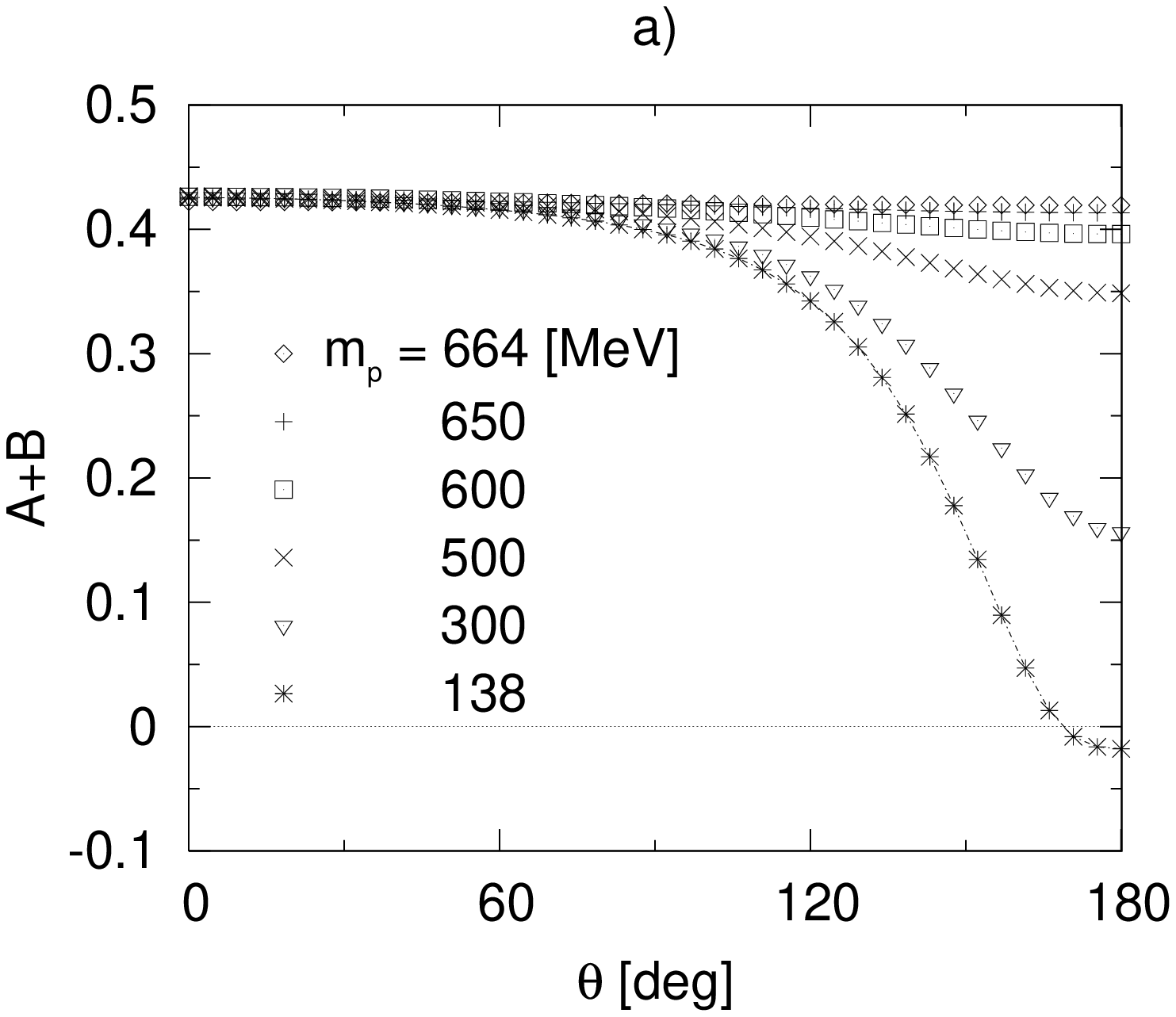}
\includegraphics[scale=\spinfigscale,trim=20  0 10  10]{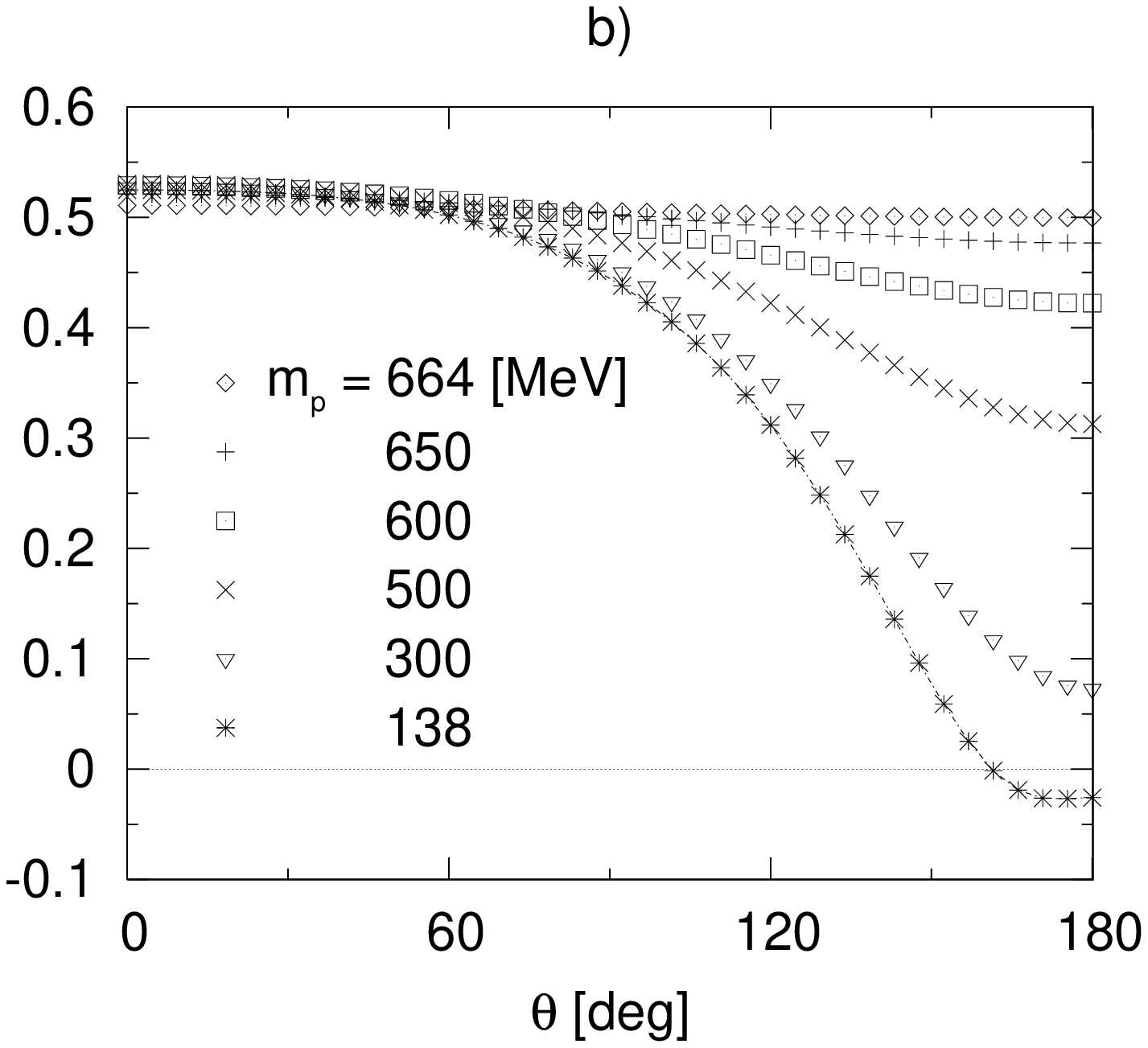}
\caption{\label{fig:scan}\label{fig:scan:first}
  The angular dependence of the decay amplitude for various masses of
  the light meson, $m_p$. The plot a) is for the hybrid spin factor
  equal $\bar{u}\gamma^\mu v \varepsilon_\mu$, and b) is for spin factor
  $\bar{u}\gamma^\mu v G_{\mu\nu} P^\nu_{123}$. 
  In both cases the hybrid meson decays into two $J^{PC} = 0^{++}$ scalar 
  mesons (spin factors $\bar{u} v$).
  The model wave functions contain factors $N_{ps}$, $N_{pm}$, 
  $N_{bs}$, $N_{bm}$ and $N_{hs}$, $N_{hm}$ that secure that 
  the normalization integrals have a non-relativistic appearance
  of integrals of plain Gaussian functions (case $N \neq 1$).
  All parameters of the wave functions are given in the first column of 
  Table~\ref{tab:scan-and-hsmin}.}
\end{figure}
\begin{figure}[!tbp]
\centering
\includegraphics[scale=\spinfigscale,trim=20  0  0  10]{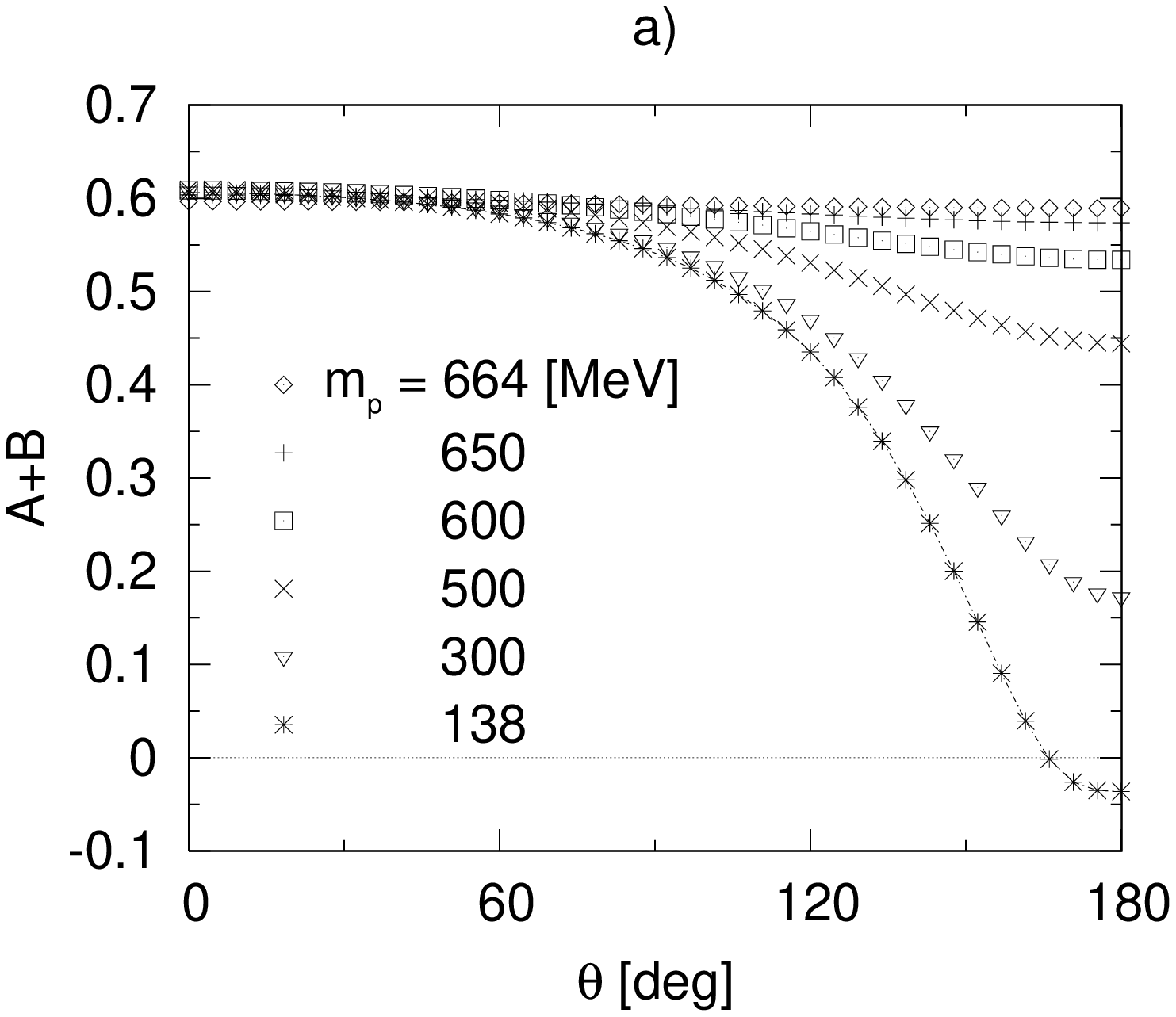}
\includegraphics[scale=\spinfigscale,trim=20  0 10  10]{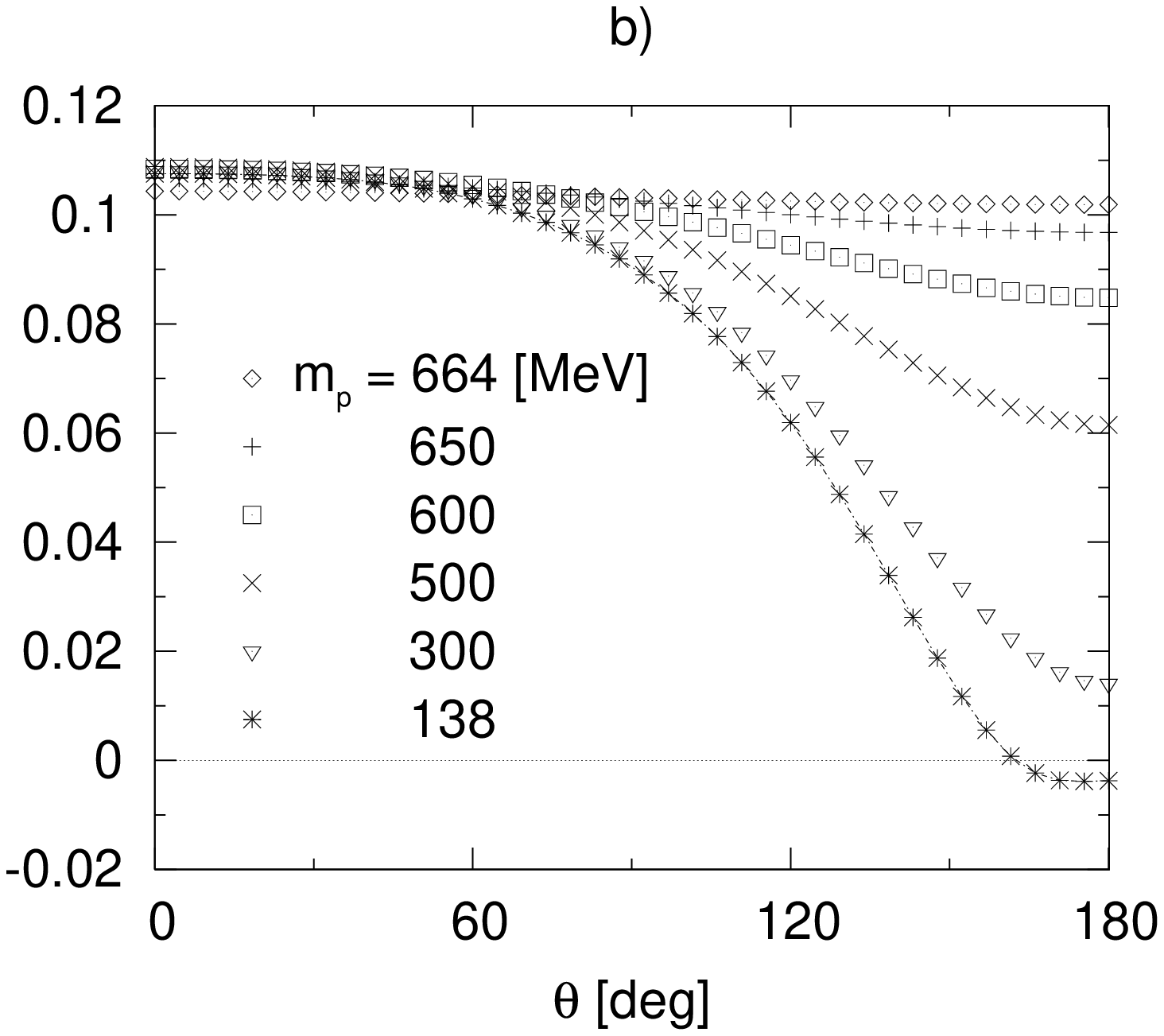}
\caption{\label{fig:nofxx:scan}
  The same as in Fig.~\ref{fig:scan}, but with the factors $N = 1$.}
\end{figure}
\begin{figure}[tbp]
\centering
\includegraphics[scale=\spinfigscale,trim=20  0  0  10]{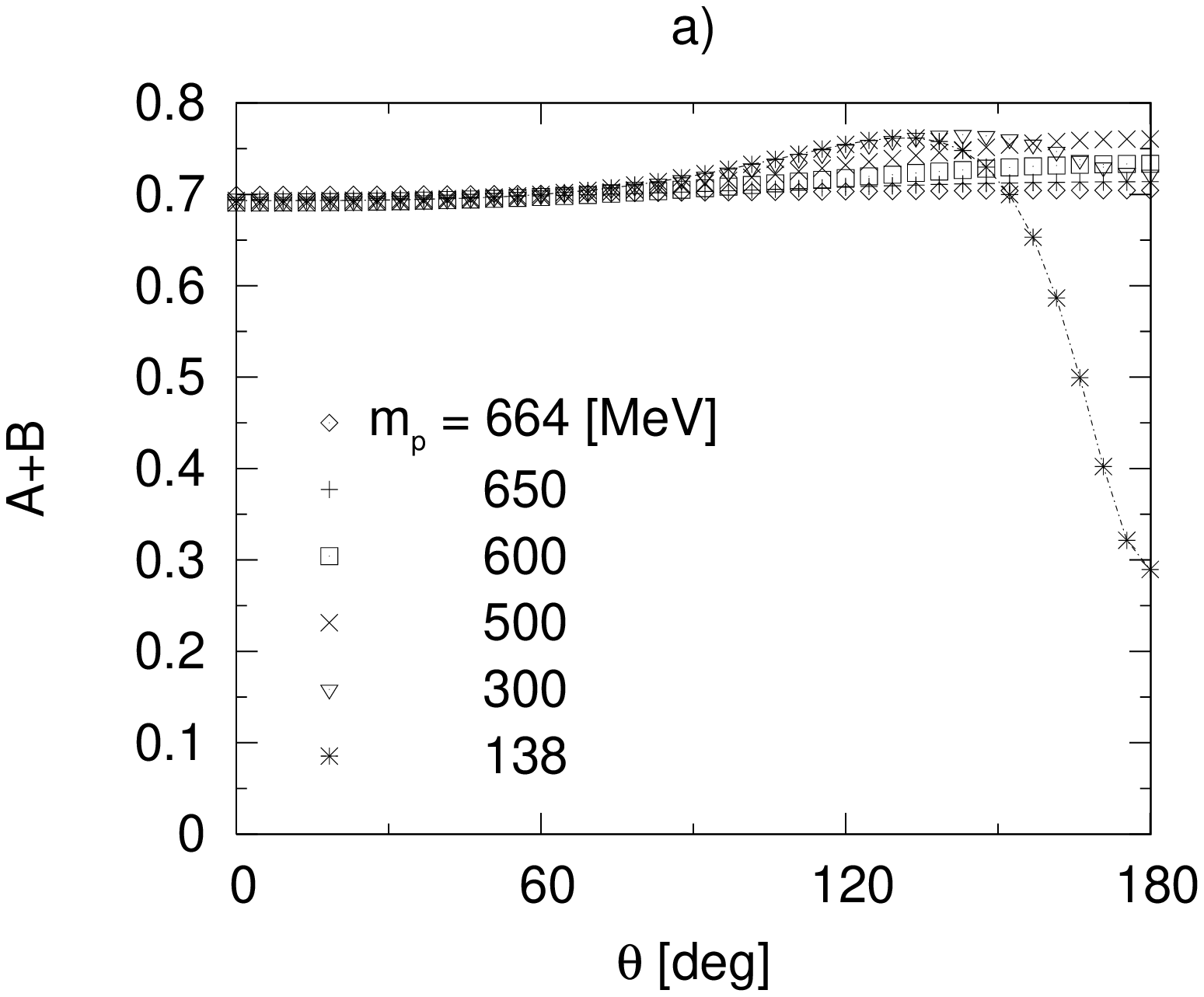}
\includegraphics[scale=\spinfigscale,trim=20  0 10  10]{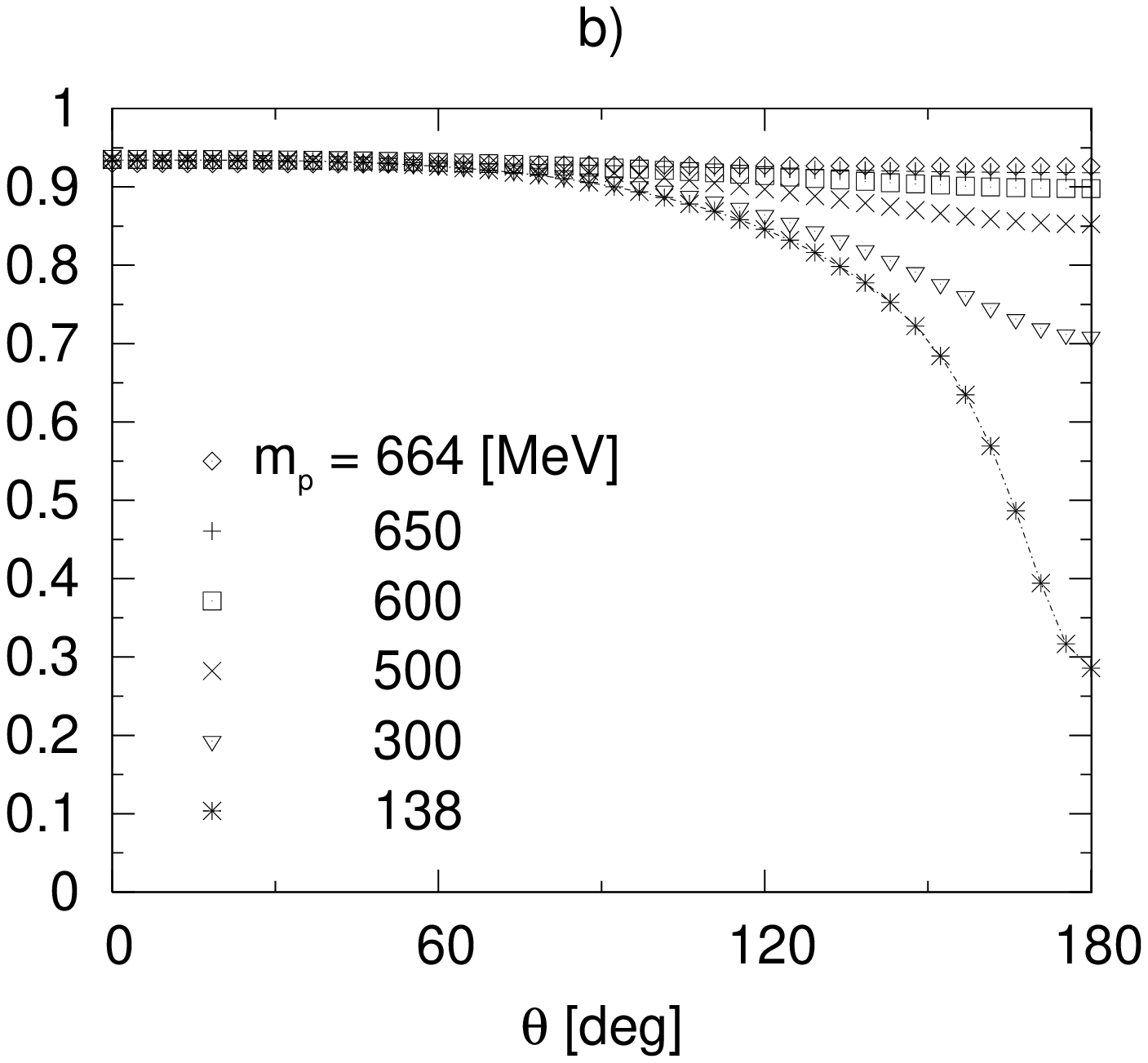}
\caption{\label{fig:u5v:scan}
  The same as in Fig.~\ref{fig:scan}, but for hybrid decay into 
  two $J^{PC} = 0^{-+}$ pseudo-scalar mesons (spin factors $\bar{u}\gamma^5 v$).}
\end{figure}
\begin{figure}[tbp]
\centering
\includegraphics[scale=\spinfigscale,trim=20  0  0  10]{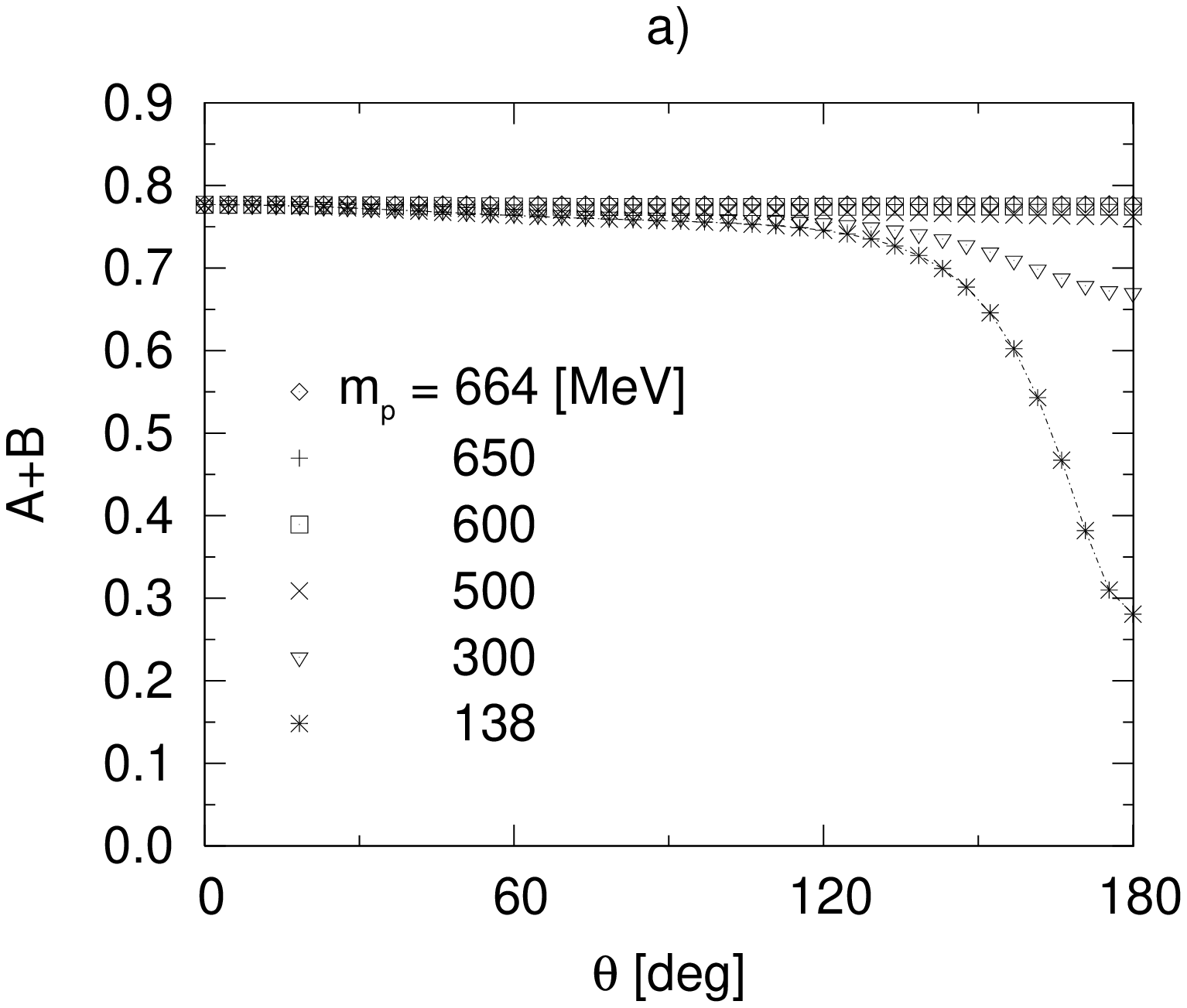}
\includegraphics[scale=\spinfigscale,trim=20  0 10  10]{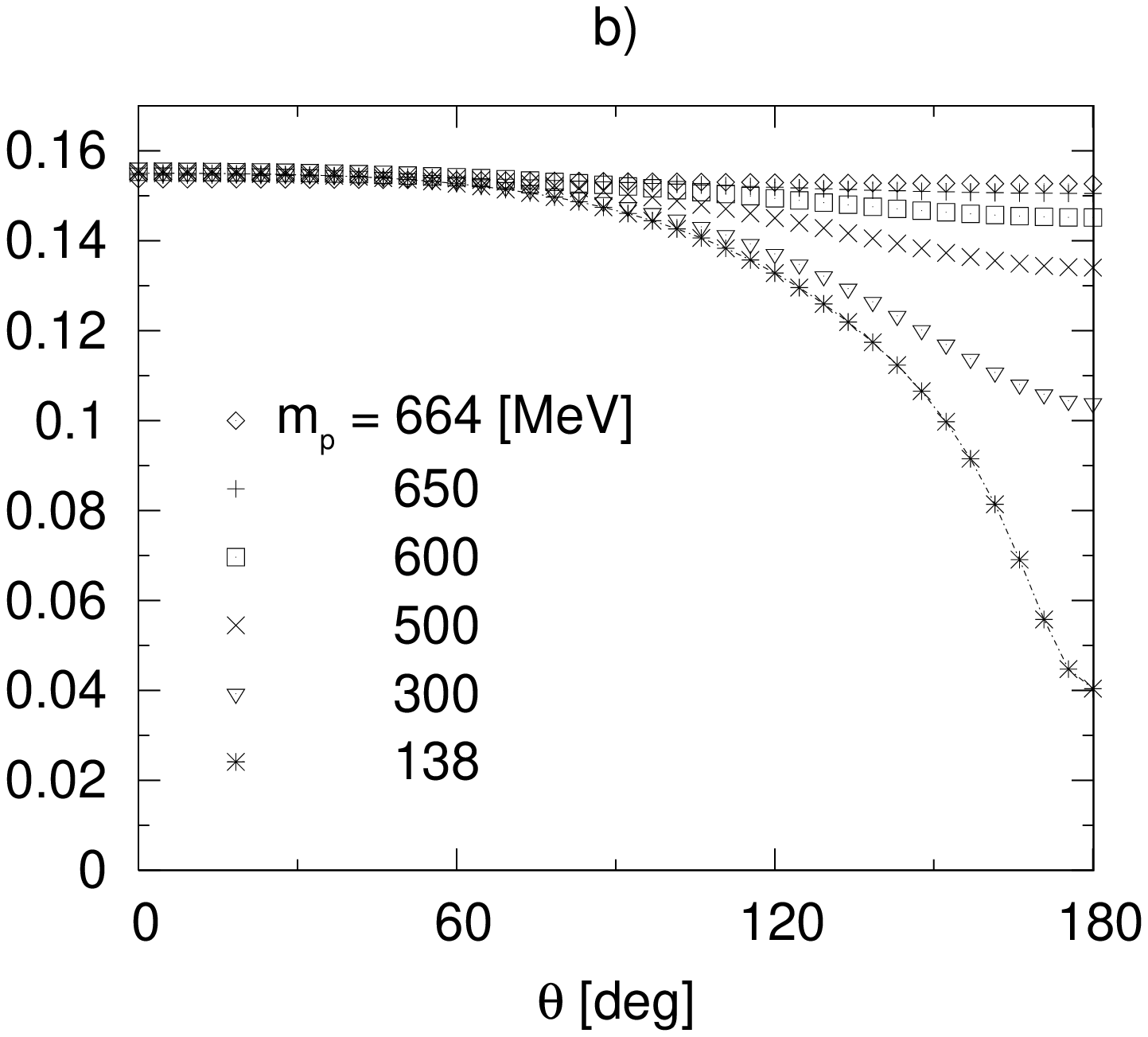}
\caption{\label{fig:nofxx-u5v:scan}\label{fig:scan:last}
  The same as in Fig.~\ref{fig:scan}, but with the combined 
  effect due to the changes from Fig.~\ref{fig:nofxx:scan}
  and~\ref{fig:u5v:scan}: $N=1$ and the hybrid decays into two 
  pseudo-scalar mesons.}
\end{figure}

Fig.~\ref{fig:scan} illustrates how badly the
rotational symmetry is violated in a decay into 
two scalar mesons (spin factors $\bar{u}v$) when 
the light meson mass varies from 664 MeV toward 
the value of 138 MeV. In this figure the factors $N_{hm}$, 
$N_{hs}$, $N_{bm}$, $N_{bs}$, $N_{pm}$, and $N_{ps}$, 
in the wave functions, are kept different from 1 
to secure non-relativistic normalization. In the
case of 664 MeV, the hybrid mass $m_h$ is just above 
the threshold of $m_b+m_p$ and the product mesons can 
barely move. The LF constituent model renders an 
amplitude that does not depend on the angle $\theta$.
In the 138 MeV case, corresponding to the $\pi$-mesons,
the hybrid mass is far above the threshold and the 
light meson has a highly relativistic velocity. 
In this case, the amplitude depends on the angle 
$\theta$ to an unacceptable degree. The effect is 
caused by the fact that when the outgoing light 
meson flies against the $z$-axis with nearly speed
of light, it must be built from quarks that have very 
small momentum $k^+$ and such quarks are suppressed
in the case of wave functions used in the calculation.
The parameters that we use are given in the first column 
of Table~\ref{tab:scan-and-hsmin} on 
page~\pageref{tab:scan-and-hsmin} (cf.~\cite{GlazekSzczepaniak}). 
Figs.~\ref{fig:nofxx:scan} to~\ref{fig:nofxx-u5v:scan} 
show the same effect, but with all factors $N_{hm}$, 
$N_{hs}$, $N_{bm}$, $N_{bs}$, $N_{pm}$, and $N_{ps}$ 
equal 1, or in the case of two pseudoscalar ($J^{PC} 
= 0^{-+}$) mesons 
instead of the scalar ones, or in the case where the 
two changes are combined. In order to satisfy constraints 
of special relativity, the decay amplitude should not 
depend on the angle $\theta$.

Broadly speaking, the violation of rotational symmetry 
results from the fact that the model wave functions 
are not constrained dynamically by any underlying
relativistic theory. Given that the model is reasonable, 
in the sense that (1) the meson states are formed using 
well-defined degrees of freedom that appear in the LF 
Hamiltonian $H_\lambda$ in QCD with a small RGPEP 
parameter $\lambda$, (2) the boost symmetry is 
preserved exactly, and (3) the decay is driven by an 
interaction term in the same Hamiltonian, the most 
questionable element of the model is the assumption 
that a small number of constituents is sufficient to 
build a solution of a relativistic theory. The model 
assumes that the number of constituents is the smallest 
possible. It may fail to produce rotational symmetry 
because the symmetry is dynamical in the LF scheme and
the interactions can change the number of 
constituents~\cite{Dirac, SGTM}. 
We see in Figs.~\ref{fig:scan} to~\ref{fig:nofxx-u5v:scan} 
that all models we test respect rotational symmetry very 
well in non-relativistic decays. But in the relativistic 
decays, all the models fail more or less equally badly
(on average, decays into pseudoscalars are a bit less 
wrong than decays into scalars because pseudoscalars are 
dominated by non-relativistic, momentum-independent components 
in their wave functions). Does this mean that all models 
based on the assumption of the smallest number of constituents 
must be entirely wrong?

We find that the answer to this question is no:
a minimal constituent model does not have to be 
wrong. Since the rotational symmetry is dynamical
in the approach we study, it is not known if the 
parameters listed in Table~\ref{tab:scan-and-hsmin} 
in the first column do correspond to a solution of a
relativistic theory. Suppose that a different set of 
parameters should be used in a reasonable model that 
approximates a solution of a relativistic theory. Can 
one find a set of parameters in the constituent wave
functions for which the required rotational symmetry 
of the decay amplitude is obtained? This question is
found to have a positive answer but the sets of parameters 
that we find point to a new picture for the hybrids. 
The picture seems to be generic in the sense that its
dominant features are independent of how the spin of 
the gluon and the spins of quarks are treated. Our
numerical studies produce examples of models with a 
smallest number of constituents in which the rotational 
symmetry is respected well when one allows the parameters 
in the wave functions and the RGPEP scale $\lambda$ in 
the Hamiltonian to vary. The reader should remember 
that the number of variable parameters in the class of
models we consider is 7 and there exists a great number 
of possibilities to check, each demanding a multidimensional 
integration for every value of the angle $\theta$.

\begin{figure}[!tbp] 
\centering 
\includegraphics[scale=\minscale,trim=95 0 0 245]{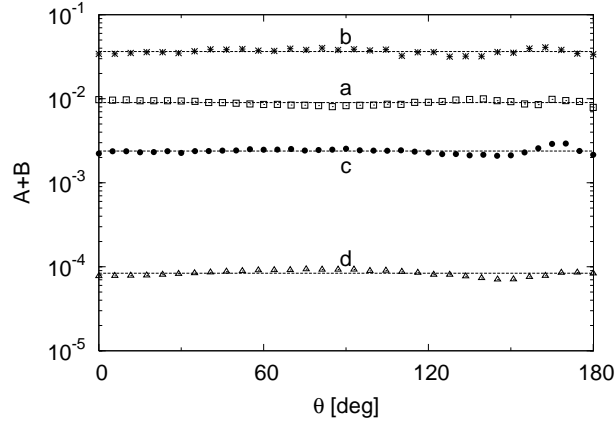}
\caption{\label{fig:hsmin} 
  The hybrid decay amplitude as function of $\theta$ in four cases: 
  a) $\uev$, $N \neq 1$ 
  and second column of Table~\ref{tab:scan-and-hsmin}, 
  b) $\uev$, $N = 1$ 
  and third column of Table~\ref{tab:scan-and-hsmin}, 
  c) $\uGPv$, $N \neq 1$
  and fourth column of Table~\ref{tab:scan-and-hsmin},
  d) $\uGPv$, $N = 1$
  and fifth column of Table~\ref{tab:scan-and-hsmin}.
  Decays into two $J^{PC} = 0 ^{++}$ mesons.
}
\end{figure}
\begin{table}[!tbp]
\def\v{\multicolumn{1}{c}{\textit{varies}}}
\centering 
\begin{tabular}{lllllll}
\hline 
\hline 
Fig. \# 
& \ref{fig:scan:first}--\ref{fig:scan:last}a,b
& \ref{fig:hsmin}a & \ref{fig:hsmin}b & \ref{fig:hsmin}c & \ref{fig:hsmin}d
& hs 
\\ 
\hline 
$m_h$        & 1.9    & $s$    & $s$  & $s$  & $s$  &  $s$   \\ 
$m_b$        & 1.235  & $s$    & $s$  & $s$  & $s$  &  $s$   \\ 
$m_p$        & \v     & 0.1375 & $s$  & $s$  & $s$  &  $s$   \\ 
$m_q$        & 0.3    & 0.68   & 0.75 & 0.67 & 0.87 &  0.365 \\ 
$m_g$        & 0.8    & 2.6    & 1.4  & 3.7  & 1.3  &  1.63  \\ 
$\beta_p$    & 0.4    & 0.19   & 0.34 & 0.14 & 0.21 &  0.375 \\
$\beta_b$    & 0.4    & 0.77   & 1.2  & 1.0  & 0.96 &  0.719 \\
$\beta_{hg}$ & 1.0    & 0.28   & 0.31 & 0.31 & 0.52 &  0.60  \\
$\beta_{hq}$ & 1.0    & 3.8    & 4.4  & 3.8  & 7.5  &  4.61  \\ 
$\lambda$    & 10000  & 4.8    & 4.4  & 4.4  & 4.4  &  4.49  \\ 
\hline 
\hline
\end{tabular} 
\caption{\label{tab:scan-and-hsmin} 
  The parameters of wave functions (in GeV) that we use in
  Figs.~\ref{fig:scan:first}-\ref{fig:hsmin} (in the same notation as
  in Ref.~\cite{GlazekSzczepaniak}). 
  $s$ means that the parameter is the same as in the left neighboring
  column.
  Columns~\ref{fig:hsmin}a--d display results of
  local minimization using Powell's procedure~\cite{Powell}, starting
  from the values given in the last column (labeled ``hs''), which
  is the same as the corresponding column in Table~I in 
  Ref.~\cite{GlazekSzczepaniak}.}
\end{table}

Fig.~\ref{fig:hsmin} shows how well the rotational symmetry 
can be restored by selecting a different set of parameters in
the wave functions (the sets corresponding to Fig.~\ref{fig:hsmin}
are given in columns~\ref{fig:hsmin}a--d in Table~\ref{tab:scan-and-hsmin}).
Curves ``a'' and ``b'' represent the decay amplitudes 
for a hybrid meson with the spin factor given in Eq.~\eqref{uev} (case
$\uev$) and the wave-function parameters given in the second column of
Table~\ref{tab:scan-and-hsmin}.  Curves ``c'' and ``d'' on the same figure
show the decay amplitudes for a hybrid with the spin factor given by
Eq.~\eqref{P123} (case $\uGPv$) and with the parameters given in the
third column of Table~\ref{tab:scan-and-hsmin}.  Curves ``a'' and ``c'' 
are obtained with $N \neq 1$, and curves ``b'' and ``d'' are obtained with 
$N=1$.  

The optimal choice of the parameters that one obtains from the
condition of rotational symmetry includes $\beta_p$ about twice smaller
than the typical value of 0.4 GeV in the first column of
Table~\ref{tab:scan-and-hsmin}, which corresponds to the size of a
real meson $\pi$ (we will return to this issue below). We could also
find other sets of parameters with even smaller $\beta_p$ and
considerably smaller quark masses, a feature observed already in
Ref.~\cite{GlazekSzczepaniak}. The cases shown here are obtained by
starting a Powell local minimization procedure~\cite{Powell, NR}, 
starting from the $hs$ (for ``heavy-scalar.'') set of parameters 
that was found in a scalar model in Ref.~\cite{GlazekSzczepaniak}.
The nomenclature refers to the relatively heavy
scalar particles that played the role of quarks in Ref. 
\cite{GlazekSzczepaniak}. 

The cases we display here are characterized by quite good spherical
symmetry in comparison to other locally optimal choices which we were
also able to identify but which displayed more variation with
$\theta$. In choosing the minimization we also adopted a
criterion that the resulting spherically symmetric amplitude should
not be many orders of magnitude smaller than in the case of the
parameters in the first column of Table~\ref{tab:scan-and-hsmin}.
Note, however, that the amplitudes in Fig.~\ref{fig:hsmin} are, in
fact, a whole order of magnitude smaller than in
Figs.~\ref{fig:scan:first}--\ref{fig:scan:last}.  This indicates how
important the constraints of relativity can be for analysis of data.
We should also add that the size of the coupling constant in the
interaction Hamiltonian that drives the decay was fixed as in
Ref.~\cite{GlazekSzczepaniak} and never changed in the fit.

\begin{table}[!htbp]
  \centering
  \def\horizorvertical{1}         
  \ifcase\horizorvertical
  \begin{tabular}{llll}
    \hline
    \hline
    param        & min   & max   \\
    \hline
    $m_q$        &0.1    &0.6    \\
    $m_g$        &0.5    &2.0    \\
    $\beta_p$    &0.1    &0.8    \\
    $\beta_b$    &0.1    &1.6    \\
    $\beta_{hg}$ &0.1    &2.0    \\
    $\beta_{hq}$ &0.1    &8.0    \\
    $\lambda$    &0.1    &8.0    \\         
    \hline
    \hline
  \end{tabular}
  \or
  \begin{tabular}{c*{7}l}
    \hline
    \hline
    param        &
    $m_q$        &
    $m_g$        &
    $\beta_p$    &
    $\beta_b$    &
    $\beta_{hg}$ &
    $\beta_{hq}$ &
    $\lambda$    \\
    \hline
    min    &
    0.1    &
    0.5    &
    0.1    &
    0.1    &
    0.1    &
    0.1    &
    0.1    \\
    max    &
    0.6    &
    2.0    &
    0.8    &
    1.6    &
    2.0    &
    8.0    &
    8.0    \\         
    \hline
    \hline
  \end{tabular}
  \fi
  \caption{\label{tab:limits} 
    The limits on parameters of the wave functions (in GeV)
    that we imposed using the Adaptive Simulated Annealing 
    algorithm~\cite{ASA}.}
\end{table}

It is clear that one should not consider an unbiased minimization of
symmetry violation as a most reasonable approach. The minimization
should include additional constraints, including restrictions such as
the radius of a meson $p$, and as much of the
dynamical constraints as possible. But in order to impose correlations
such as the ones coming from the radius, one has to be very careful
about how one calculates the radius and if that calculation does obey
requirements of special relativity, which is a problem in itself.
Concerning the dynamical constraints, we were not able and not even
interested in imposing any such constraints at this stage, because we
were only searching for the answer to the question if any choice of
the parameters could produce spherical symmetry, and it is interesting
that even a minimal model can produce the symmetry of the quality as
good as shown in Fig.~\ref{fig:hsmin}. We remind the reader that the
gluon spin introduces functions of momenta that vary rapidly with
angles and it was not clear at all that any choice of parameters could
lead to a constant amplitude. But once it is established that such
result is possible, one can make further observations based on a
systematic search through the space of the parameters.

\begin{figure}[!tbp]
\centering
\includegraphics[scale=\minscale,trim=95 0 5 75]{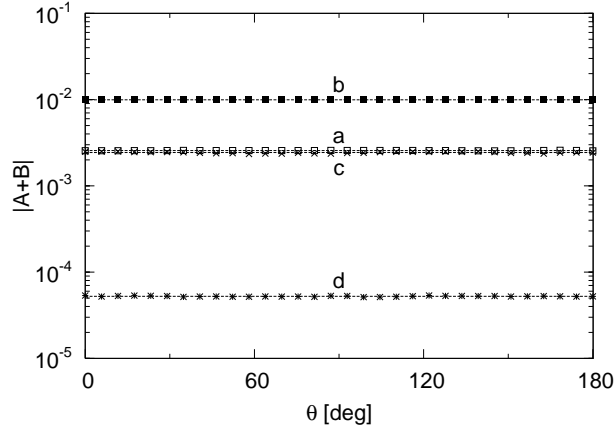}
\caption{\label{fig:ASA:devsqrp-1}\label{fig:ASA:first}
  The hybrid decay amplitude as function of $\theta$ for the decay
  into two $J^{PC} = 0^{++}$ mesons in different cases:
  a) \uev, $N \neq 1$, first 
  column of Table~\ref{tab:ASA:devsqrp-1},
  b) \uev, $N=1$, second column of
  Table~\ref{tab:ASA:devsqrp-1},
  c) \uGPv, $N \neq 1$, third column of
  Table~\ref{tab:ASA:devsqrp-1},
  d) \uGPv, $N=1$, fourth column of
  Table~\ref{tab:ASA:devsqrp-1}.}
\end{figure}
\begin{table}[!tbp]
\def\c#1{\multicolumn{1}{c}{#1}}
\centering
\begin{tabular}{lllll}
\hline
\hline
  Fig. \#     &\ref{fig:ASA:devsqrp-1}a 
              &\ref{fig:ASA:devsqrp-1}b 
              &\ref{fig:ASA:devsqrp-1}c 
              &\ref{fig:ASA:devsqrp-1}d \\
  \hline
  \c{spin}    &\uev    &\uev     &\uGPv   &\uGPv    \\
  \c{term}    &\fxx    &\nofxx   &\fxx    &\nofxx  \\
  \hline
  $m_h$       &1.9     &$s$      &$s$     &$s$     \\
  $m_b$       &1.235   &$s$      &$s$     &$s$     \\ 
  $m_p$       &0.1375  &$s$      &$s$     &$s$     \\
  $m_q$       &0.152   &0.21     &0.17    &0.155   \\
  $m_g$       &1.28    &1.07     &1.70    &1.90    \\
  $\beta_p$   &0.132   &0.219    &\m{0.1} &\m{0.1} \\
  $\beta_b$   &0.320   &0.536    &0.321   &0.371   \\
  $\beta_{hg}$&0.766   &1.05     &0.267   &0.263   \\
  $\beta_{hq}$&\M{8.0} &7.83     &4.59    &5.73    \\
  $\lambda$   &7.68    &6.13     &2.86    &7.98    \\
  \hline
  \hline
\end{tabular}
\caption{\label{tab:ASA:devsqrp-1} 
  The optimal parameters of the wave
  functions (in GeV) for a decay into two $J^{PC} = 0^{++}$ mesons.
  These are results of a global minimization using ASA~\cite{ASA}, 
  minimizing standard deviation from the average value of the amplitude, 
  for parameters within limits in Table~\ref{tab:limits}. The resulting 
  amplitudes are shown in Fig.~\ref{fig:ASA:devsqrp-1}.  
  The bold face numbers are on the limit of the
  allowed range.}
\end{table}

\begin{figure}[tbp!]
\centering
\includegraphics[scale=\minscale,trim=95 0 5 75]{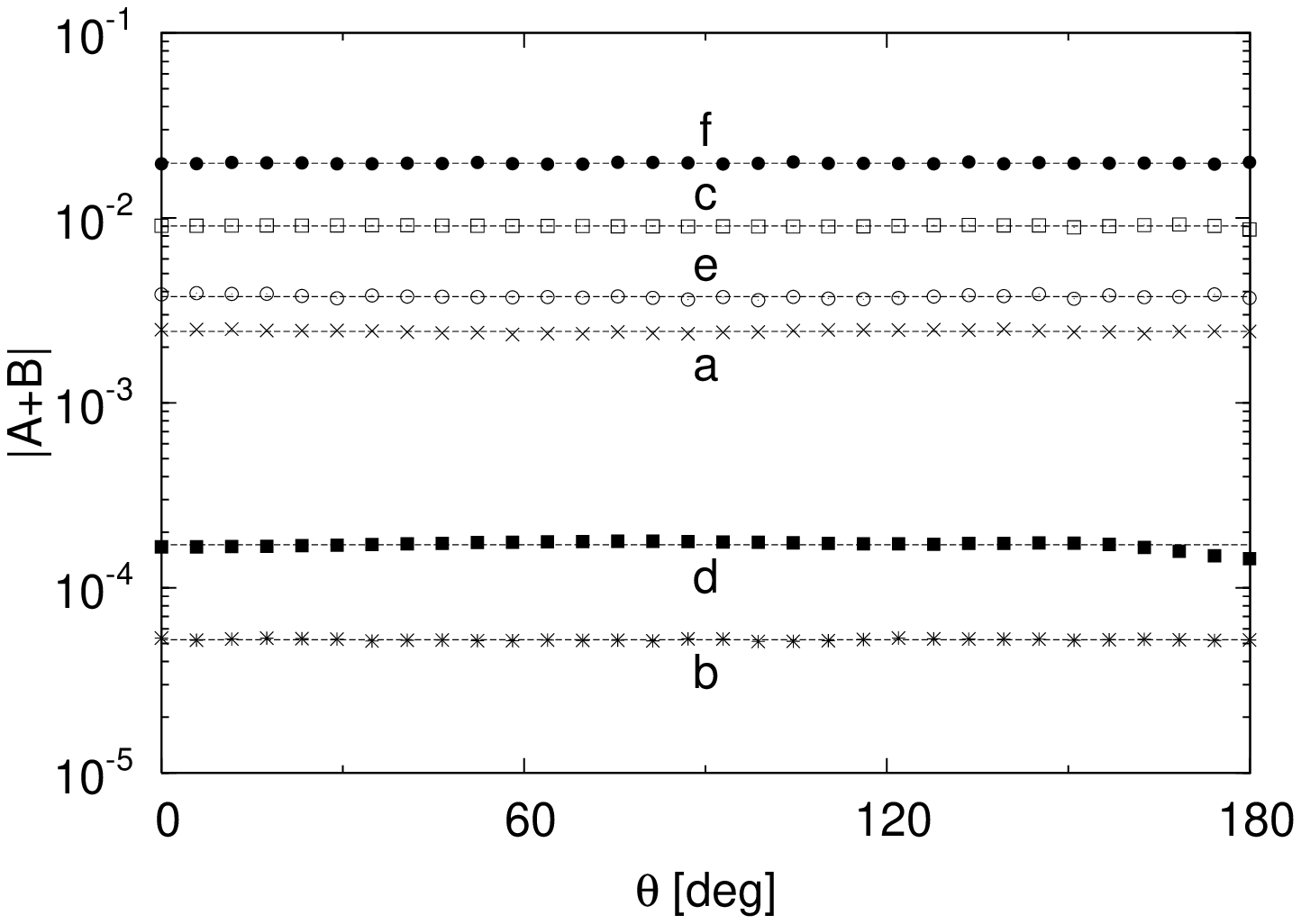}
\caption{\label{fig:ASA:uGPv:devsqrp-1}
  The hybrid decay amplitude as function of $\theta$ for the decay
  into two $J^{PC} = 0^{++}$ mesons in different cases:
  a) $\bar{u} G P v$ with $N \neq 1$,
  b) $\bar{u} G P v$ with $N  =   1$,
  c) $\bar{u} \tilde{G} P v$ with $N \neq 1$,
  d) $\bar{u} \tilde{G} P v$ with $N  =   1$,
  e) $\bar{u} \tilde{G} \tilde{P} v$ with $N \neq 1$,
  f) $\bar{u} \tilde{G} \tilde{P} v$ with $N  =   1$.
  Parameters are given in Table~\ref{tab:ASA:uGPv:devsqrp-1}.
}
\end{figure}
\begin{table}[tbp!]
\def\c#1{\multicolumn{1}{c}{#1}}
\centering
\begin{tabular}{lllllll}
\hline
\hline
  Fig. \#     &\ref{fig:ASA:uGPv:devsqrp-1}a 
              &\ref{fig:ASA:uGPv:devsqrp-1}b 
              &\ref{fig:ASA:uGPv:devsqrp-1}c 
              &\ref{fig:ASA:uGPv:devsqrp-1}d 
              &\ref{fig:ASA:uGPv:devsqrp-1}e 
              &\ref{fig:ASA:uGPv:devsqrp-1}f \\
  \c{spin}    &\uGPv   &\uGPv   &\ugPv   &\ugPv   &\ugpv   &\ugpv   \\
  \c{term}    &\fxx    &\nofxx  &\fxx    &\nofxx  &\fxx    &\nofxx  \\
  \hline
  $m_h$       &1.9     &$s$     &$s$     &$s$     &$s$     &$s$     \\
  $m_b$       &1.235   &$s$     &$s$     &$s$     &$s$     &$s$     \\
  $m_p$       &0.1375  &$s$     &$s$     &$s$     &$s$     &$s$     \\
  $m_q$       &0.17    &0.155   &0.373   &\m{0.1} &0.214   &0.268   \\
  $m_g$       &1.70    &1.90    &1.82    &\M{2.0} &1.82    &1.46    \\
  $\beta_p$   &\m{0.1} &\m{0.1} &0.286   &\m{0.1} &0.211   &0.411   \\
  $\beta_b$   &0.321   &0.371   &0.209   &1.23    &0.365   &0.470   \\
  $\beta_{hg}$&0.267   &0.263   &0.565   &1.02    &0.244   &0.434   \\
  $\beta_{hq}$&4.59    &5.73    &4.44    &\M{8.0} &7.60    &3.22    \\
  $\lambda$   &2.86    &7.98    &3.57    &\M{8.0} &4.11    &3.86    \\
  \hline
  \hline
\end{tabular}
\caption{\label{tab:ASA:uGPv:devsqrp-1} 
  The optimal parameters of the wave
  functions (in GeV) for a decay into two $J^{PC} = 0^{++}$ mesons.
  These are results of global minimization using ASA~\cite{ASA}, 
  minimizing standard deviation from the average value of the amplitude, 
  for parameters within limits in Table~\ref{tab:limits}. The resulting 
  amplitudes are shown in Fig.~\ref{fig:ASA:uGPv:devsqrp-1}.  
  The bold face numbers are on the limit of the
  allowed range.}
\end{table}

\begin{figure}[tbp!]
\centering
\includegraphics[scale=\minscale,trim=95 0 5 75]{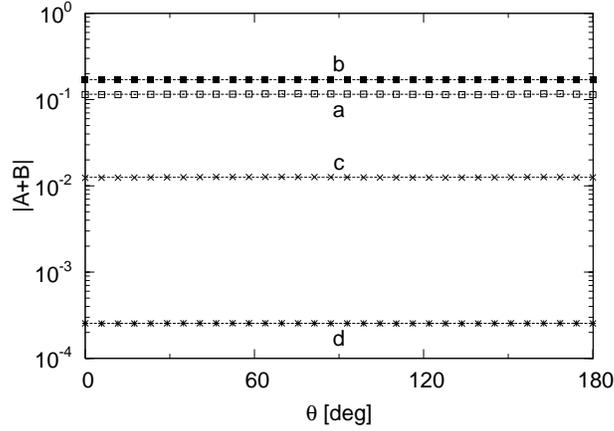}
\caption{\label{fig:ASA:u5v:devsqrp-1}
  The hybrid decay amplitude as a function of $\theta$, the same as in
  Fig.~\ref{fig:ASA:devsqrp-1}, but for a decay into two $J^{PC} =
  0^{-+}$ mesons (spin factor $\bar{u}\gamma^5 v$).  The optimal 
  wave function parameters are given in the correspondingly marked 
  columns of Table~\ref{tab:ASA:u5v:devsqrp-1}.}
\end{figure}
\begin{table}[tbp!]
\def\c#1{\multicolumn{1}{c}{#1}}
\centering
\begin{tabular}{lllll}
  \hline
  \hline
  Fig. \#     &\ref{fig:ASA:u5v:devsqrp-1}a 
              &\ref{fig:ASA:u5v:devsqrp-1}b 
              &\ref{fig:ASA:u5v:devsqrp-1}c 
              &\ref{fig:ASA:u5v:devsqrp-1}d \\
  \hline
  \c{spin}    &\uev    &\uev     &\uGPv    &\uGPv    \\
  \c{term}    &\fxx    &\nofxx   &\fxx    &\nofxx  \\
  \hline
  $m_h$       &1.9     &$s$      &$s$     &$s$     \\
  $m_b$       &1.235   &$s$      &$s$     &$s$     \\ 
  $m_p$       &0.1375  &$s$      &$s$     &$s$     \\
  $m_q$       &0.15    &0.16     &0.18    &0.19    \\
  $m_g$       &1.08    &0.88     &1.69    &1.79    \\
  $\beta_p$   &0.21    &0.25     &0.16    &0.22    \\
  $\beta_b$   &0.59    &0.59     &0.42    &0.30    \\
  $\beta_{hg}$&0.59    &0.72     &0.68    &0.47    \\
  $\beta_{hq}$&2.40    &2.62     &6.96    &6.80    \\
  $\lambda$   &3.97    &3.71     &2.29    &2.39    \\
  \hline
  \hline
\end{tabular}
\caption{\label{tab:ASA:u5v:devsqrp-1} 
  The optimal parameters of the wave functions (in GeV) for 
  a decay into two $J^{PC} = 0^{-+}$ mesons. 
  These are results of global minimization using ASA~\cite{ASA}, 
  minimizing standard deviation from the average value of the 
  amplitude, for parameters within limits in Table~\ref{tab:limits}. 
  The resulting amplitudes are shown in Fig.~\ref{fig:ASA:u5v:devsqrp-1}.
}
\end{table}

\begin{figure}[tbp!]
\centering
\includegraphics[scale=\minscale,trim=95 0 5 75]{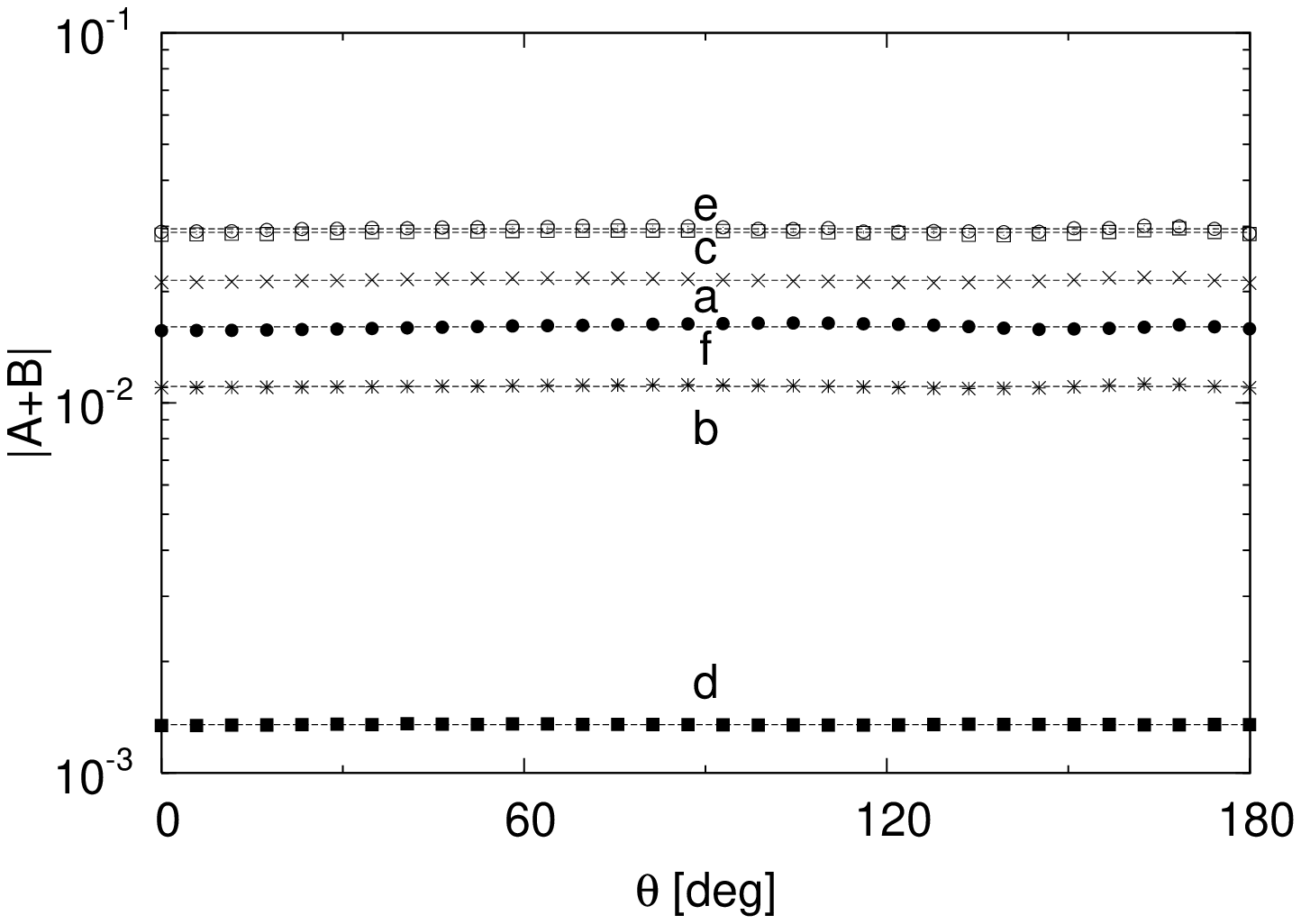}
\caption{\label{fig:ASA:uGPv_m5:devsqrp-1}
  The hybrid decay amplitude as function of $\theta$ for the decay
  into two pseudoscalar $J^{PC} = 0^{-+}$ mesons in different cases:
  a) $\bar{u} G P v$ with $N \neq 1$,
  b) $\bar{u} G P v$ with $N  =   1$,
  c) $\bar{u} \tilde{G} P v$ with $N \neq 1$,
  d) $\bar{u} \tilde{G} P v$ with $N  =   1$,
  e) $\bar{u} \tilde{G} \tilde{P} v$ with $N \neq 1$,
  f) $\bar{u} \tilde{G} \tilde{P} v$ with $N  =   1$.
  Parameters are given in Table~\ref{tab:ASA:uGPv_m5:devsqrp-1}.
}
\end{figure}
\begin{table}[tbp!]
\def\c#1{\multicolumn{1}{c}{#1}}
\centering
\begin{tabular}{lllllll}
\hline
\hline
  Fig. \#     &\ref{fig:ASA:uGPv_m5:devsqrp-1}a 
              &\ref{fig:ASA:uGPv_m5:devsqrp-1}b 
              &\ref{fig:ASA:uGPv_m5:devsqrp-1}c 
              &\ref{fig:ASA:uGPv_m5:devsqrp-1}d 
              &\ref{fig:ASA:uGPv_m5:devsqrp-1}e 
              &\ref{fig:ASA:uGPv_m5:devsqrp-1}f \\
  \c{spin}    &\uGPv   &\uGPv   &\ugPv   &\ugPv   &\ugpv   &\ugpv   \\
  \c{term}    &\fxx    &\nofxx  &\fxx    &\nofxx  &\fxx    &\nofxx  \\
  \hline
  $m_h$       &1.9     &$s$     &$s$     &$s$     &$s$     &$s$     \\
  $m_b$       &1.235   &$s$     &$s$     &$s$     &$s$     &$s$     \\
  $m_p$       &0.1375  &$s$     &$s$     &$s$     &$s$     &$s$     \\
  $m_q$       &0.26    &0.42    &0.29    &0.18    &0.31    &0.31    \\
  $m_g$       &1.34    &1.85    &1.21    &1.62    &1.22    &0.61    \\
  $\beta_p$   &0.21    &0.21    &0.25    &0.19    &0.28    &0.54    \\
  $\beta_b$   &0.48    &0.21    &0.59    &0.40    &0.58    &0.56    \\
  $\beta_{hg}$&0.57    &1.40    &0.47    &0.57    &0.43    &0.96    \\
  $\beta_{hq}$&7.53    &6.88    &7.80    &4.75    &7.99    &4.58    \\
  $\lambda$   &2.26    &7.66    &2.41    &2.71    &2.37    &2.61    \\
  \hline
  \hline
\end{tabular}
\caption{\label{tab:ASA:uGPv_m5:devsqrp-1} 
  The optimal parameters of the wave functions (in GeV) for a decay
  into two $J^{PC} = 0^{-+}$ mesons.  These are
  results of global minimization using ASA~\cite{ASA},  minimizing
  standard deviation from the mean amplitude for parameters within
  limits in Table~\ref{tab:limits}. 
  The resulting amplitudes are shown in Fig.~\ref{fig:ASA:uGPv_m5:devsqrp-1}.  
}
\end{table}

The simplest and least restrictive way of setting bounds on the
parameters of the models we test is to limit all of the parameters 
to fixed intervals around values that are considered reasonable.  
Such least restrictive parameter bounds adopted in the minimizations 
described below are given in Table~\ref{tab:limits}.

We performed a global minimization of departures from rotational
symmetry using different measures of how much a decay amplitude
differs from a constant as a function of the angle $\theta$: a standard
deviation from the average value (sum of squares of deviations from
the average value, labeled ``stddev'' in Appendix~\ref{sec:examples}), 
or maximum of the modulus of the deviations from the average value 
(labeled ``maxdev'' in the Appendix~\ref{sec:examples}), both measured 
relative to the average. Our global minimization within the assumed 
bounds is done using Adaptive Simulated Annealing (ASA)~\cite{ASA}.

Fig.~\ref{fig:ASA:devsqrp-1} shows results obtained using
Eq.~\eqref{uev} for the spin factor of a hybrid meson: curve ``a'' 
in case $N \neq 1$ and ``b'' in case $N = 1$, and using Eq.~\eqref{P123}: 
curve ``c'' in case $N \neq 1$ and ``d'' in case $N = 1$, all 
cases for a decay into two scalar ($\bar{u} v$) mesons. The factors
$N$ have considerable impact on the magnitude of the amplitudes
and can compensate or dramatically enhance the effect of changing
the spin factors. This means that one should probably not trust 
models of hybrids that are based solely on non-relativistic 
intuitions.

Results for decays into two $J^{PC} = 0^{-+}$ mesons ($\bar{u}\gamma^5
v$ spin factor) are shown in Fig.~\ref{fig:ASA:u5v:devsqrp-1}. The
corresponding values of the wave function parameters are given in
Table~\ref{tab:ASA:u5v:devsqrp-1}. Similarly, Figs.~\ref{fig:ASA:uGPv:devsqrp-1}
and~\ref{fig:ASA:uGPv_m5:devsqrp-1} show results for alternate
choices of hybrid spin factor: cases referred as $\uGPv$, $\ugPv$ 
and $\ugpv$.

In Ref.~\cite{GlazekSzczepaniak}, there were
found two locally best sets of values of
parameters in each of the two cases: one case
with all constituents being scalars, and
another one with fermionic quarks and a
scalar gluon. These good sets were
characterized by either light or heavy mass
of the quarks (Table~I and Fig.~5 in Ref.
\cite{GlazekSzczepaniak}). In our studies,
including the spin of the gluon, the complete
ASA algorithm finds only one best set of the
parameters that minimizes deviation from
rotational symmetry in every case we
consider. These best sets appear with small
quark masses and small $\beta_p$. Such small $\beta_p$ 
implies a too large size of the meson $p$, apparently
corresponding to a too weak binding of too
light quarks, as if the number of such light
effective constituents could not be only
minimal. But we can find different minima
when we impose an additional restriction
that the quark mass, $m_q$, is ``heavy'',
i.e. greater than 300 MeV. Other parameters
are still limited to the intervals given in
Table~\ref{tab:limits}.

\begin{figure}[!tbp]
\centering
\includegraphics[scale=\minscale,trim=95 0 5 75]{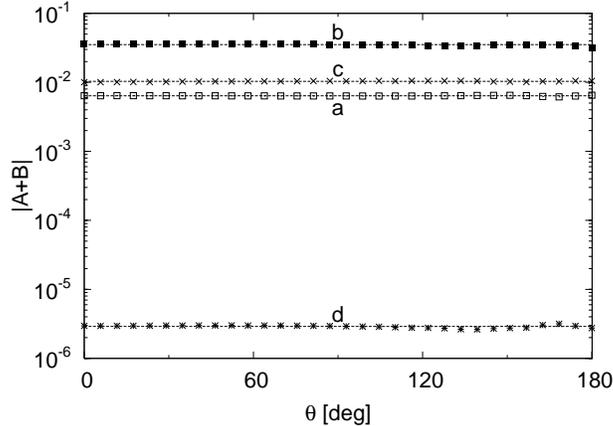}
\caption{\label{fig:ASA:devsqrp-2}\label{fig:ASA:last}
  The hybrid decay amplitude, like in Fig.~\ref{fig:ASA:devsqrp-1},
  for the decay into two scalar mesons, as function of $\theta$.  
  Parameters are given in the corresponding columns of
  Table~\ref{tab:ASA:devsqrp-2}: the lower limit on the quark
  mass is $m_q \geq 300\;\text{MeV}$.
}
\end{figure}
\begin{table}[!tbp]
\def\c#1{\multicolumn{1}{c}{#1}}
\centering
\begin{tabular}{lllll}
\hline
\hline
Fig. \#   &\ref{fig:ASA:devsqrp-2}a &\ref{fig:ASA:devsqrp-2}b & 
           \ref{fig:ASA:devsqrp-2}c &\ref{fig:ASA:devsqrp-2}d \\
\hline
\c{spin}  &\uev    &\uev     &\uGPv    &\uGPv    \\
\c{term}  &\fxx    &\nofxx   &\fxx    &\nofxx  \\
\hline
$m_h$       &1.9     &$s$      &$s$     &$s$     \\
$m_b$       &1.235   &$s$      &$s$     &$s$     \\ 
$m_p$       &0.1375  &$s$      &$s$     &$s$     \\
$m_q$       &\m{0.3} &0.459    &\m{0.31}&\m{0.31}\\
$m_g$       &\M{1.95}&1.37     &1.90    &1.89    \\
$\beta_p$   &0.211   &0.353    &0.344   &\m{0.1} \\
$\beta_b$   &0.295   &0.722    &0.263   &0.169   \\
$\beta_{hg}$&0.894   &0.754    &0.450   &1.21    \\
$\beta_{hq}$&5.83    &7.75     &7.69    &\M{8.0} \\
$\lambda$   &4.67    &7.75     &4.44    &7.96    \\
\hline
\hline
\end{tabular}
\caption{\label{tab:ASA:devsqrp-2} 
  The optimal parameters of the wave functions (in GeV) from a global
  minimizaton using ASA~\cite{ASA}, minimizing standard deviation 
  relative to the mean amplitude with all parameters within limits
  like in Table~\ref{tab:limits} except for the lower limit on the quark
  mass, $m_q \geq 300\;\text{MeV}$. 
  The resulting amplitudes are shown in Fig.~\ref{fig:ASA:devsqrp-2}. 
}
\end{table}

An example of such restricted minimization
for ``heavy'' effective constituent quarks is
shown in Fig.~\ref{fig:ASA:devsqrp-2}. The
corresponding optimal parameters are given in
Table~\ref{tab:ASA:devsqrp-2}. In all cases
except the case ``d,'' the size of the light
meson $p$ is now much closer to the size of
the real mesons $\pi$. This shows that the
size (radius) of the light meson may not be as 
big an issue as one might think on the basis 
of a search for the best parameters allowing
quarks to be much lighter than 300 MeV (this
case was discussed earlier). 

Note that the assumption of case ``d,'' that the 
spin factor in the hybrid wave function contains 
a four-momentum of a massive gluon, leads to a 
small optimal quark mass and a very small decay 
amplitude. The constraints of rotational symmetry 
promise to be very useful in future studies of
reasonable models because when they are combined 
with inspection of observables such as radii one 
immediately obtains large differences between 
predictions based on different models.

\section{Discussion}
\label{sec:discussion}

First of all, let us note that the inclusion of spin 
of a constituent gluon does not change the previously 
obtained result for scalar ``gluons'' \cite{GlazekSzczepaniak}
that rotational symmetry is restored when the quark-antiquark 
pair momentum-space width in the hybrid, $\beta_{hq}$, is 
about the same in size as the width $\lambda$ in the vertex 
form factor in the renormalized interaction Hamiltonian
$H_\lambda$ in LF QCD, and both are on the order of 4-6 GeV, 
much larger than all other parameters. On the basis of our
study of many cases with different ways of including the
gluon spin and different cases of meson spin factors, we
can state that the required relative momentum-space structure 
of the wave functions appears to be qualitatively independent 
of the spin of the effective constituents. Quite generally, 
the parameters $\beta_{hq}$ and $\lambda$ appear to
have to be about 5 times larger than the other parameters, 
see Tables~\ref{tab:scan-and-hsmin} 
and~\ref{tab:ASA:devsqrp-1}--~\ref{tab:ASA:devsqrp-2}.  
This result suggests that the quark pair should be thought 
about as spatially small in comparison to the size of the 
hybrid. It also suggests that the pair could originate from 
a gluon that belonged to a gluonium before the interaction 
changed one gluon into the pair.

One may worry that this result is obtained by minimizing 
just one observable in a space of seven parameters. Here 
comes the strength of the continuous symmetry condition 
on a reasonable model: sepcial relativity provides infinitely
many conditions instead of just one -- the just one decay 
amplitude must not be a function of the angle. It was highly 
questionalble that it was possible to even come close to a 
constant function of the angle before we carried our study
including the singular spin factors for gluons. The remarkable 
fact is that the symmetry cannot be satisfied with a minimal
constituent picture unless the hybrid built from a pair and a 
constituent gluon looks differently than expected assuming
that the gluons follow quarks and this picture does not 
seem to depend on any particular detail but only on the major 
assumption that a minimal constituent model can approximate 
the effective LF dynamics.

Another worry concerns the small spatial size of the 
$q \bar q$-pair, about 4 to 5 times smaller than the 
distance between the gluon and the octet diquark, 
correlated with $\lambda$ on the order of 3 or even 5 
GeV. The real question is for what values of $\lambda$
a constituent picture of hybrids may work. In principle,
if RGPEP equations were solved exactly, no physical result
should depend on $\lambda$ and no matter what $\lambda$
is used one should obtain rotationally symmetric decay
amplitude provided that the decay is calculated exactly 
using exact solutions for the participating hadrons. But 
we know \cite{GlazekMlynik} that in order to approximate
the full dynamics of an asymptotically free theory by a 
simple picture one has to lower $\lambda$ to values that 
are about twice above the scale of eigenvalues one is 
seeking to describe. One cannot lower $\lambda$ to smaller
values using perturbation theory in RGPEP because the 
resulting $H_\lambda$ would begin to contain too large 
errors due to cutting into the mechanism of binding.
Thus, the scale we obtain from the heuristic fit is
quite reasonable. On the other hand, one may worry that 
no tight diquark clusters are seen in the proton deep 
inelastic structure. One possible explanation of such
special feature of hybrids could be that they contain
octet diquarks interacting with constituent gluons,
while in the proton we have primarily triplets and 
diquark antitriplet and no counterpart of the constituent
gluon structure. Our symmetry study in hybrids should
be seen as groping into the sectors of effective color 
dynamics that are squeezed out of and cannot be seen 
in nucleons. 

There exist small differences between the case
of decays into two scalar and two pseudoscalar mesons,
and between various choices for handling the spin of 
an effective gluon in the hybrid. But the outstanding 
feature that $\beta_{hq}$ is about the same as $\lambda$ 
and both are larger than the rest of parameters is common 
to all the cases we studied. Also, the mass of the gluon, 
$m_g$, appears to have to be much greater than the masses 
of the quarks, $m_q$.  

The hybrid spin wave function from Eq.~\eqref{P123}, 
inspired by the lattice operators but with a massive 
four-momentum for the gluon, gives minima with a much 
smaller decay amplitude $\A$ than the simplest hybrid 
spin wave function of Eq.~\eqref{uev}. This result 
shows that one has to be very careful about treatment 
of spin of gluons in model building. 

Let us stress that the absolute size of the
amplitude is not under control in our 
calculation because we did not include the
dynamical constraints between the coupling
constant $g$, the RGPEP parameter $\lambda$, 
and the wave function parameters. But there 
exists a systematic trend in all our results,
which requires further study and is not
understood here. Namely, when the parameters 
of the wave functions are varied from the values 
in the first column of Table~\ref{tab:scan-and-hsmin}, 
to the values required by rotational symmetry, 
like in columns 2--5 of Table~\ref{tab:scan-and-hsmin}, 
or the values in 
Tables~\ref{tab:ASA:devsqrp-1}--\ref{tab:ASA:uGPv_m5:devsqrp-1}, 
and~\ref{tab:ASA:devsqrp-2}, the size of the 
amplitude changes from about 0.9 to 0.1
in Figs.~\ref{fig:scan:first}--\ref{fig:scan:last} to
the much smaller values of $10^{-2}$ or
$10^{-3}$ in Fig.~\ref{fig:hsmin}, or $10^{-2}$ 
or $10^{-4}$ in Figs.~\ref{fig:ASA:first} to~\ref{fig:ASA:last}.

These are considerable changes in the order
of magnitude. It seems unlikely that the
coupling constant can vary by that much and
compensate this change. The change is so
large only because of the relativistic motion
of the light outgoing meson. If the outgoing
mesons are slow, rotational symmetry in our
model is respected very accurately for
typical values of parameters in non-relativistic
constituent models, exemplified in the first 
column of Table~\ref{tab:scan-and-hsmin}. We 
are forced to conclude that a relativistic hybrid 
decay (including fast mesons $p$) may involve 
relativistic effects that are not accounted
for in the non-relativistic phenomenology,
cf.~\cite{CloseDudek}.

Another feature worth mentioning is that the
minima for the wave functions motivated by
lattice operator structures, Eq.~\eqref{P123}, 
are narrower than in the case of Eq.~\eqref{uev}. 
If the range of parameters for which a violation 
of rotational invariance is small is very
narrow, the symmetry itself becomes a source 
of detailed information about the necessary
values of the parameters even if the
corresponding dynamical equations are too
difficult to solve with comparable precision. 
By the same token, one obtains a very strict 
criterion for judgement of dynamical models
that attempt to produce the relevant wave 
functions.

\section{Conclusion}
\label{sec:concl}

The example of a simple model described here
shows that the wave function parameters for
hadrons involved in a relativistic decay of a
hybrid must be strongly correlated in order
that the decay amplitude satisfies
requirements of special relativity. Thanks to
the use of the LF scheme, boost symmetry is
respected exactly and the parameters are
constrained by the condition of rotational
symmetry. In the example, they have to take
values that do not correspond to the picture
based on the non-relativistic intuition that
the gluons are mainly between two quarks.
Instead, the relativistic effective constituent 
picture almost universally points toward the 
structure in which a heavy gluon is accompanied 
by a quark-antiquark pair that resembles a
relatively small octet diquark. This is a 
stunning result because it suggests that the
picture with gluons playing a role of a relatively 
light chain, or a vibrating flux, or string between 
relatively heavy quarks may be not as realistic 
as one hopes for on the basis of non-relativistic
intuition. 

The alternative hybrid structure occurs 
in a variety of cases that differ in details 
of the spin factors for gluons and quarks. 
But the requirement of relativistic symmetry 
turns out to be very restrictive when one demands
that only sectors with the smallest possible
number of constituents are important.
Therefore, we conclude that the effective
constituent dynamics in QCD should be always
considered including constraints of special
relativity. These constraints appear capable
of forcing us to consider hadrons with
significant gluon content not as if the
gluons were just added to quarks and antiquarks,
but as if gluons could actually dominante the
dynamics of hybrids and force the quarks to
adjust.

Since the hybrid structure we are forced to
seriously consider by the results of this 
analysis contains a spatially tight octet 
diquark pair, as if the pair emerged from a 
constituent gluon through a single interaction 
in an effective LF QCD, one may ask if it is 
possible that such effective quark-antiquark-gluon 
states with a tight pair can mediate decays of
usual mesons. A decay of a usual meson may 
proceed by an emission of a gluon from one quark
and subsequent decay of the gluon into a new pair
of quarks. The emerging two quarks and two antiquarks
can form the mesons that are produced in the decay
of the usual meson. But if the pair accompanied
by the intermediate gluon has to be small in size, 
as if the three effective particles had to form 
a structure similar to our finding for a hybrid,
the intermediate quark configuration would have 
to have a small overlap with the initial usual  
meson configuration. The decay mechanism through
an intermediate hybrid meson would have small
contribution to the total strong decay width.
Would not this width be too small if the hybrid 
structure were as we obtain? Not necessarily, since 
in the effective theory there must exist other 
interactions that are capable of producing four
effective quarks from two effective quarks. These
interactions do not correspond to the intermediate 
excitation of a massive effective gluon and they 
are not characterized by the coupling of such gluons 
to quark-antiquark pairs. Examples of such interactions
are present already in the canonical Hamiltonian
in LF QCD. The canonical interactions are not mediated 
by emission or absorption of gluons, and they must 
contribute to the mechanism of strong decay of usual 
mesons in the effective theory characterized by width 
$\lambda$ on the order of hadronic masses. In addition,
the RGPEP procedure generates more interactions that 
can turn a quark-antiquark pair into two such pairs 
without explicit creation and decay of a massive
effective gluon corresponding to small $\lambda$. 
Unfortunately, our study is not telling us anything
about the dynamical structure of ordinary mesons 
and interactions that mediate their decays. It is 
limited to a preliminary study of symmetry constraints 
in simplest models with a dominant hybrid component. 

The wave function parameters that we find to be 
preferred by the condition of rotational symmetry 
of the decay amplitude of a model hybrid, may turn 
out to be invalid when the actual dynamics is 
included in the analysis. For example, it may turn 
out that the approximation by the Fock sectors with 
only the smallest possible number of constituents 
does not apply. But it is clear that the relativistic 
constraints cannot be ignored in the search
for a leading constituent picture.

Our discussion was limited to $0^{++}$ hybrids for
simplicity, while the most interesting from practical point
of view is the structure of exotics~\cite{lattice, lattice2,
charmoniumhybrids, massesinmodels, Kalashnikova}. One can
change factors in the wave functions and change the quantum
numbers of the hybrid states to the exotic values. For
example, one can replace the color electric field by a color
magnetic field, or introduce $p$-wave wave functions. Simple
introduction of $p$-wave for gluon (introducing a factor of
$\vec k_g$ in the hybrid wave function) changes the hybrid
states we consider to the $J^{PC} = 1^{-+}$ exotic hybrid
mesons, those of most interest experimentally. In such
cases, the decay amplitudes calculated in a LF scheme should
exhibit the required angular dependence in the CMF of an
exotic hybrid. In calculations using the standard form of
dynamics, one should amake sure that the boost symmetry is
respected. However, already on the basis of our analysis of
the non-exotic $0^{++}$ hybrid decays, we suggest that no
matter what scheme one uses, a complete set of constraints
of special relativity should be seriously taken into account
in searches for a suitable constituent picture.

\subsubsection*{Acknowledgements}
The authors acknowledge several helpful discussions with 
Adam Szczepaniak and Nikodem Pop{\l}awski. 
This work was supported in part by MEiN BST-975/BW-1640.

\appendix

\section{Basic definitions}
\label{sec:appendix}
Spinors we use are defined as
  \begin{align}
  \label{eq:u-and-v-mpl}
    u_{mp\lambda} &= B(p,m)\, u_{0\lambda}, &
    v_{mp\lambda} &= B(p,m)\, v_{0\lambda},
  \end{align}
where the operator $B(p,m)$
\begin{equation}
  \label{eq:Bkm}
  B(p,m) = \frac{1}{\sqrt{m p^{+}}}
  \bigl[ \Lambda_+ p^+ 
  + \Lambda_- (m + \alpha^{\perp} p^{\perp}) \bigr]
\end{equation}
represents a boost that changes the mass $m$
at rest into the four-momentum $p$. Spinors 
of fermions at rest $u_{0\lambda}$ are 
  \begin{align}
  \label{eq:u}
    u_{0\uparrow} &= \sqrt{2m}
    \begin{pmatrix}
      \chi_{+} \\ 0 
    \end{pmatrix} \, , &
    u_{0\downarrow} &= \sqrt{2m}
    \begin{pmatrix}
      \chi_{-} \\ 0 
    \end{pmatrix} \, ,  
  \end{align}
and for anti-fermions at rest $v_{0\lambda}$ are
  \begin{align}
    \label{eq:v}
    v_{0\uparrow} &= \sqrt{2m}
    \begin{pmatrix}
       0 \\ \chi_{-}  
    \end{pmatrix} \, , &
    v_{0\downarrow} &= \sqrt{2m}
    \begin{pmatrix}
      0 \\ - \chi_{+} 
    \end{pmatrix} \, , 
  \end{align} 
where $\chi_{\pm}$ are two-component spinors, 
$\chi_{+} = (1, 0)$ and $\chi_{-} = (0, 1)$.
We use the convention $\Lambda_\pm = \frac{1}{2}\gamma_0 \gamma^\pm$, 
where $\gamma^{\pm} = \gamma^0 \pm \gamma^3$.

The integration measure over momenta in a meson, denoted by 
$[k_1 k_2]$ or just $[1 2]$, is 
\begin{equation}
  \label{eq:dk1dk2}
  \frac{d\!k^+_1 d^2\!k^\perp_1}{2 (2 \pi)^3 k^+_1}
  \frac{d\!k^+_2 d^2\!k^\perp_2}{2 (2 \pi)^3 k^+_2}
  = 
  \frac{d\!x_{12}   d^2\!k^\perp_{12}}{2 (2 \pi)^3 x_{12}(1-x_{12})} 
  \frac{d\!P^+_{12} d^2\!P^\perp_{12}}{2 (2 \pi)^3 P^+_{12}} \, .
\end{equation}
In terms of the three-vector $\vec k_{12}$, the integration measure 
for two constituents with the same mass $m_q$, is given by
\begin{equation}
  \label{eq:dveck}
  \frac{d\!x_{12} d^2\!k^\perp_{12}}{2 (2 \pi)^3 x_{12}(1-x_{12})} 
  =
  \frac{4 d^3\!\vec{k}_{12}}{2 (2 \pi)^3 \mathcal{M}_{12}} \; .
\end{equation}
Therefore, 
\begin{equation}
  \label{eq:f_pm}
  N_{pm}({\vec k}_{12}) =
  \sqrt{\frac{\mathcal{M}_{12}}{2 m_q}} \, .
\end{equation}
Similarly,
\begin{equation}
  \label{eq:f_ps}
  N_{ps}({\vec k}_{12}) =
  \sqrt{\frac{1}{\Tr\left[S_p^\dag(1,2) S_p(1,2)\right]}} \;.
\end{equation}

In the normalization equation for a hybrid meson, 
we have a three-particle integration measure
\begin{equation}
  \label{eq:dvkqdvkg}
  \prod_{i=1}^3 \frac{dk^+_i d^2k^\perp_i}{16\pi^3 k^+_i} 
  = 
  \frac{4 d^3\!k_q}{2 (2 \pi)^3 \mathcal{M}_q}
  \frac{d^3\!k_g \mathcal{M}_{qg}}
  {2 (2 \pi)^3 \sqrt{m_g^2 + k_g^2} \,
\sqrt{\mathcal{M}_q^2 + k_g^2}} \; .
\end{equation}
Therefore,
\begin{equation}
  \label{eq:f_hm}
  \begin{split}
    N_{hs}(\vec{k}_q, \vec{k}_g)
    =
    \sqrt{\frac{\mathcal{M}_q}{2 m_q}} \,\,
    \sqrt{\frac{2 m_q + m_g}{\mathcal{M}_{qg}}} \,
    \sqrt{\frac{\sqrt{m_g^2 + \vec{k}^{\,2}_g}}{m_g}} \,\,
    \sqrt{\frac{\sqrt{\mathcal{M}_q^2 +
          \vec{k}^{\,2}_g}}{2 m_q}} \, ,
  \end{split}
\end{equation}
and
\begin{equation}
  \label{eq:N_hs}
  N_{hs}(\vec{k}_q, \vec{k}_g) =
  \sqrt{\frac{1}{\sum\limits_{\text{pol}}\Tr\left[S_h^\dag(1,2,3) S_h(1,2,3)\right]}} \; ,
\end{equation}
where $\sum_{\text{pol}}$ means a sum over two transverse 
polarizations of a gluon.

\begin{figure}[!htb]
  \centering
  \includegraphics[scale=0.43,viewport=0  110 630 430]{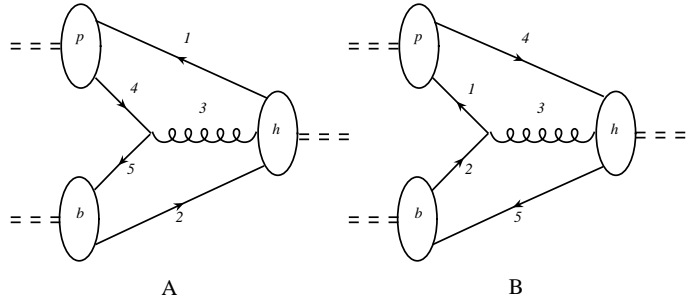}
  \caption{\label{fig:decay} 
    Hybrid meson decay amplitude into two non-exotic mesons $p$ (the
    light one) and $b$ (the heavy one).}
\end{figure}
The decay amplitude of a hybrid into two mesons is 
\begin{align*}
  &\mathcal{A}(p,b,h) = 
  (-1) \frac{2}{3} \frac{1}{\sqrt{2}} \frac{g_\lambda}{(16 \pi^3)^2} 
  \int \frac{dx_{14} d^2\kappa^\perp_{14}}{x_{14}(1-x_{14})}
  \int \frac{dx_{52} d^2\kappa^\perp_{52}}{x_{52}(1-x_{52})} 
  \\ 
  & \times N_p N_b N_h \psi^*_p(1,4) \psi^*_b(5,2) 
  \left[ \mathfrak{A}(1,2,3,4,5) + \mathfrak{B}(1,2,3,4,5) \right]  
  \\
  &= 
  - \frac{16}{3} \frac{g_\lambda}{(16 \pi^3)^2} 
  \int \frac{d^3 k_{52}}{{\cal M}_b} 
  \int \frac{d^3 k_{14}}{{\cal M}_p} 
  N_p N_b N_h \psi^*_p(\vec k_{14}) \psi^*_b(\vec k_{52}) \\
  & \times \left[ \mathfrak{A}(1,2,4,5) + \mathfrak{B}(1,2,4,5) \right] \; ,
\end{align*}
where 
\begin{align}
  \label{eq:AA(1,2,4,5)}
  \mathfrak{A}(1,2,4,5) &= \left.
  \frac{1}{x_3} T_A(1,2,3,4,5) A(1,2,3,4,5)
  \right\rvert_{k_3 = k_4 + k_5} \, ,
  \\
  \label{eq:BB(1,2,4,5)}
  \mathfrak{B}(1,2,4,5) &= \left.
  \frac{1}{x_3} T_B(1,2,3,4,5) B(1,2,3,4,5)
  \right\rvert_{k_3 = k_1 + k_2} \, ,
\end{align}
\begin{align}
  \label{eq:A(1,2,3,4,5)}
  A(1,2,3,4,5) &= 
  \psi_h(1,2,3) f_\lambda({\cal M}^2_{45}) \; ,
  \\
  \label{eq:B(1,2,3,4,5)}
  B(1,2,3,4,5) &= 
  \psi_h(5,4,3) f_\lambda({\cal M}^2_{12}) \; ,
\end{align}
and the spin factors in the decay amplitude
are
\begin{align}
  \label{eq:T_A(1,2,3,4,5)}
  T_A(1,2,3,4,5) &=
  \Tr\left[ S^\dagger_p(1,4) S_h(1,2,3)
            S^\dagger_b(5,2) S_{QCD}(5,4,3) \right] \, ,
  \\
  \label{eq:T_B(1,2,3,4,5)}
  T_B(1,2,3,4,5) &=
  \Tr\left[ S^\dagger_p(1,4) S_{QCD}(1,2,3)
            S^\dagger_b(5,2) S_h(5,4,3) \right] \, .
\end{align}
Parts A and B refer to the two arrangements of quarks 
shown in Fig.~\ref{fig:decay}. $S_{QCD}$ is the spin factor 
coming from the interaction  Hamiltonian of QCD: 
\begin{equation}
  \label{eq:S_QCD}
  \chi^\dag_{s_1} S_{QCD}(1,2,3) \chi^{\vphantom{\dag}}_{s_2} 
  = \bar{u}_1 \varepsilon^\mu_3 \gamma_{\mu} v_2 \, .
\end{equation}

\smallskip
In both parts of the amplitude, $A$ and $B$, one 
has $x_p = p^+/h^+$, $x_b = b^+/h^+ = 1 - x_p$.
In meson $p$, one has $\vec{k}_{14} \equiv \vec{k}_p$, 
so that ${\cal M}_{14} \equiv {\cal M}_p = 2 \tsqrt{m_q^2 + \vec{k}_p^{\,2}}$,
and the following relations hold: 
\begin{gather}
  x_{14} = (\tsqrt{m_q^2 + \vec{k}^{\,2}_p} + k_p^3)/{\cal M}_p,\\
  \begin{aligned}
    x_1   &= x_{14} x_p, &
    k_1^+ &= x_{14} p^+, &
    k_1^\perp &=    x_{14}  p^\perp + k_p^\perp, 
    \\
    x_4   &= (1-x_{14}) x_p, &
    k_4^+ &= (1-x_{14}) p^+, &
    k_4^\perp &= (1-x_{14}) p^\perp - k_p^\perp.  
  \end{aligned}
\end{gather}
In meson $b$, one has $\vec{k}_{52} \equiv \vec{k}_b$, 
so that ${\cal M}_b \equiv {\cal M}_{52} = 2 \tsqrt{m_q^2 + \vec{k}_b^{\,2}} $,
and the analogous relations are: 
\begin{gather}
  x_{52} = (\tsqrt{m_q^2 + \vec{k}^{\,2}_b} + k_b^3)/{\cal M}_b,\\
  \begin{aligned}
    k_5^+ &= x_{52} b^+, &
    x_5   &= x_{52} x_b, &
    k_5^\perp &=    x_{52}  b^\perp + k_b^\perp, 
    \\
    k_2^+ &= (1-x_{52}) b^+, &
    x_2   &= (1-x_{52}) x_b, &
    k_2^\perp &= (1-x_{52}) b^\perp - k_b^\perp.  
  \end{aligned}
\end{gather}

Evaluating the quarks invariant masses in the hybrid and 
in the decay vertex, one obtains
\begin{equation}
  \label{eq:M12s}
  \begin{split}
    {\cal M}^2_{12} &= (x_1 + x_2)
    \left[
      \frac{k_1^{\perp \, 2} + m^2_q}{x_1} +
      \frac{k_2^{\perp \, 2} + m^2_q}{x_2}
    \right] 
    - 
    (k_1^\perp + k_2^\perp)^2  \, ,
  \end{split}
\end{equation}
and
\begin{equation}
  \label{eq:M54s}
  \begin{split}
    {\cal M}^2_{54} &= (x_5 + x_4)
    \left[
      \frac{k_5^{\perp \, 2} + m^2_q}{x_5} +
      \frac{k_4^{\perp \, 2} + m^2_q}{x_4}
    \right] 
    -
    (k_5^\perp + k_4^\perp)^2 \, .
  \end{split}
\end{equation}

The three-vectors: $\vec k_{12}$ in mesons, and $\vec k_{hq}$ and 
$\vec k_{hg}$ in the hybrid, are defined
using
\begin{subequations}
  \label{eq:qvec:def}
  \begin{align}
    \label{eq:qvec:def:M_12^2}
    \left(\sqrt{\vec{k}^2 + m_1^2} + \sqrt{\vec{k}^2 + m_2^2}\: \right)^2
    &= 
    \frac{\kappa^2 + m_1^2}{x} + \frac{\kappa^2 + m_2^2}{1 - x}
    = \mathcal{M}^2 \, ,
    \\
    \label{eq:qvec:def:perp}
    \vec{k}^\perp 
    &= 
    \kappa \, .
  \end{align}
\end{subequations}
For $m_1 = m_2 = m$, one has
\begin{subequations}
  \label{eq:qvec:def:simpl}
  \begin{align}
    \label{eq:qvec:def:M_12^2:simpl}
    4 \left(\vec{k}^2 + m^2\right)
    &= 
    \frac{\kappa^2 + m^2}{x (1 - x)}
    = \mathcal{M}^2 \, ,
    \\
    \label{eq:qvec:def:perp:simpl}
    \vec{k}^\perp 
    &= 
    \kappa \, .
  \end{align}
\end{subequations}

In the part $A$ of the decay amplitude, one has 
\begin{equation}
  \label{eq:A:kqs}
  \vec{k}^{\,2}_q = {\cal M}^2_{12}/4 - m_q^2 \; ,
\end{equation}
and
\begin{equation}
  \label{eq:A:kgs}
  \vec{k}^{\, 2}_g = 
  \frac{
    \left[{\cal M}^2_{123} - \left({\cal M}_{12} + m_g \right)^2\right]   
    \left[{\cal M}^2_{123} - \left({\cal M}_{12} - m_g \right)^2\right]
  }{
    4{\cal M}^2_{123} 
  } \; ,
\end{equation}
where
\begin{equation}
  \label{eq:M123s}
  \begin{split}
    {\cal M}^2_{123} 
    &=
    \frac{k_1^{\perp \, 2} + m^2_q}{x_1} + 
    \frac{k_2^{\perp \, 2} + m^2_q}{x_2} 
    + 
    \frac{(k_1 + k_2)^{\perp \, 2} + m^2_g}{1 - x_1 - x_2}  
    - (k_1 + k_2 + k_3)^{\perp \, 2} \; . 
  \end{split}
\end{equation}  
Similarly, in the part $B$, one has
\begin{equation}
  \label{eq:B:kqs}
  \vec{k}^{\,2}_q = {\cal M}^2_{54}/4 - m_q^2 \; ,
\end{equation}
and
\begin{equation}
  \label{eq:B:kgs}
  \vec k^{\,2}_g = 
  \frac{
    \left[{\cal M}^2_{543} - \left({\cal M}_{54} + m_g \right)^2\right]   
    \left[{\cal M}^2_{543} - \left({\cal M}_{54} - m_g \right)^2\right]
  }{
    4{\cal M}^2_{543} 
  } \; ,
\end{equation}
where
\begin{equation}
  \label{eq:M543s}
  \begin{split}
    {\cal M}^2_{543} 
    &=
    \frac{k_5^{\perp \, 2} + m^2_q}{x_5} + 
    \frac{k_4^{\perp \, 2} + m^2_q}{x_4} 
    + 
    \frac{(k_5 + k_4)^{\perp \, 2} + m^2_g}{1 - x_5 - x_4}  
    - (k_5 + k_4 + k_3)^{\perp \, 2} \; . 
  \end{split}
\end{equation}  

\section{Cross-checks}
\label{sec:check}

All our calculations of spin factors were done using two 
independent methods.
One method was to first reduce the spin factors to $2 \times 2$ 
matrices sandwiched between two component spinors 
$\chi_{\sigma}$, using Eq.~\eqref{eq:eps} for the gluon 
polarization four-vector $\varepsilon^\mu$ in $A^+ \equiv 0$ 
gauge. The spin factors were obtained from the trace of the 
product of $2 \times 2$ matrices. The other method made use of the 
expressions for $\sum u \bar{u}$, $\sum v \bar{v}$ and 
$\sum \varepsilon^{\ast\,\mu} \varepsilon^\nu$. The spin 
factors were obtained using properties of traces of products 
of $\gamma$ matrices. In the evaluation of spin factors, we
assume that the gluon is massless and has only two degrees 
of freedom. But the gluon acquires an effective mass 
dynamically. Therefore, we used $k_g^2 = m_g^2$ in the 
momentum-dependent factor of the wave function.  In 
$k_3^\mu$ and $P_{123}^\nu $ in the lattice-inspired 
spin factors, we checked what happens in both cases, i.e.,
when one inserts $k_3^2 = m_g^2$ or $k_3^2 = 0$.

The six-dimensional integrals were carried out using Monte 
Carlo integration (using the procedure VEGAS~\cite{vegas}).
The accuracy of the results of integration (standard
deviation output from VEGAS) is shown as error bars in
plots, unless the error is smaller than the size of a 
point on a plot. The VEGAS calculations using C were 
checked against iterative Gauss quadrature, also in C, 
and against a separate FORTRAN program performing the 
same calculations in several representative (but not all
regular) cases.

\section{Illustrative examples of minimization}
\label{sec:examples}

This Appendix provides examples of numerical evidence 
that we have gathered in all cases (many more than given 
here) we studied. The figures show how stddev (standard
deviation) or maxdev (maximal deviation), both in ratio
to the amplitude averaged over the angle $\theta$, change
around a global minimum when one changes just one parameter 
in the wave functions or the effective Hamiltonian width
$\lambda$. The varied parameter is on the horizontal axis,
and the deviation on the vertical axis, stddev marked
on the right-hand scale (plotted with black squares) and 
the maxdev marked on the left-hand scale (plotted with circles).

The examples demonstrate the dominant feature that the
parameters $\beta_{hq}$ and $\lambda$ are strongly correlated
with each other and both much larger than all other parameters.
The first example with the hybrid wave function with the spin factor 
$\uev$ and $N \neq 1$. The example illustrates our argument for
that the parameters $\beta_{hq}$ and $\lambda$ must be both much 
larger than all other parameters: the minimum resembles one side 
of a broad valley open toward large values. The remaining four 
examples concern the cases with various spin factors and always 
$N= 1$. The examples illustrate that $\beta_{hq}$ and $\lambda$ 
are both strongly correlated and much larger than all other 
parameters. 

\begin{figure*}[p]
  \centering
  \includegraphics[scale=\scanscale, trim=5 0 0 0]{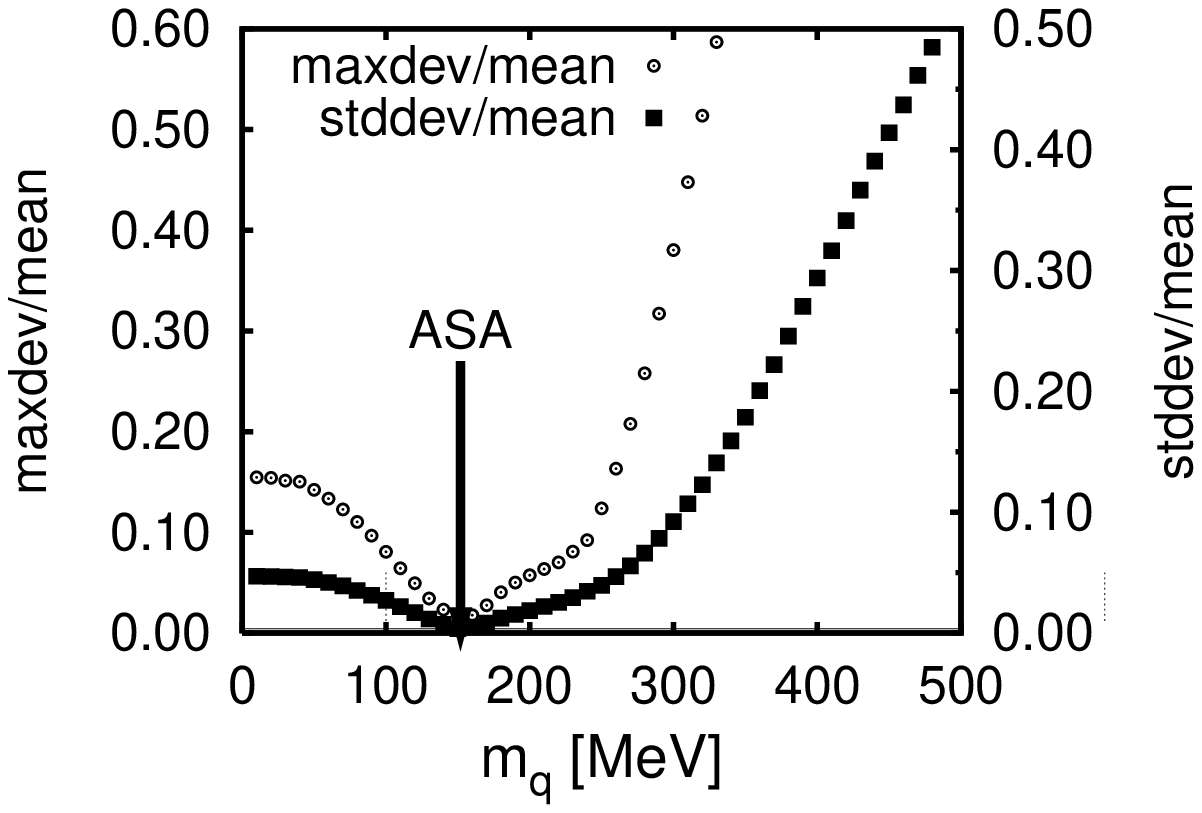}
  \includegraphics[scale=\scanscale, trim=5 0 0 0]{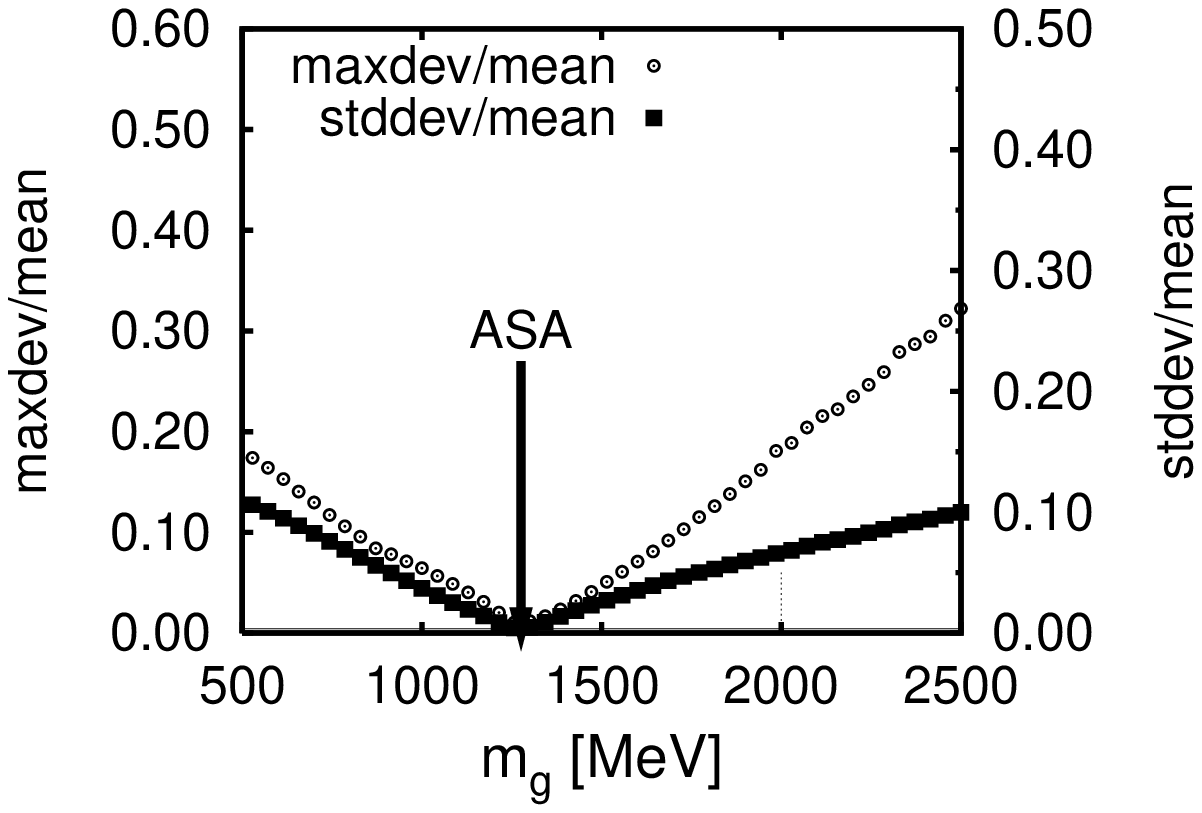}\\
  \includegraphics[scale=\scanscale, trim=5 0 0 0]{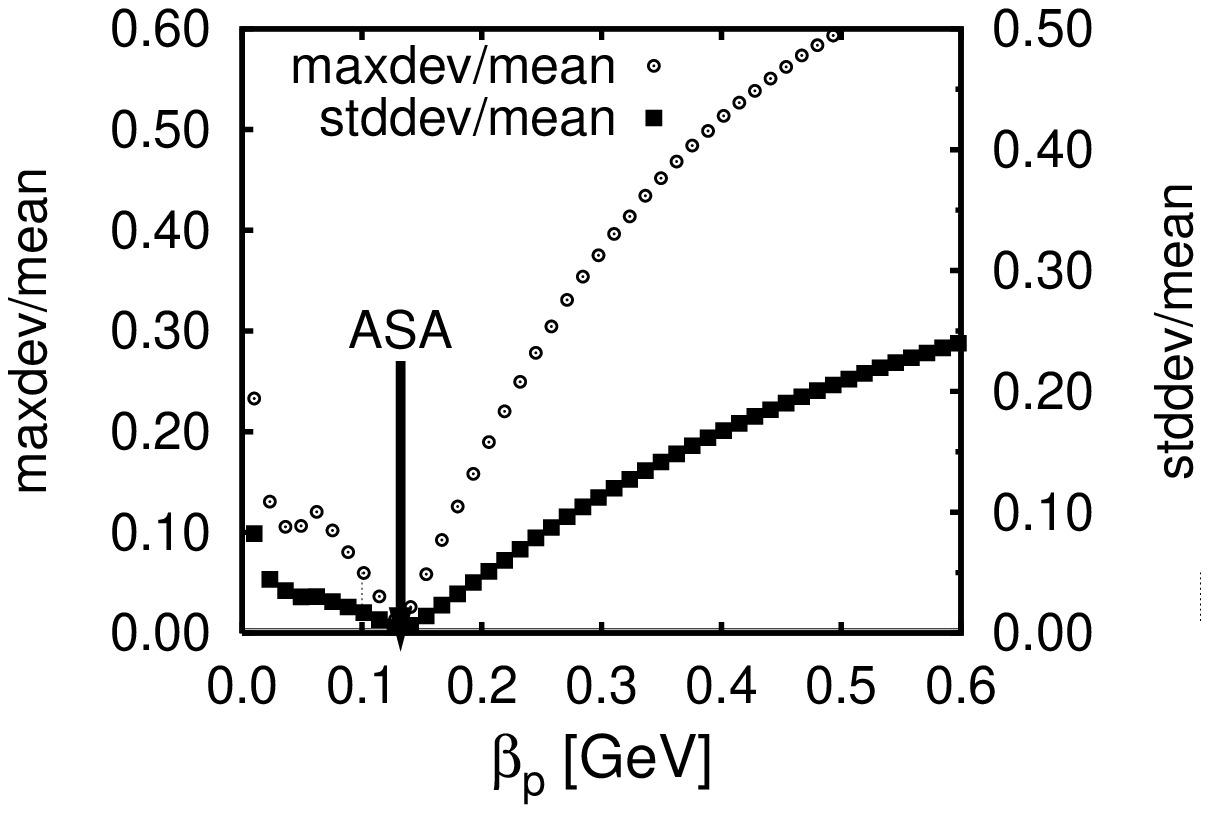}
  \includegraphics[scale=\scanscale, trim=5 0 0 0]{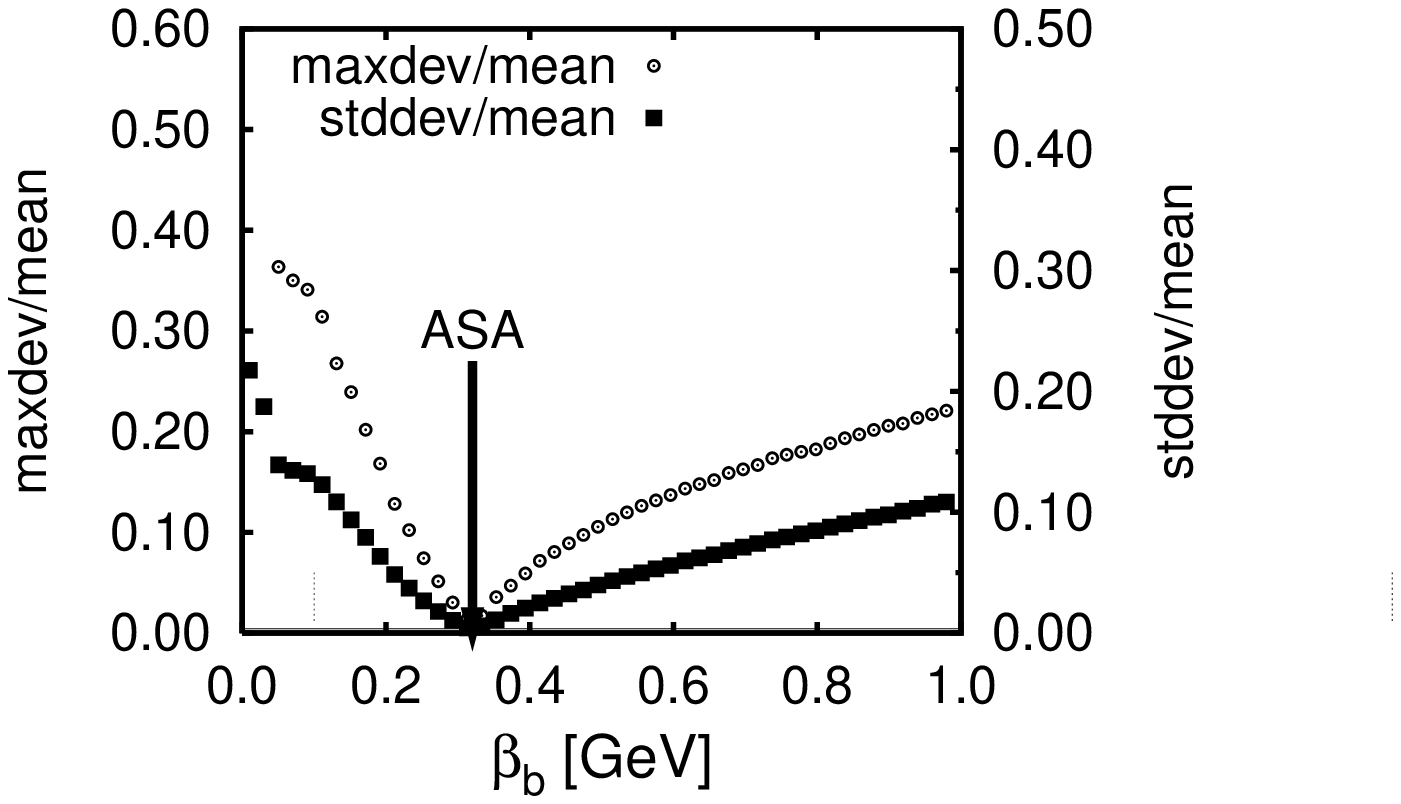}\\
  \includegraphics[scale=\scanscale, trim=5 0 0 0]{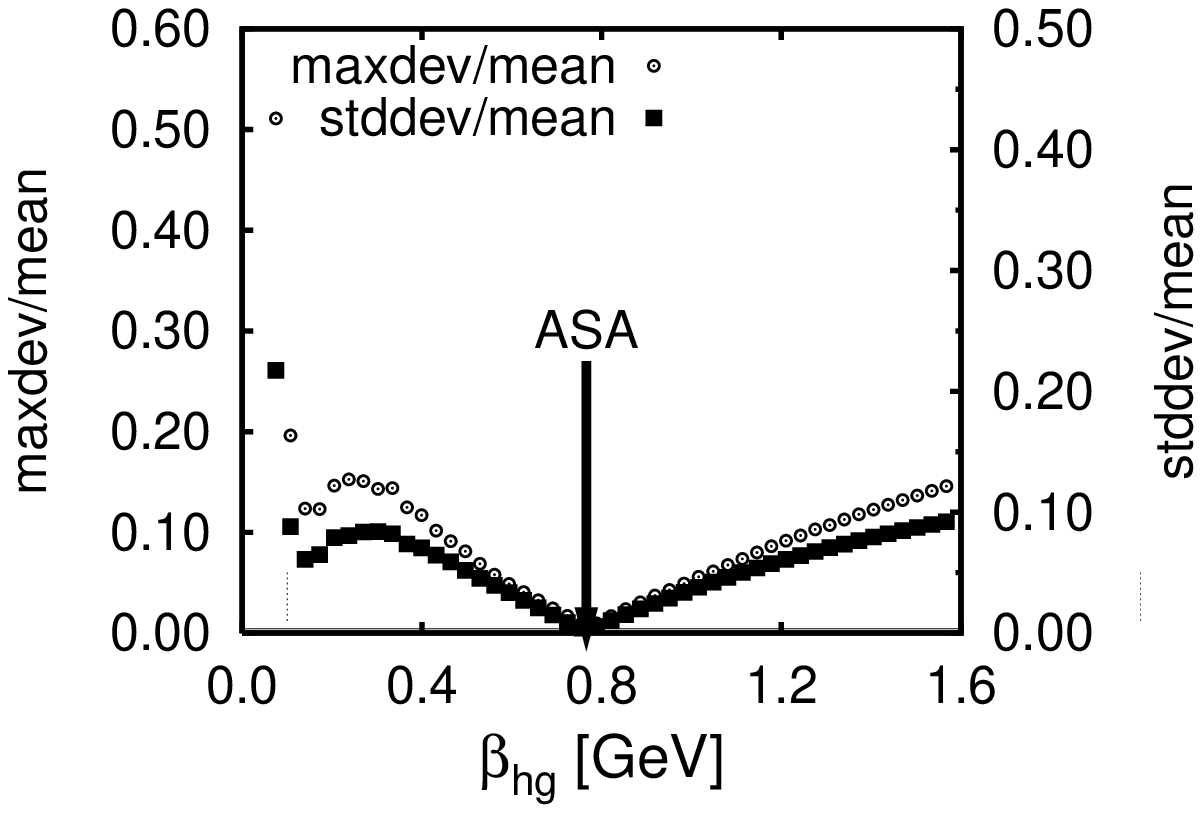}
  \includegraphics[scale=\scanscale, trim=5 0 0 0]{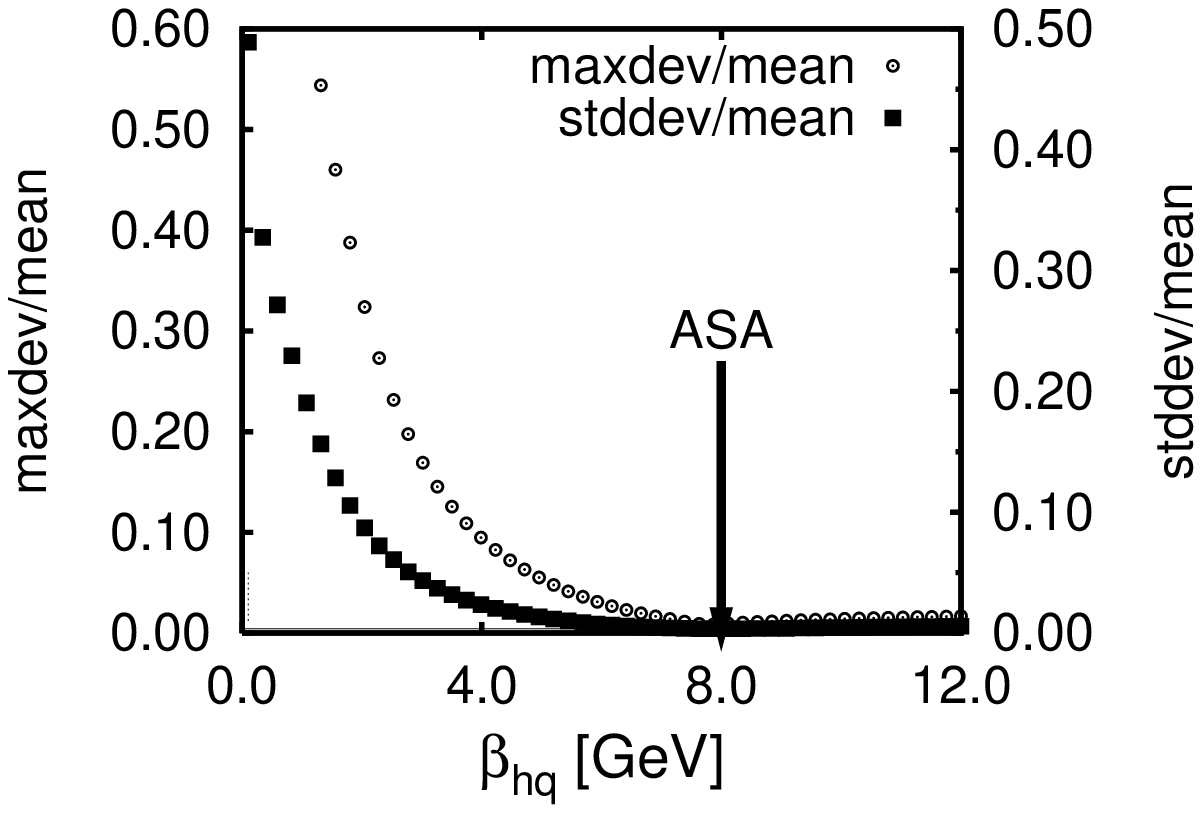}\\
  \includegraphics[scale=\scanscale, trim=5 0 0 0]{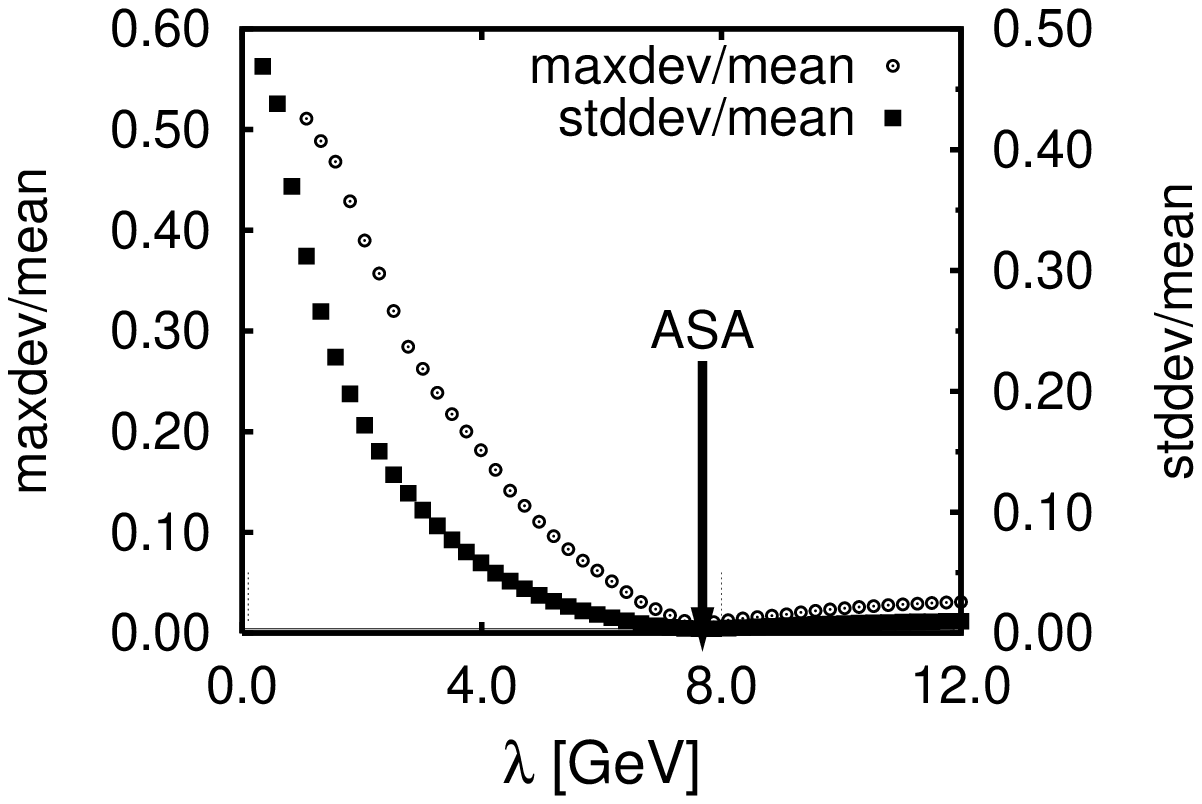}
  \includegraphics[scale=\scanscale, trim=45 0 -40 0]{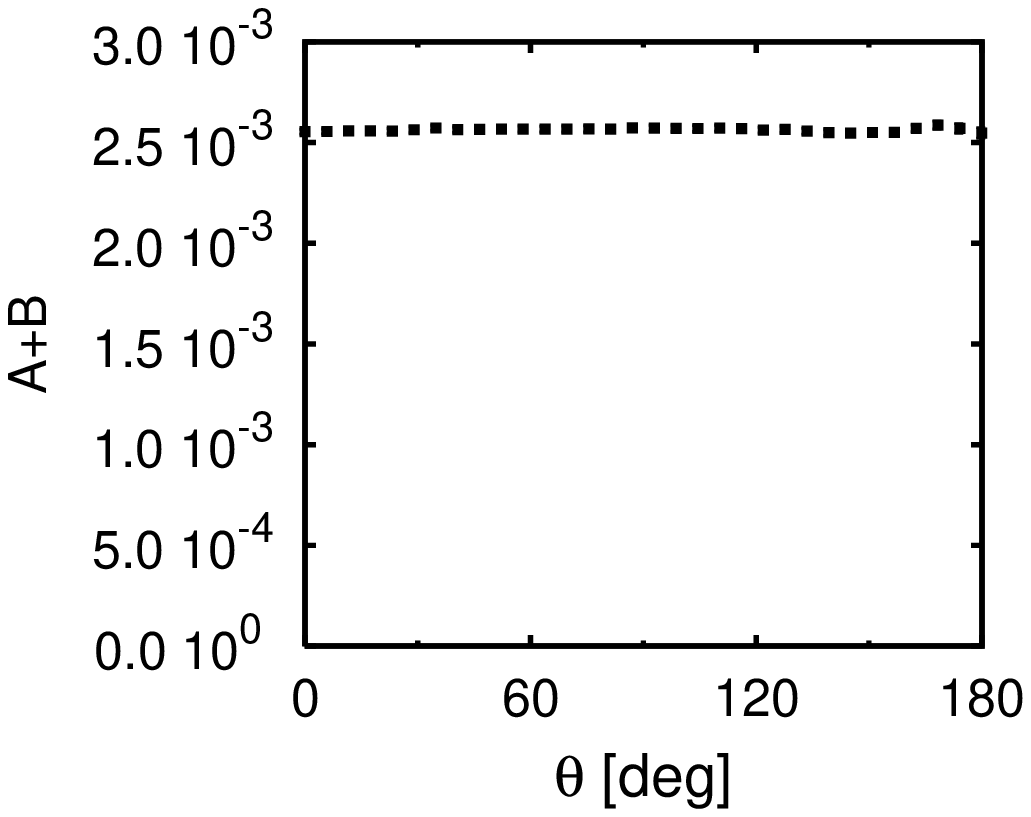}
  \caption{\label{fig:ASA:uev}\label{fig:ASA:a}\label{fig:scanmin:first}
    Variation of rotational symmetry violation
    versus changes of parameters in the wave functions 
    in the case a) in Fig.~\ref{fig:ASA:devsqrp-1}, i.e. 
    $S_h = \bar{u}\slashed{\eps} v$, with extra factors $N \neq 1$,
    decay into two $J^{PC} = 0^{++}$ mesons.     
    The optimal values of the parameters are given in the 
    first column in Table~\ref{tab:ASA:devsqrp-1}.  
    The arrow marked ``ASA'' points toward the optimal value of a 
    parameter. The last plot shows the amplitude itself for the
    optimal parameters.
  }
\end{figure*}
\begin{figure*}[p]
  \centering
  \includegraphics[scale=\scanscale, trim=5 0 0 0]{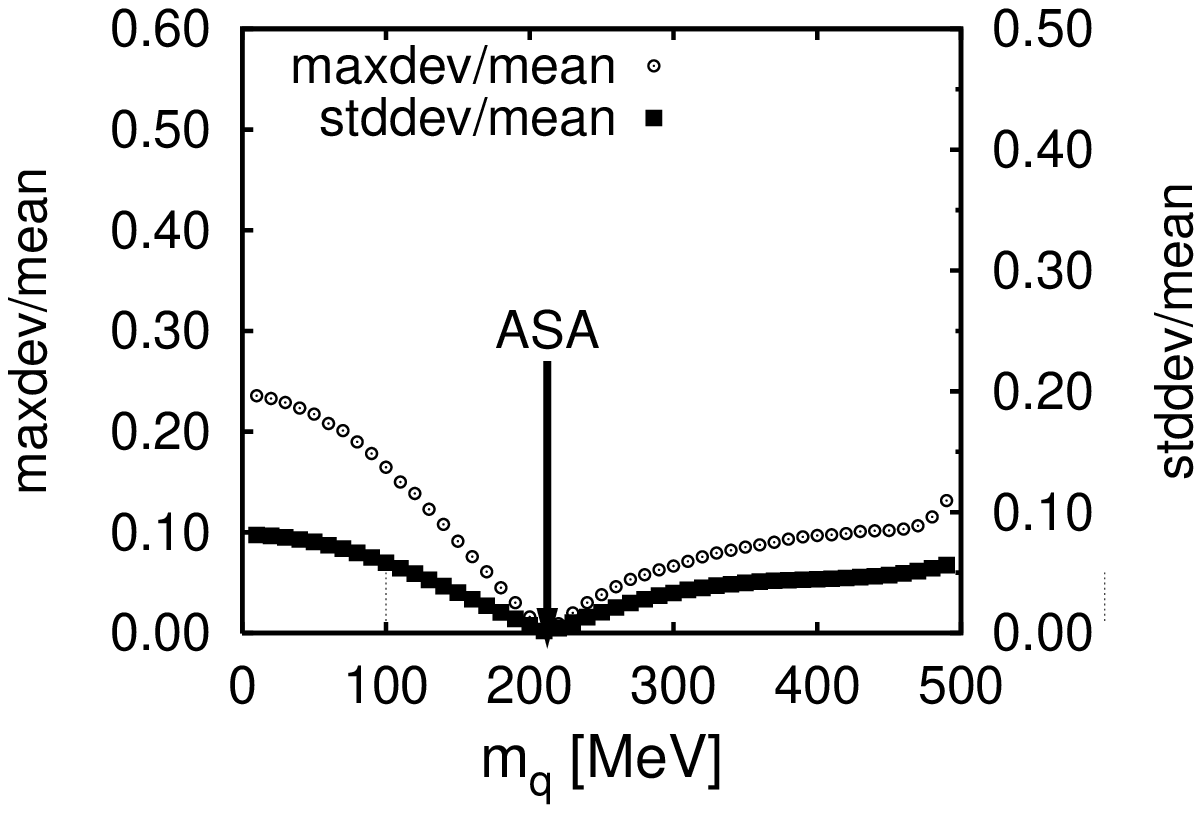}
  \includegraphics[scale=\scanscale, trim=5 0 0 0]{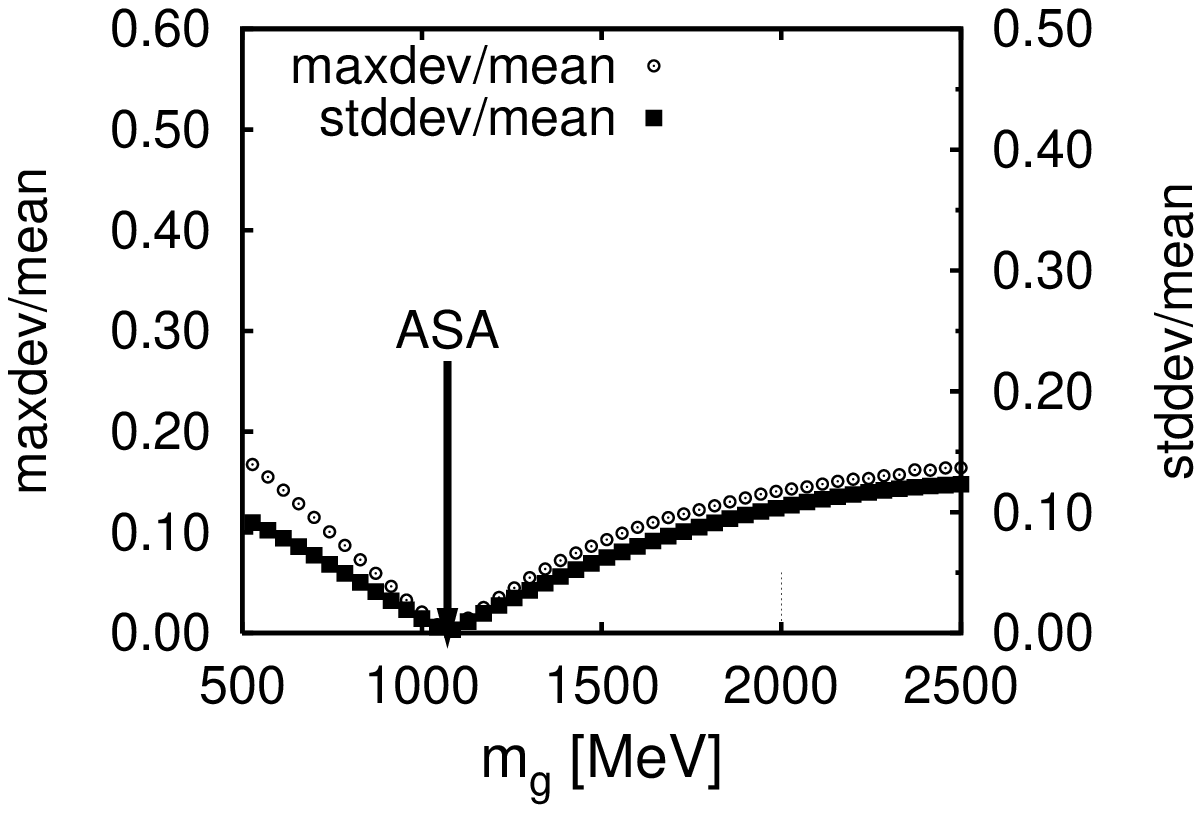}\\
  \includegraphics[scale=\scanscale, trim=5 0 0 0]{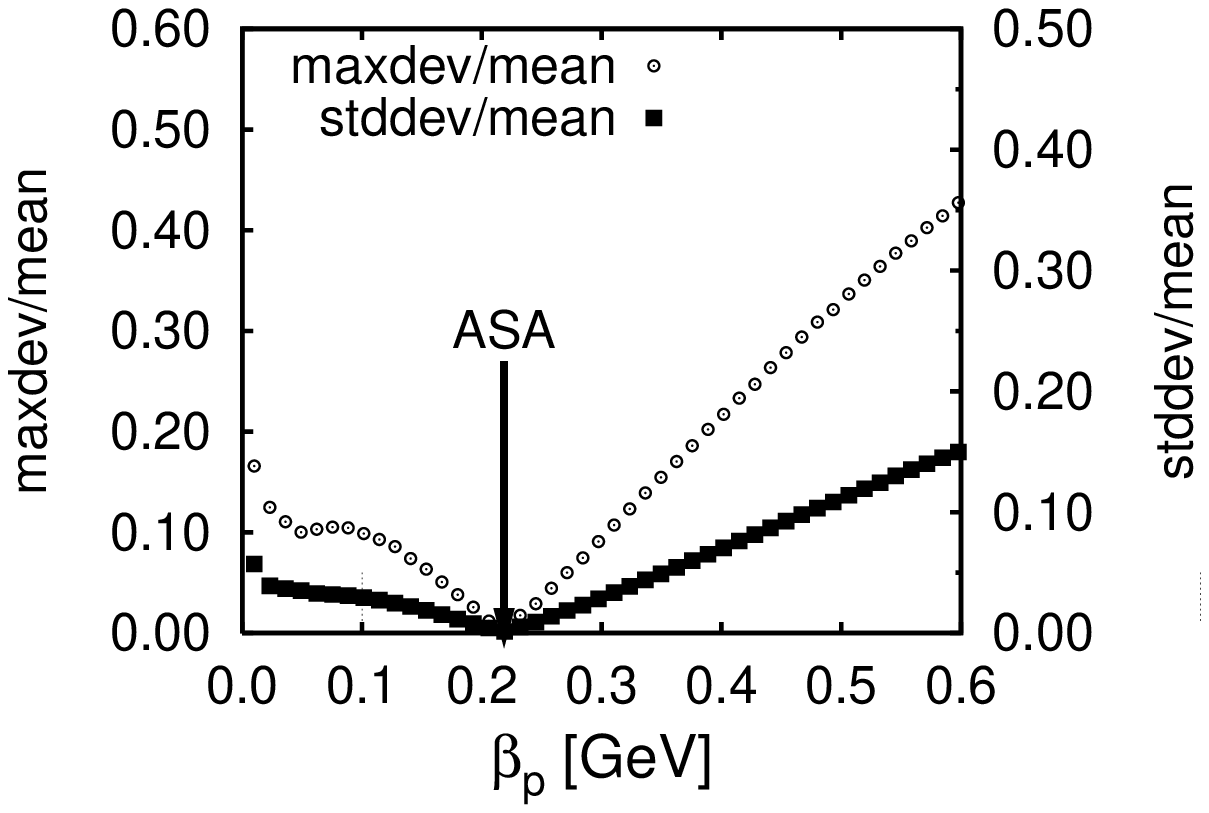}
  \includegraphics[scale=\scanscale, trim=5 0 0 0]{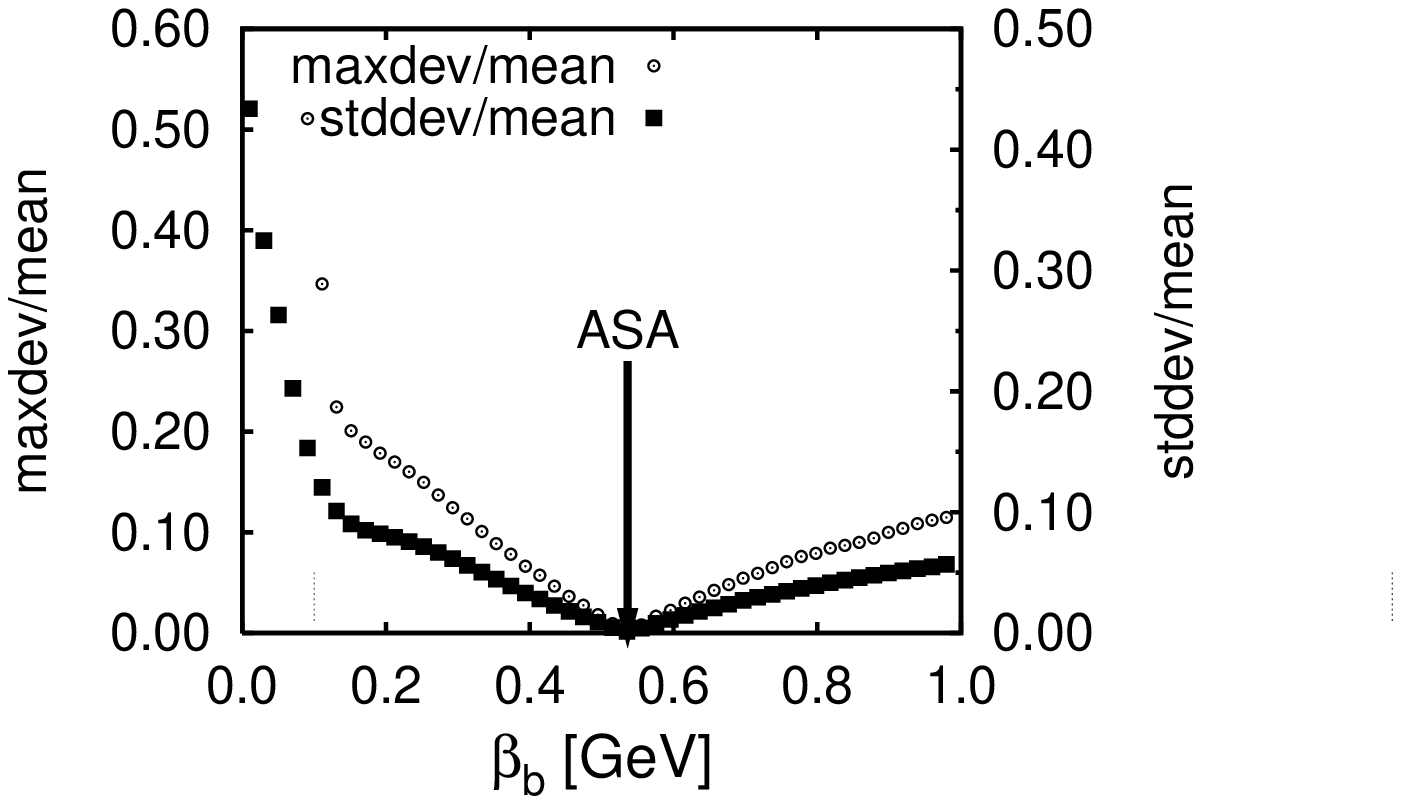}\\
  \includegraphics[scale=\scanscale, trim=5 0 0 0]{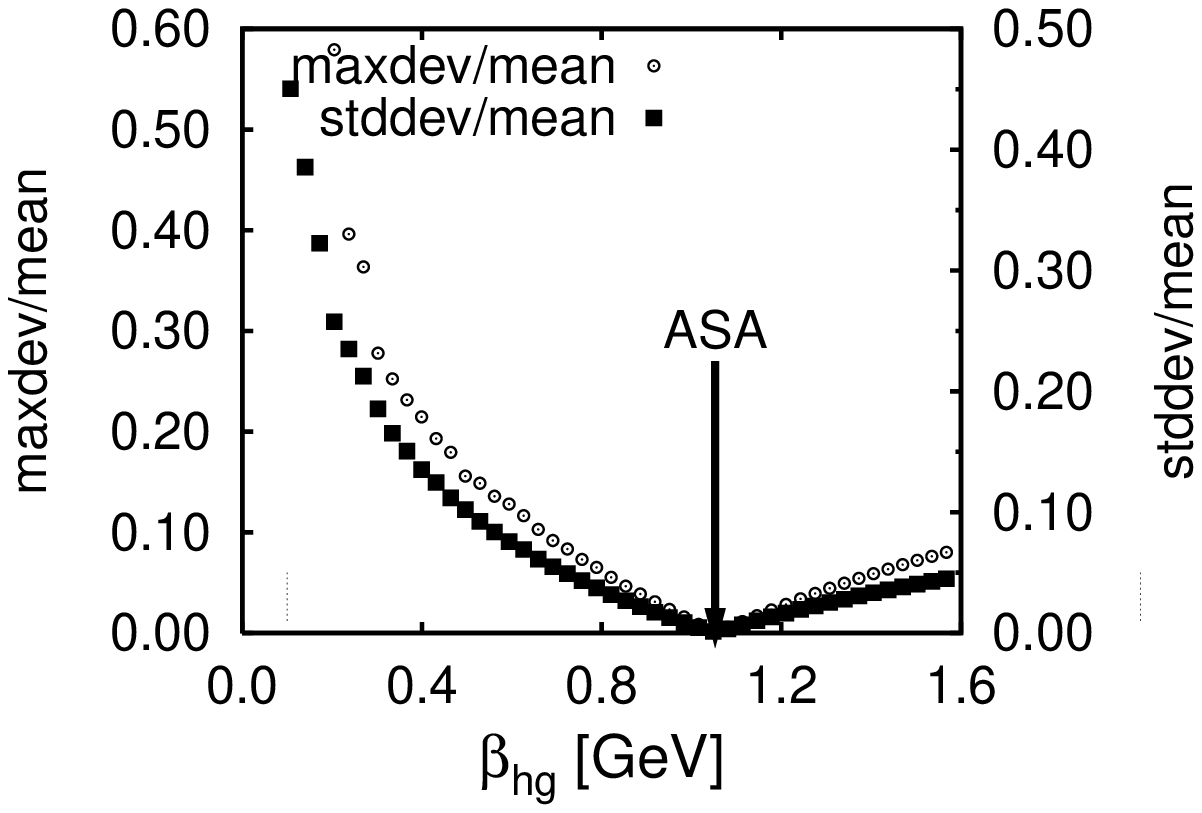}
  \includegraphics[scale=\scanscale, trim=5 0 0 0]{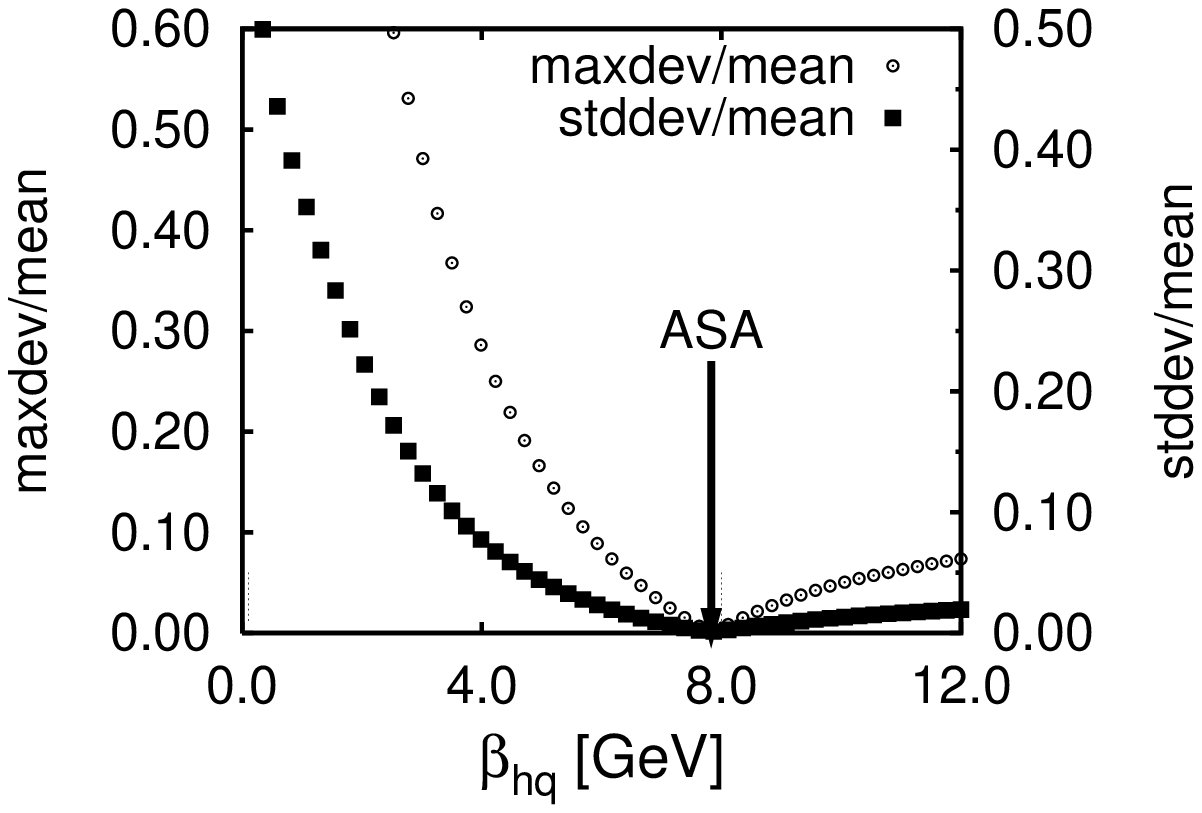}\\
  \includegraphics[scale=\scanscale, trim=5 0 0 0]{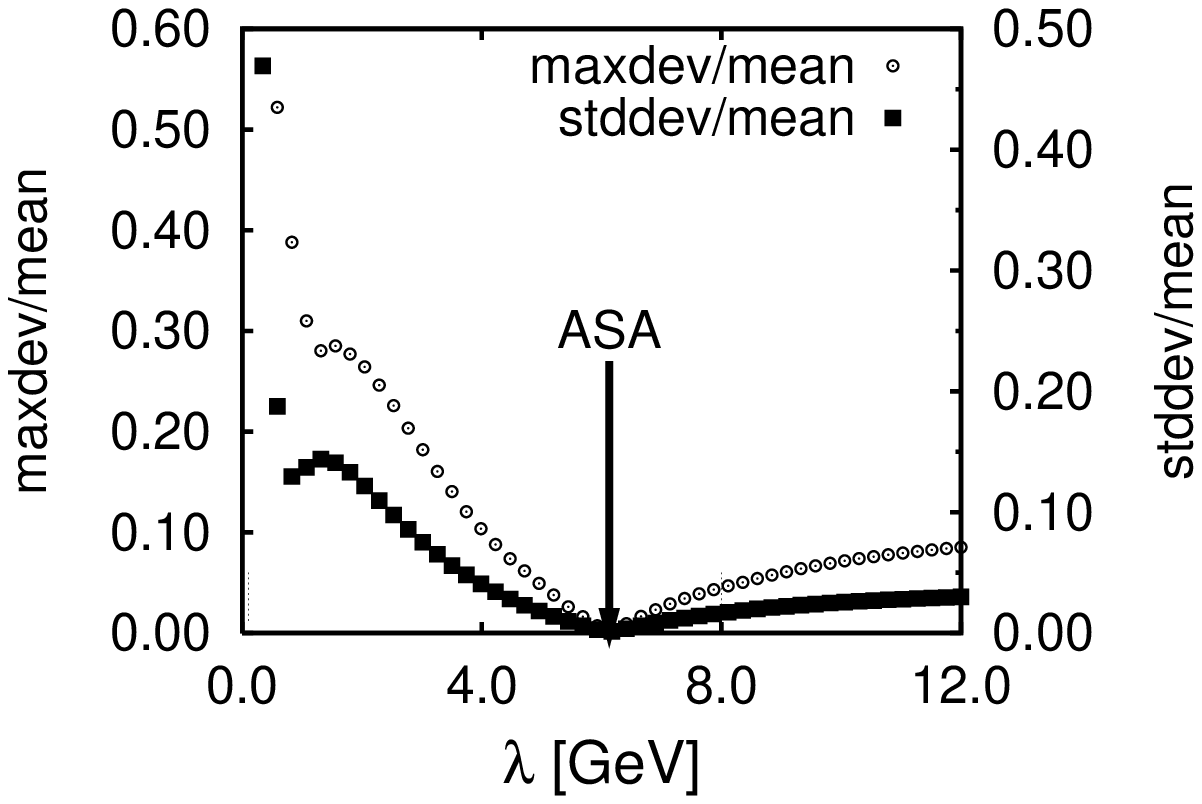}
  \includegraphics[scale=\scanscale, trim=45 0 -40 0]{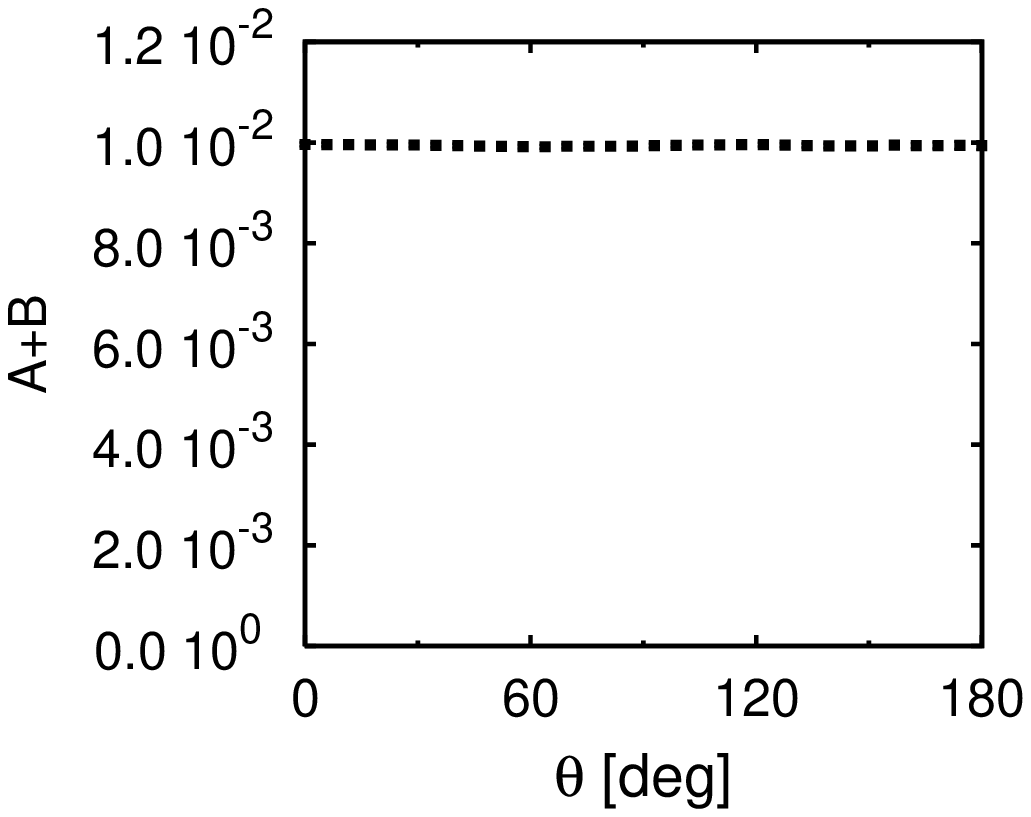}
  \caption{\label{fig:ASA:uev-nofxx}\label{fig:ASA:b}
    Variation of rotational symmetry violation
    versus changes of parameters in the wave functions 
    in the case b) in Fig.~\ref{fig:ASA:devsqrp-1}, 
    i.e. $S_h = \bar{u}\slashed{\eps} v$, with $N = 1$,
    decay into two $J^{PC} = 0^{++}$ mesons.   
    The optimal values of the parameters are given in the 
    second column in Table~\ref{tab:ASA:devsqrp-1}.  
    The arrow marked ``ASA'' points toward the optimal value of a 
    parameter. The last plot shows the amplitude itself for the
    optimal parameters.
  }
\end{figure*}
\begin{figure*}[p]
  \centering
  \includegraphics[scale=\scanscale, trim=5 0 0 0]{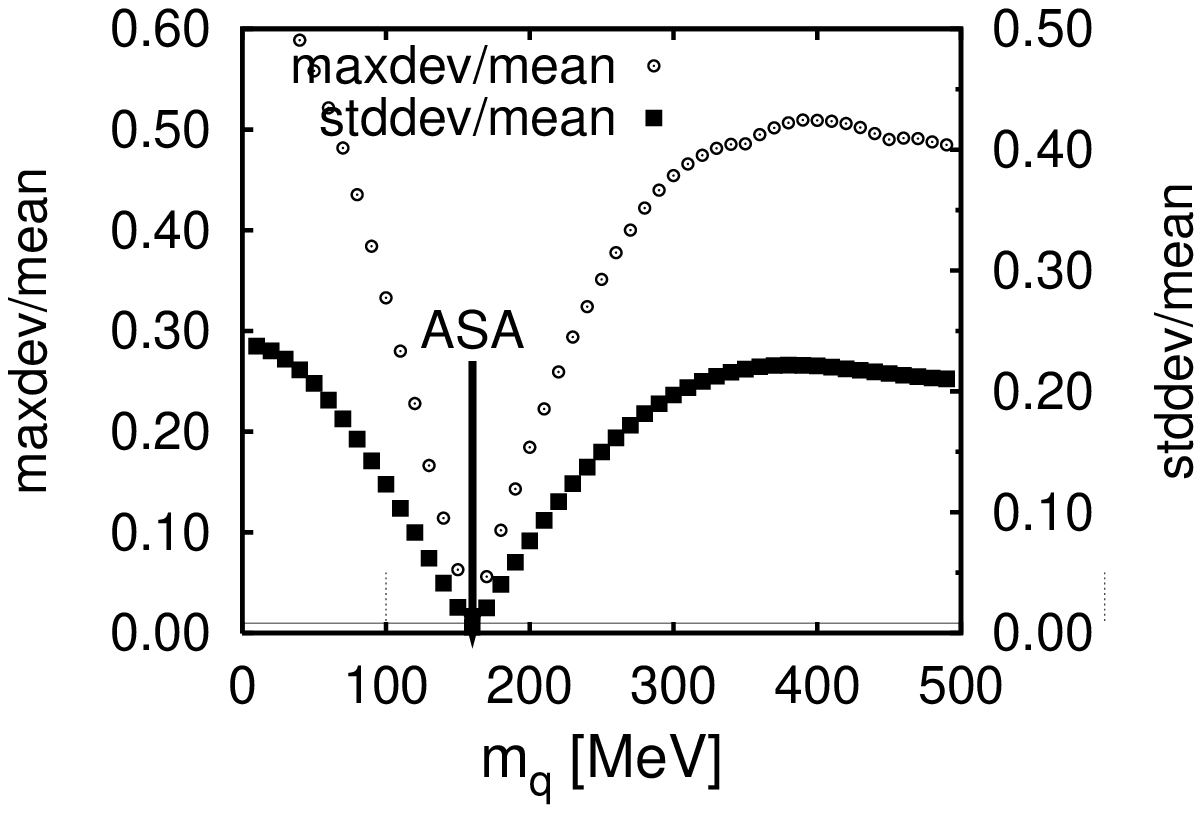}
  \includegraphics[scale=\scanscale, trim=5 0 0 0]{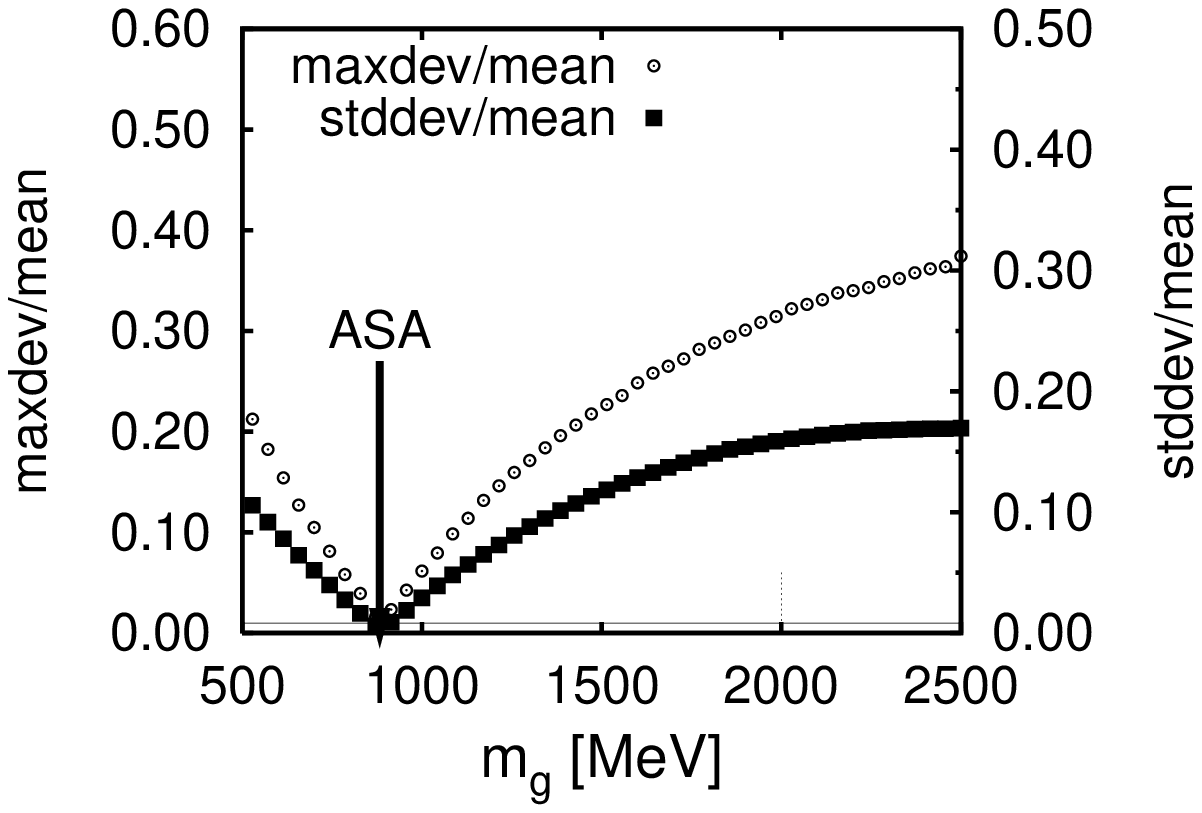}\\
  \includegraphics[scale=\scanscale, trim=5 0 0 0]{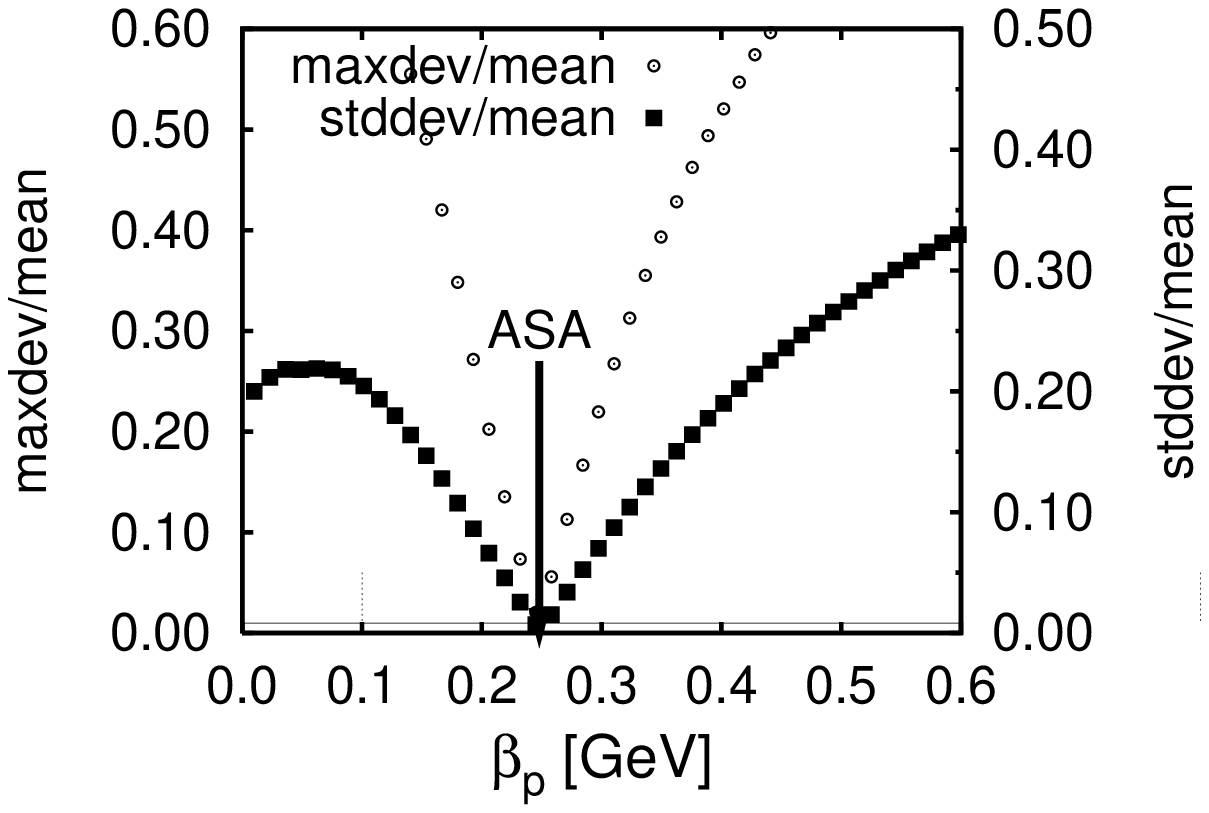}
  \includegraphics[scale=\scanscale, trim=5 0 0 0]{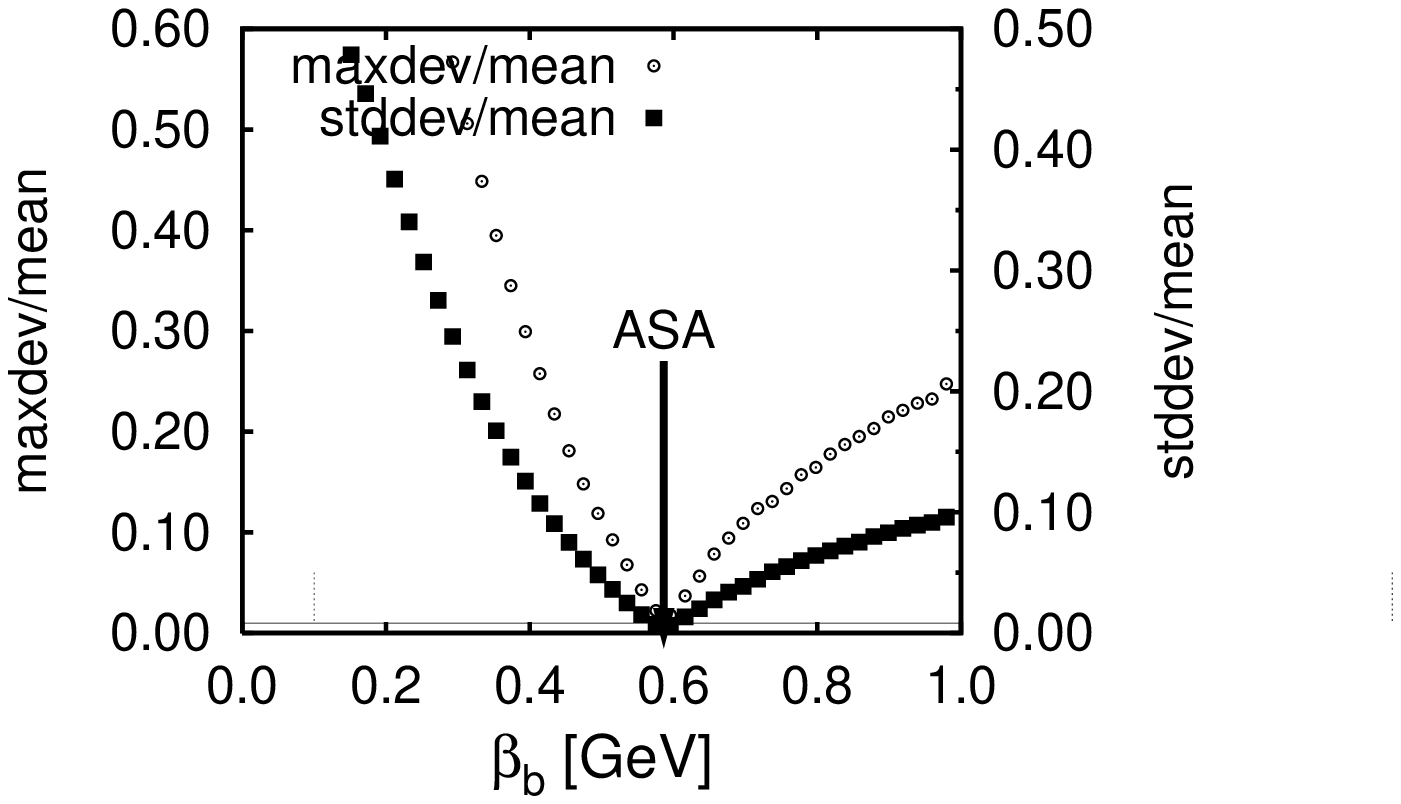}\\
  \includegraphics[scale=\scanscale, trim=5 0 0 0]{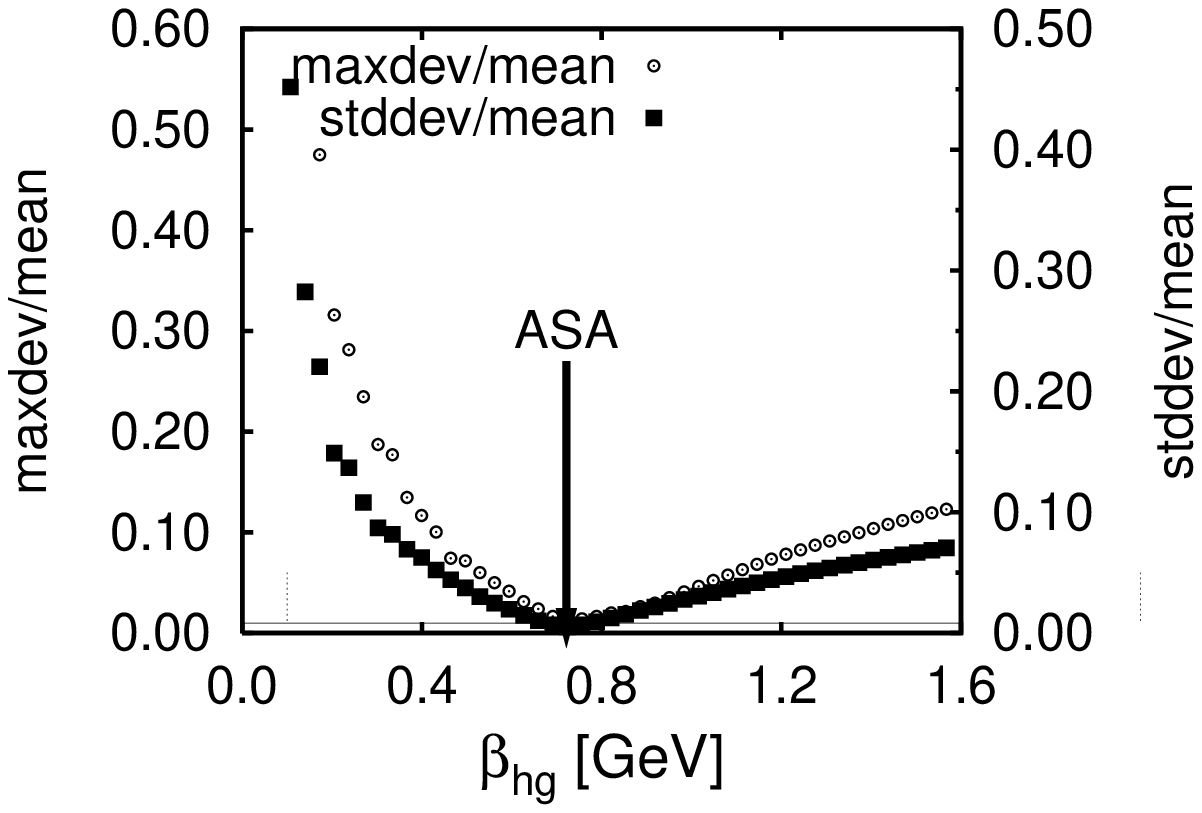}
  \includegraphics[scale=\scanscale, trim=5 0 0 0]{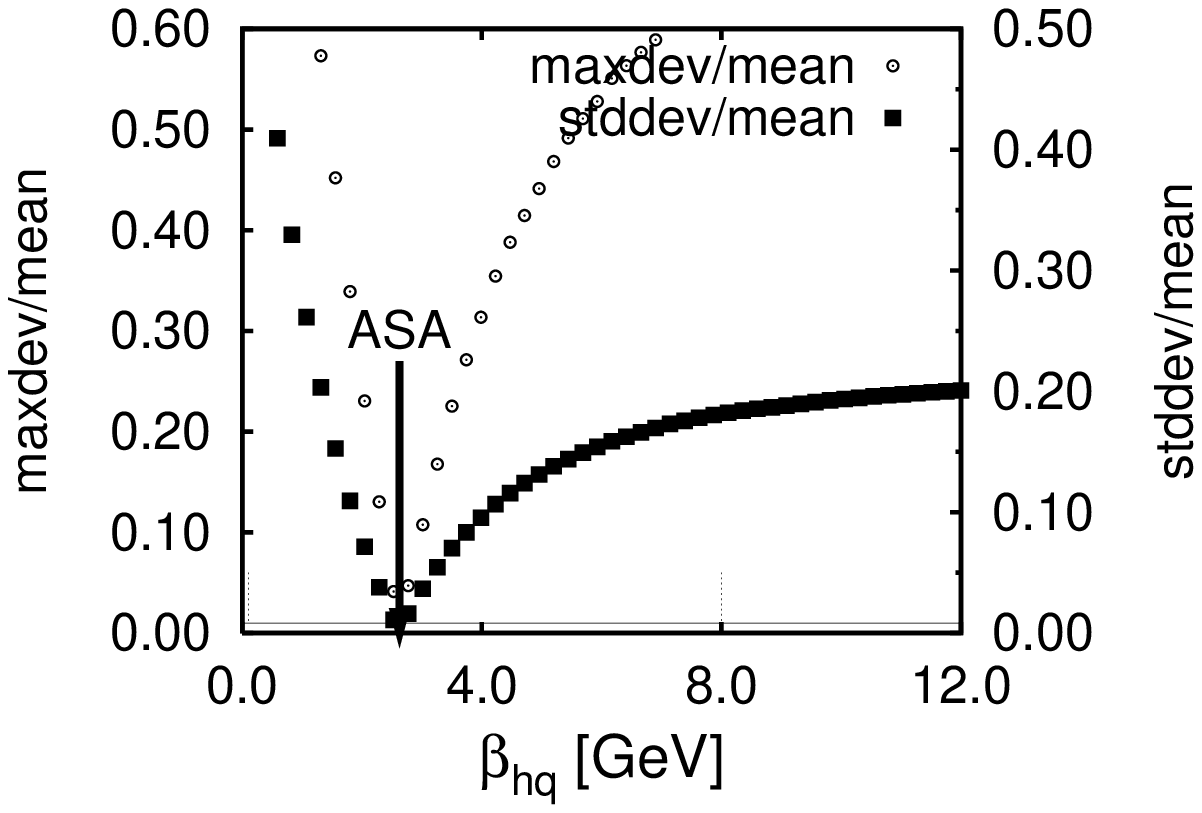}\\
  \includegraphics[scale=\scanscale, trim=5 0 0 0]{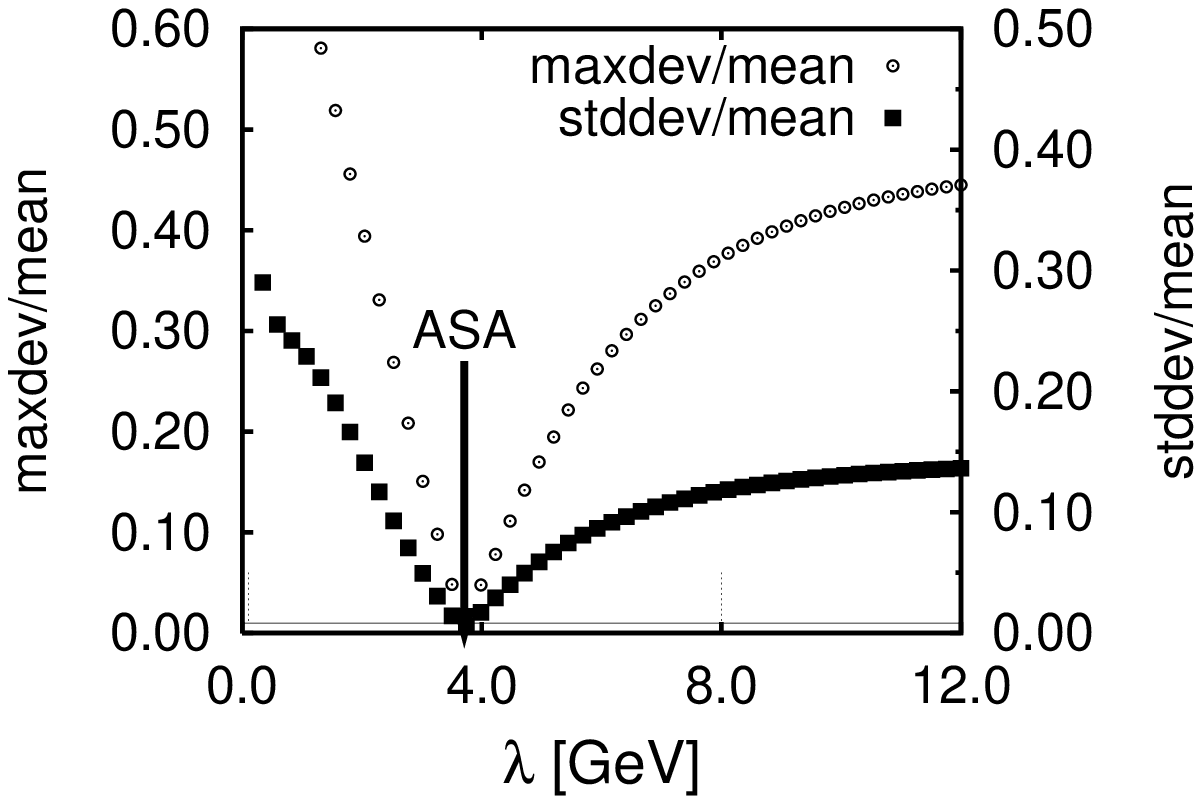}
  \includegraphics[scale=\scanscale, trim=45 0 -40 0]{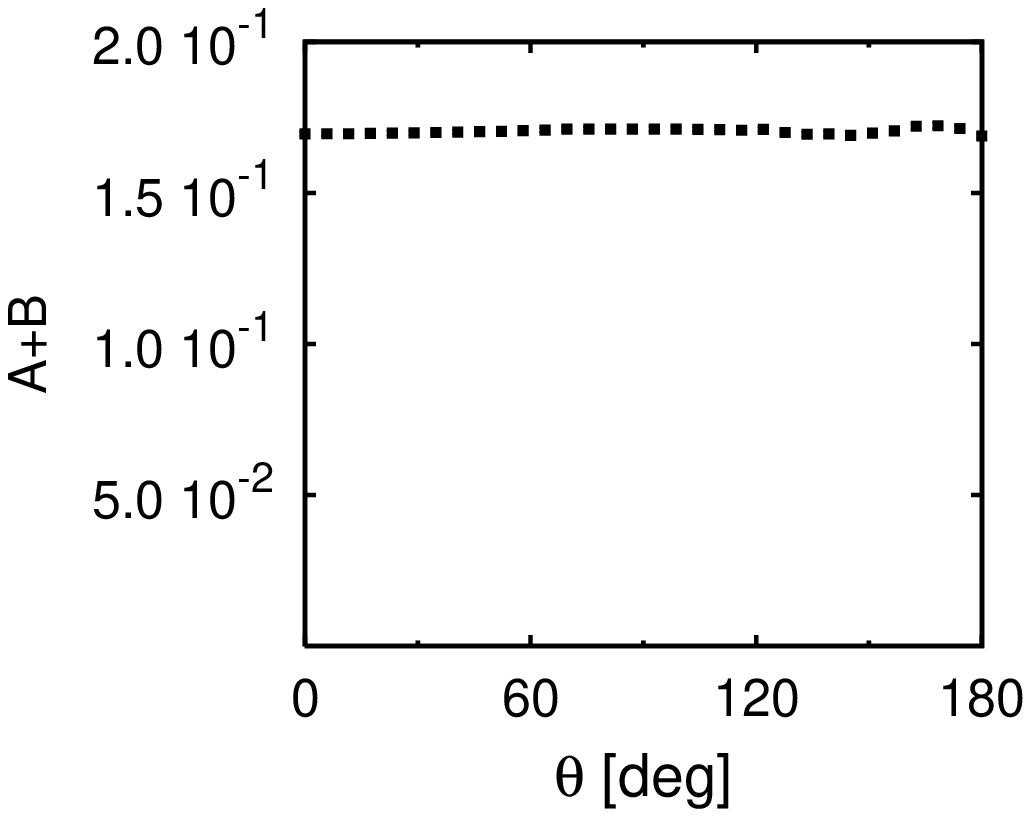}
  \caption{\label{fig:scan:uev_m5-nofxx}\label{fig:scan:ASA:u5v:b}
    Variation of rotational symmetry violation
    versus changes of parameters in the wave functions 
    in the case b) in Fig.~\ref{fig:ASA:u5v:devsqrp-1}, 
    i.e. $S_h = \bar{u}\slashed{\eps} v$, with $N = 1$, 
    decay into two $J^{PC} = 0^{-+}$ mesons.
    The optimal values of the parameters are given in the
    second column in Table~\ref{tab:ASA:u5v:devsqrp-1}. 
    The arrow marked ``ASA'' shows value of parameter found in
    minimization. The last plot shows the amplitude itself for the
    optimal parameters.
  }    
\end{figure*}
\begin{figure*}[p]
  \centering
  \includegraphics[scale=\scanscale, trim=5 0 0 0]{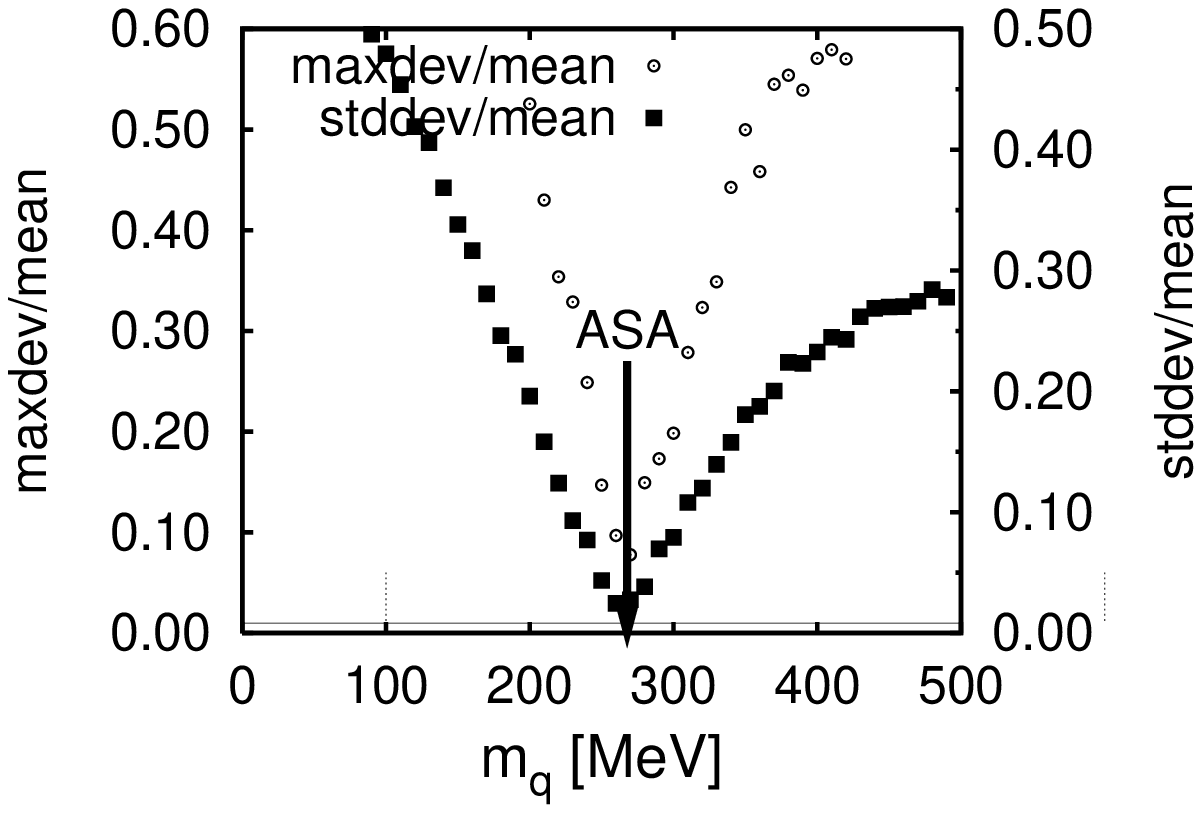}
  \includegraphics[scale=\scanscale, trim=5 0 0 0]{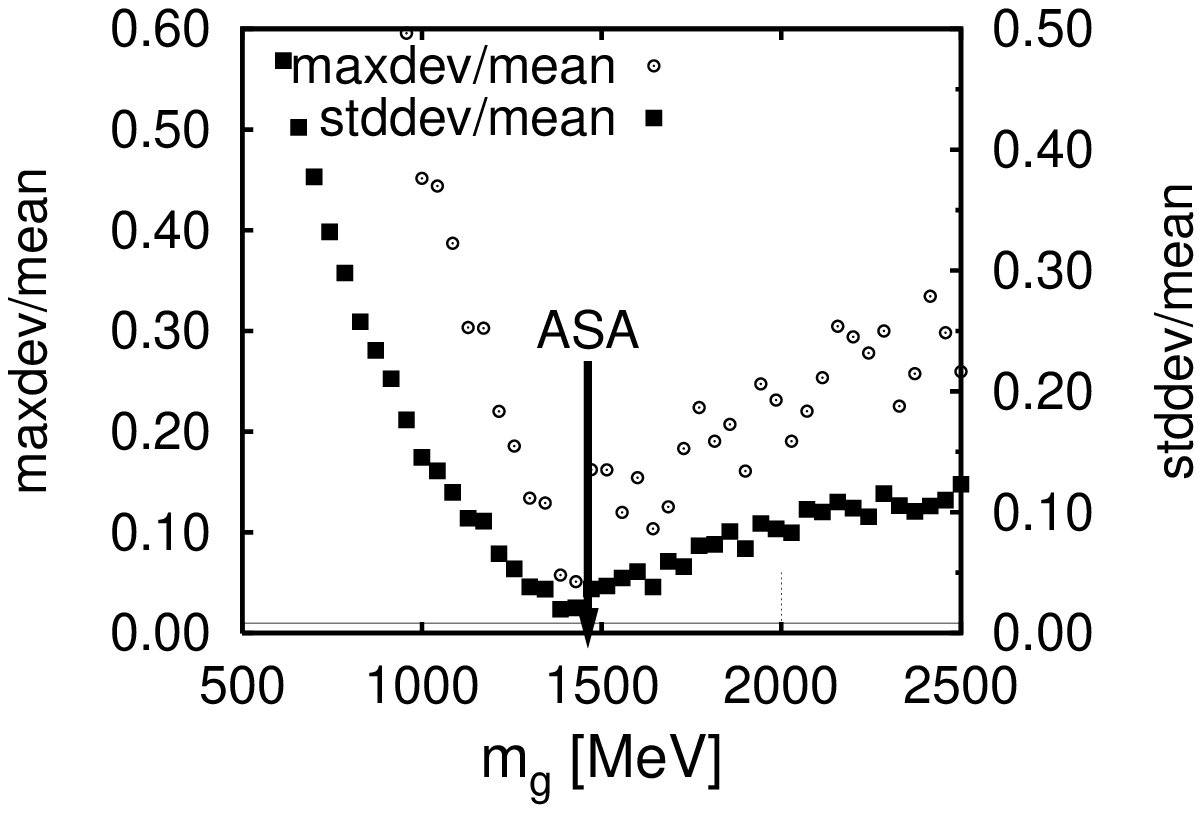}\\
  \includegraphics[scale=\scanscale, trim=5 0 0 0]{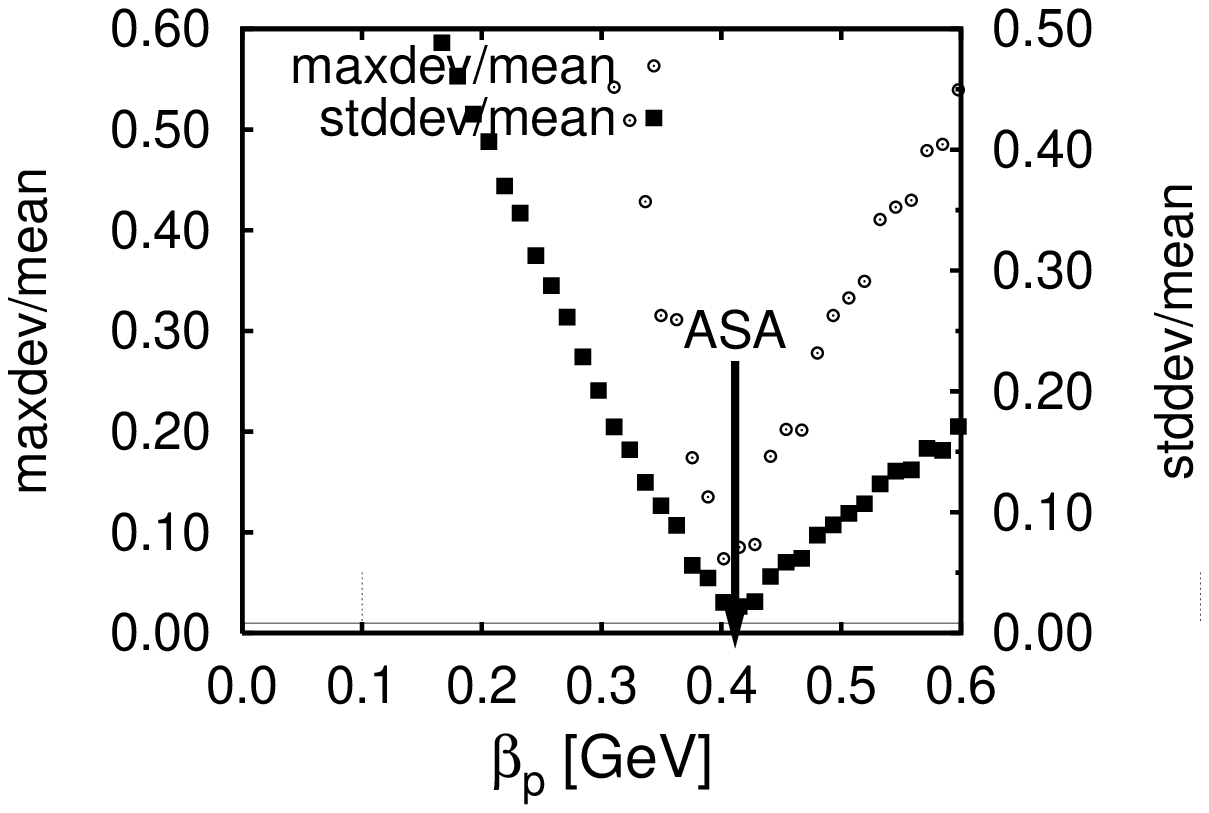}
  \includegraphics[scale=\scanscale, trim=5 0 0 0]{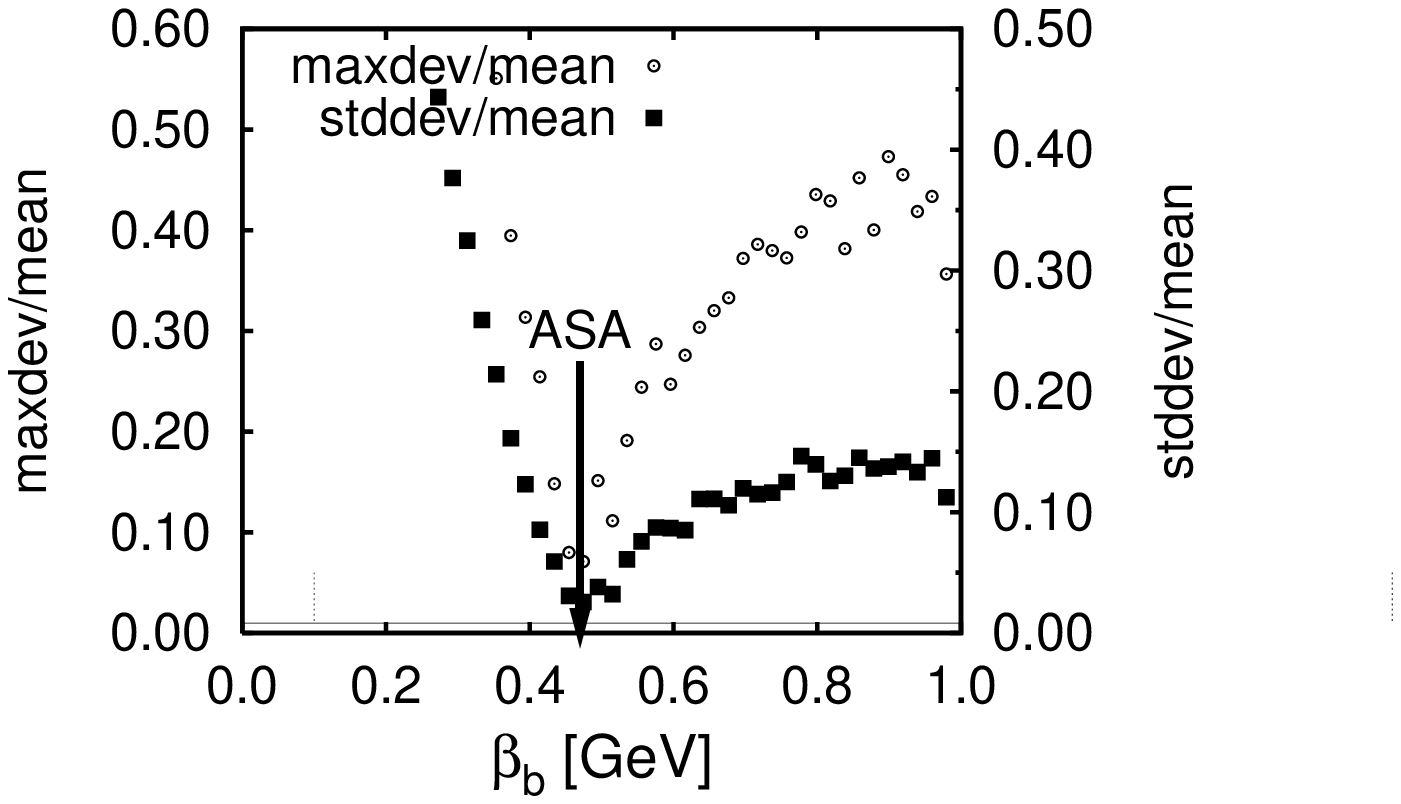}\\
  \includegraphics[scale=\scanscale, trim=5 0 0 0]{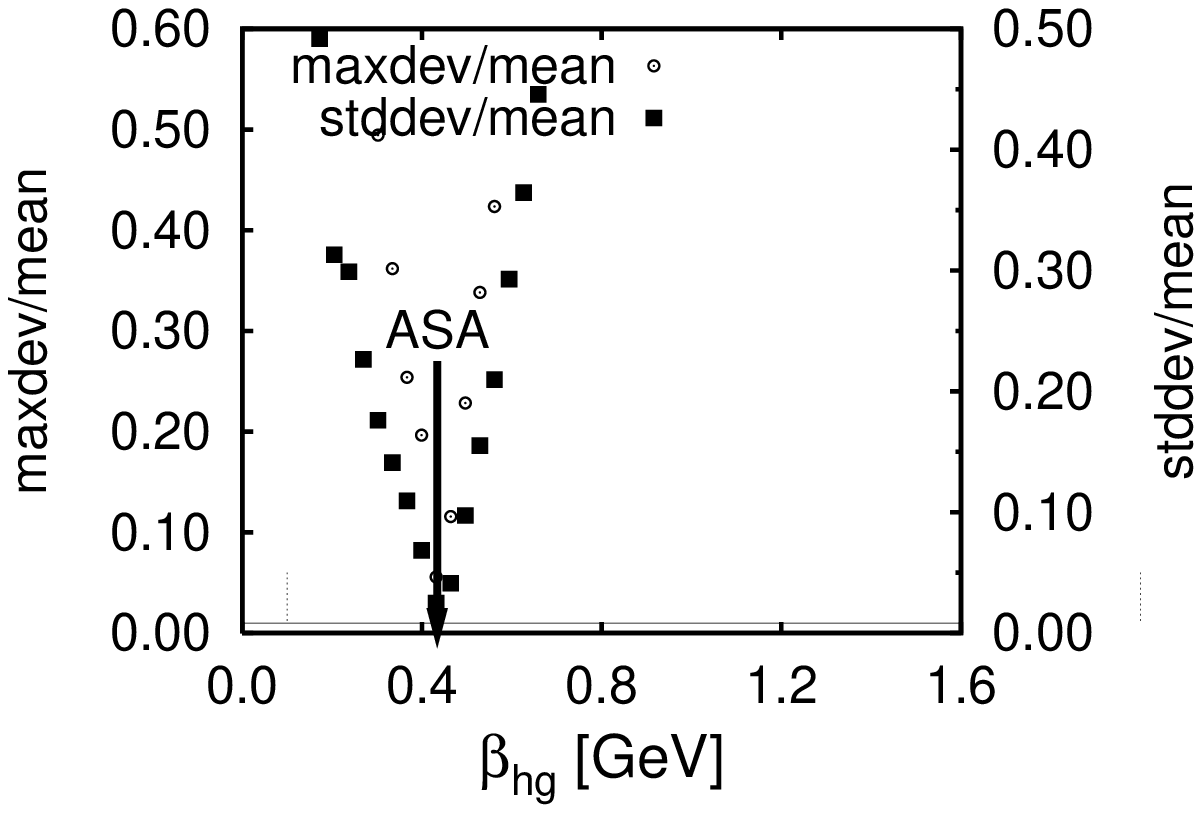}
  \includegraphics[scale=\scanscale, trim=5 0 0 0]{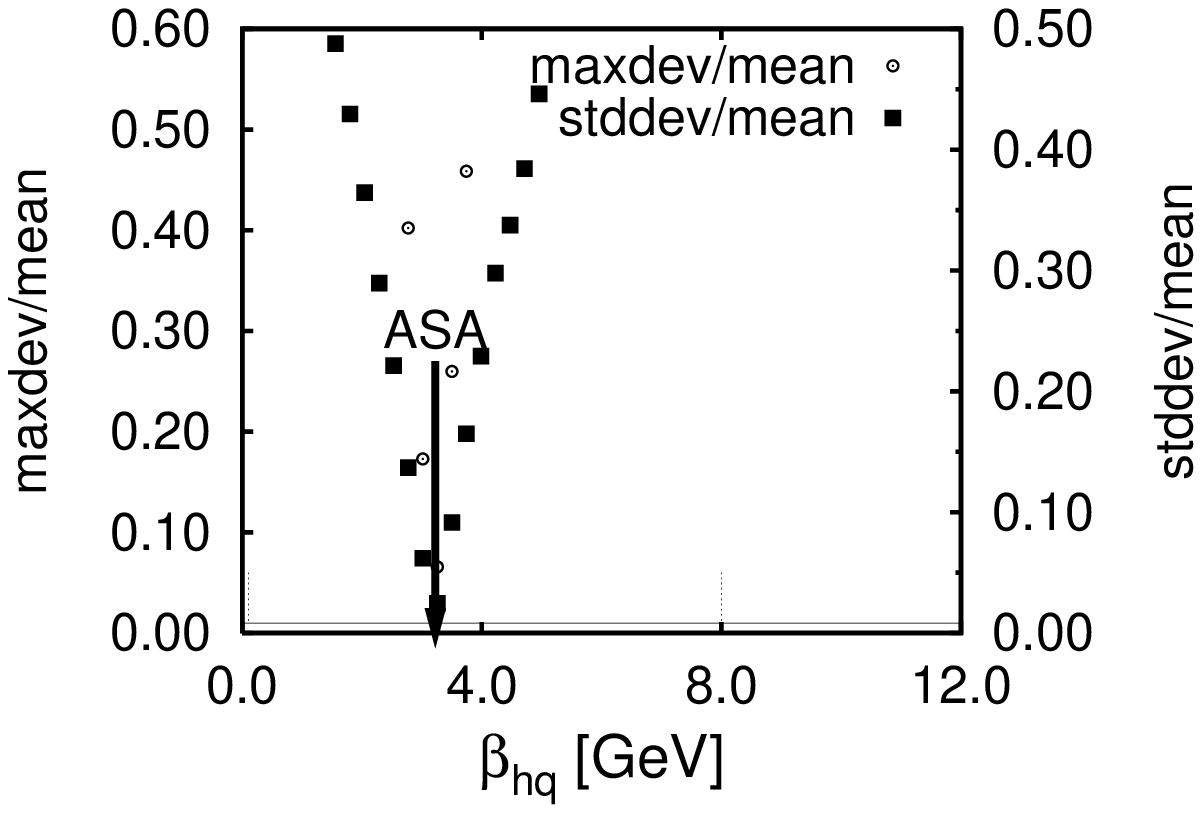}\\
  \includegraphics[scale=\scanscale, trim=5 0 0 0]{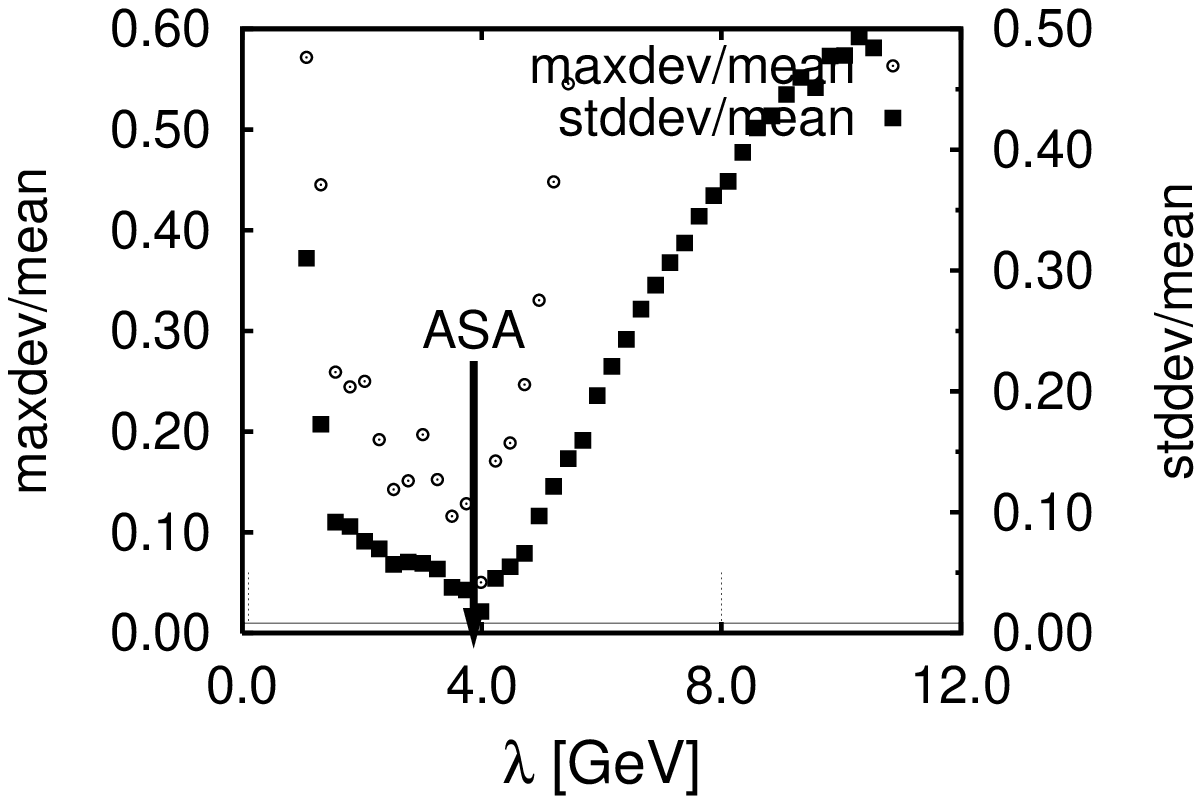}
  \includegraphics[scale=\scanscale, trim=45 0 -40 0]{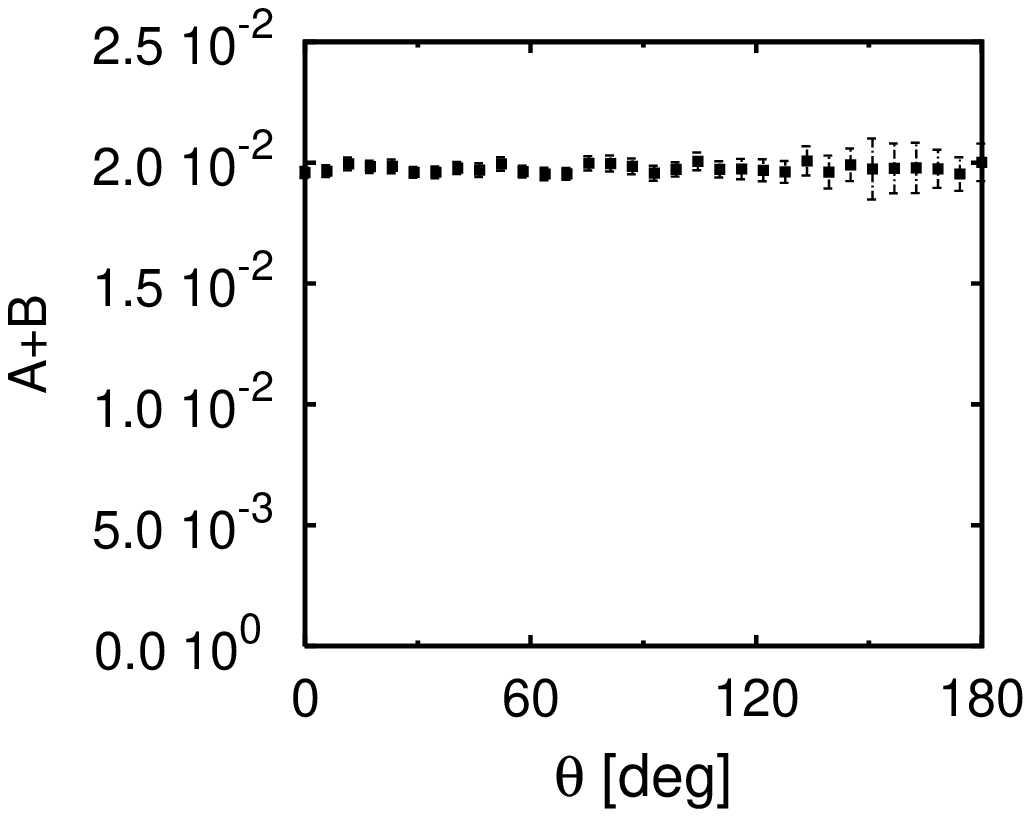}
  \caption{\label{fig:ugpv-nofxx}\label{fig:ASA:uGPv:f}
    Variation of rotational symmetry violation
    versus changes of parameters in the wave functions 
    in the case f) in Fig.~\ref{fig:ASA:uGPv:devsqrp-1}, i.e.,
    $S_h = \bar{u} \tilde{G} \tilde{P} v$, with $N = 1$,
    decay into two $J^{PC} = 0^{++}$ mesons.
    The optimal values of the parameters are given in the 
    sixth column in Table~\ref{tab:ASA:uGPv:devsqrp-1}.
    The arrow marked ``ASA'' points toward the optimal value of a 
    parameter. The last plot shows the amplitude itself for the
    optimal parameters.
  }
\end{figure*}
\begin{figure*}[p]
  \centering
  \includegraphics[scale=\scanscale, trim=5 0 0 0]{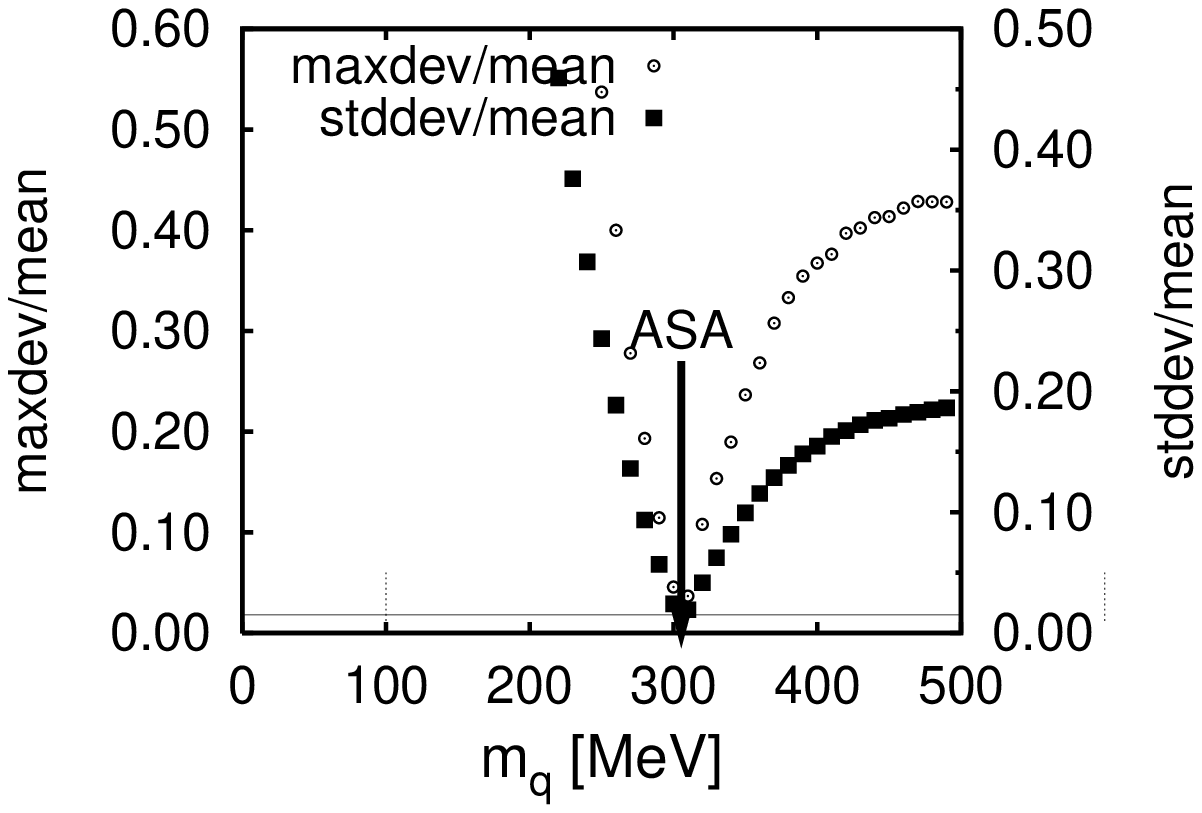}
  \includegraphics[scale=\scanscale, trim=5 0 0 0]{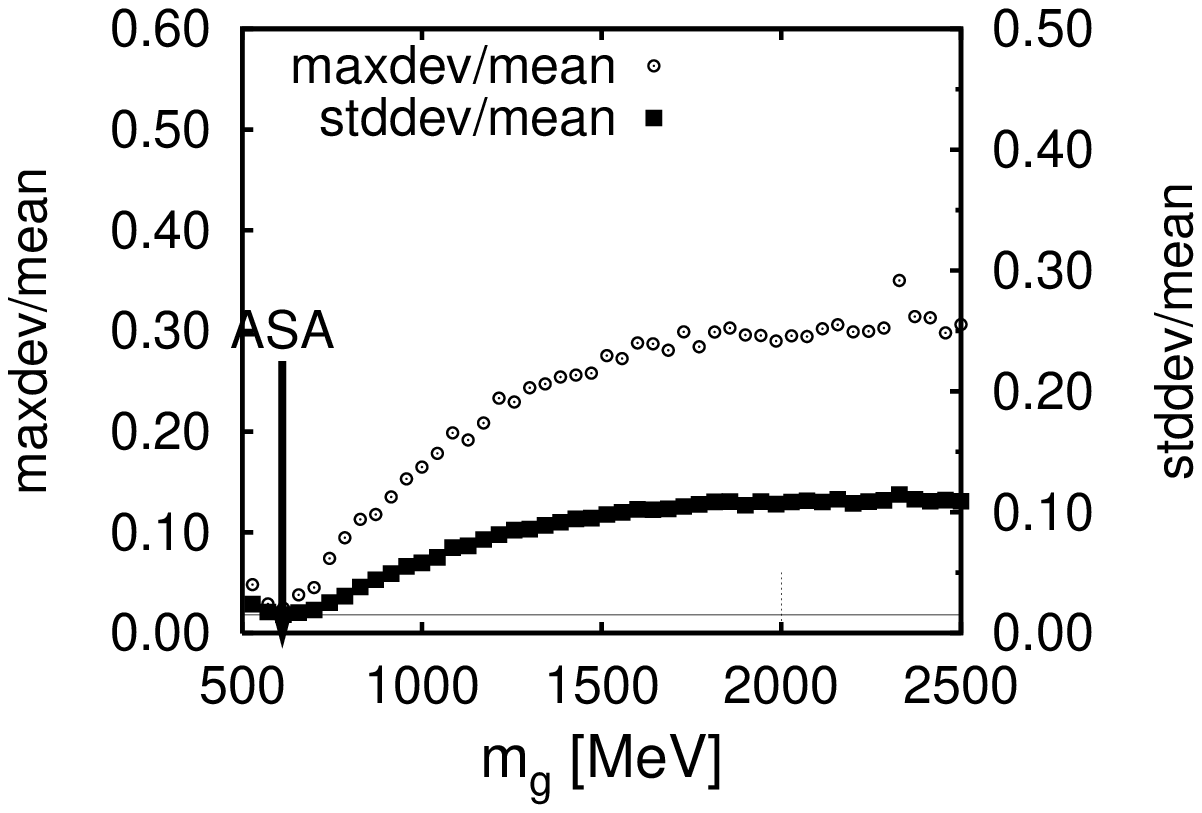}\\
  \includegraphics[scale=\scanscale, trim=5 0 0 0]{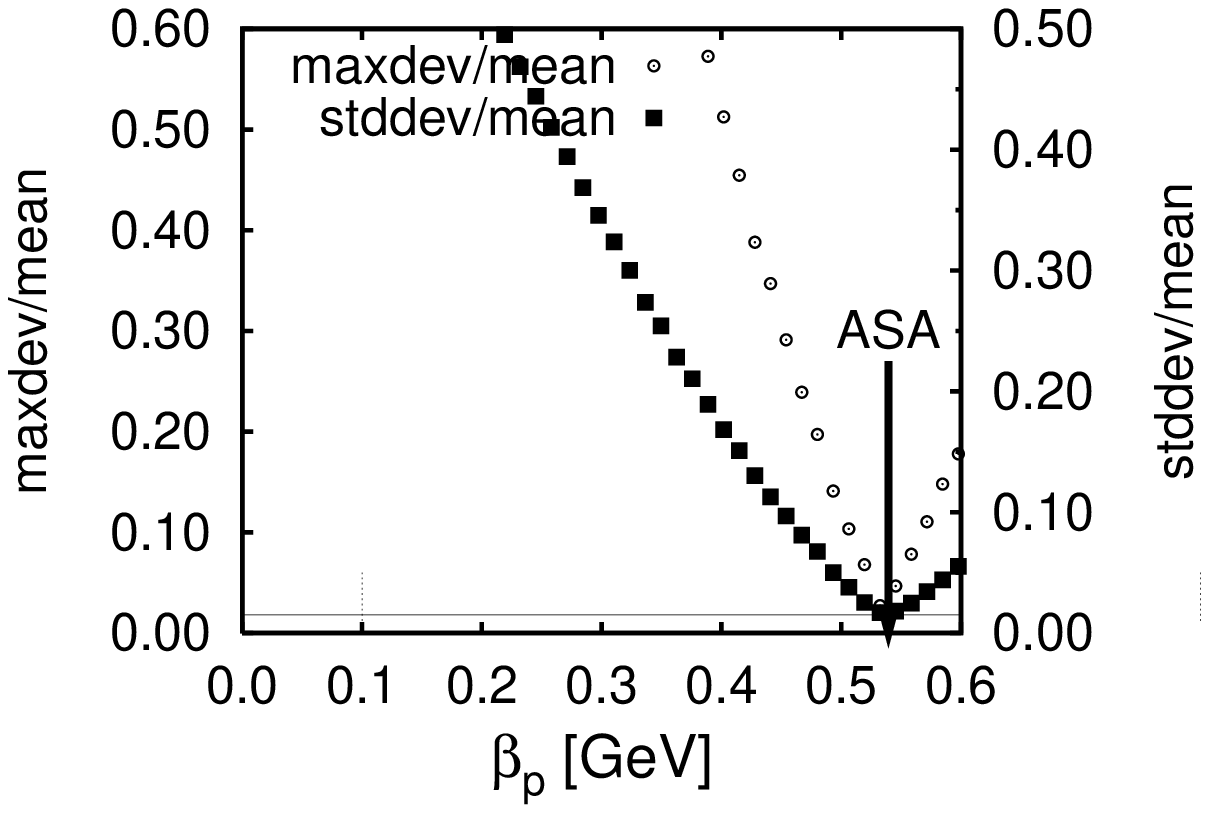}
  \includegraphics[scale=\scanscale, trim=5 0 0 0]{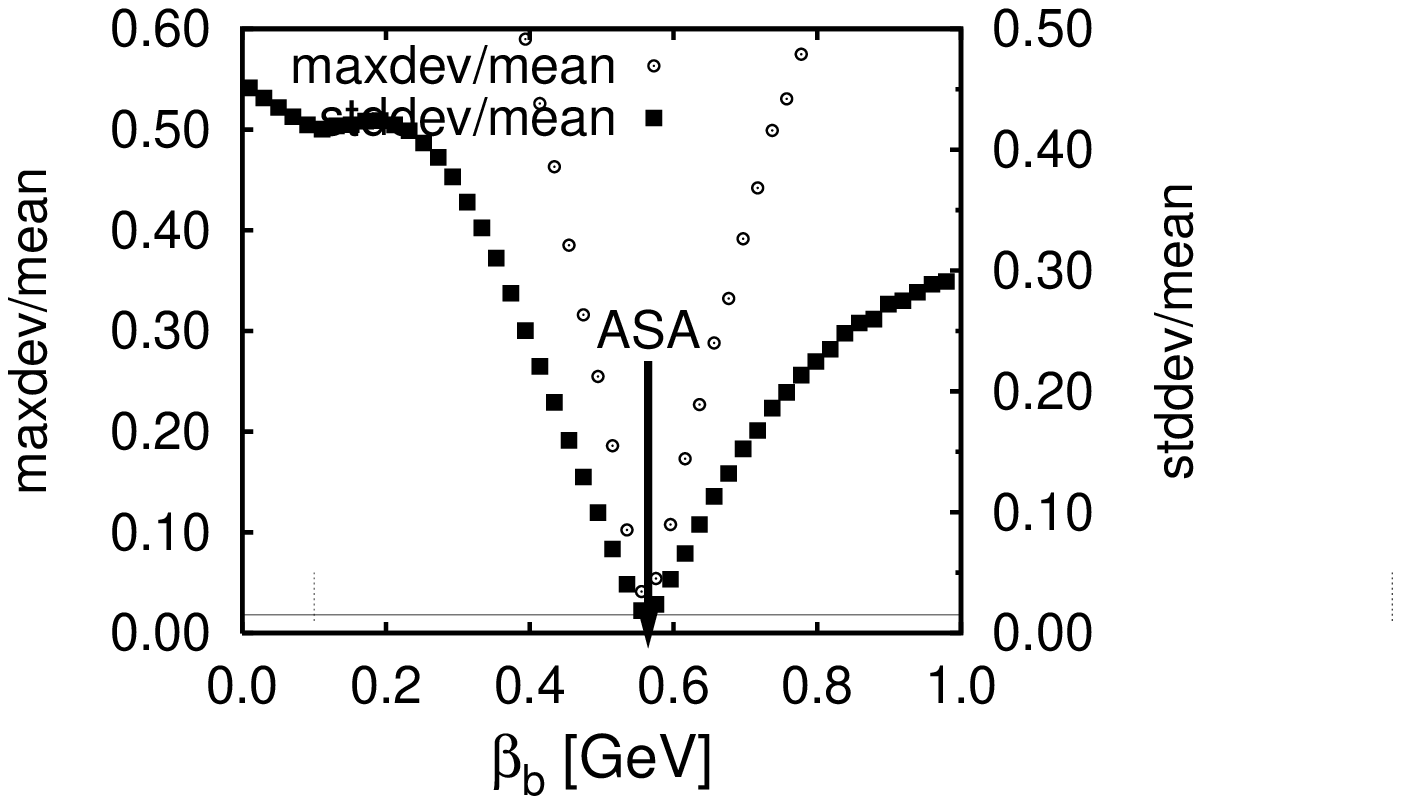}\\
  \includegraphics[scale=\scanscale, trim=5 0 0 0]{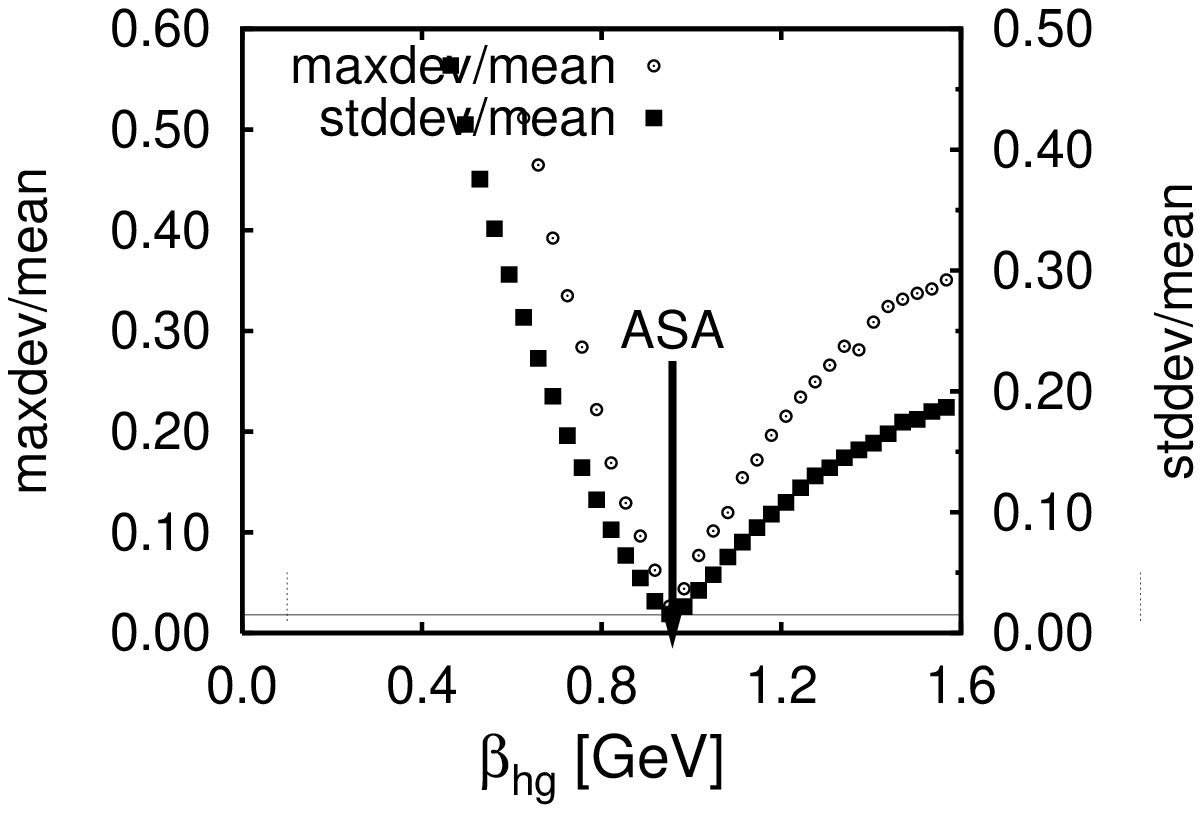}
  \includegraphics[scale=\scanscale, trim=5 0 0 0]{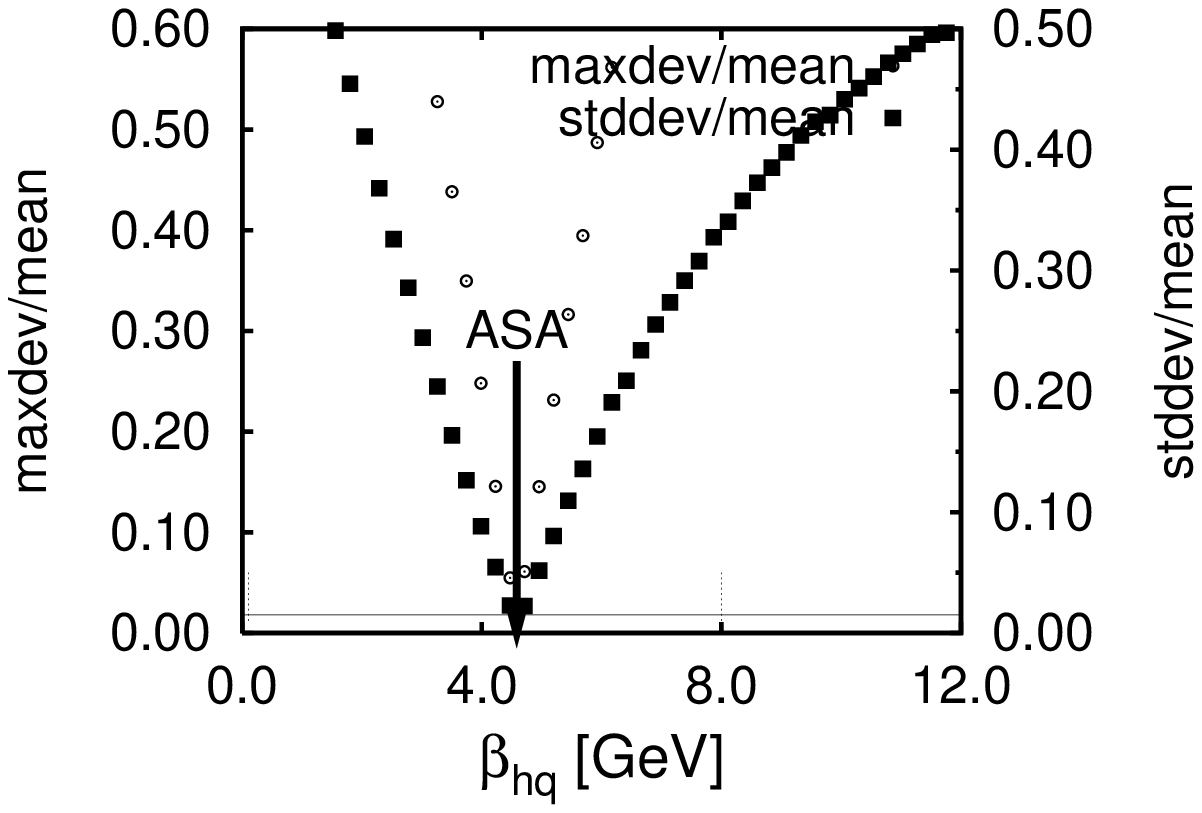}\\
  \includegraphics[scale=\scanscale, trim=5 0 0 0]{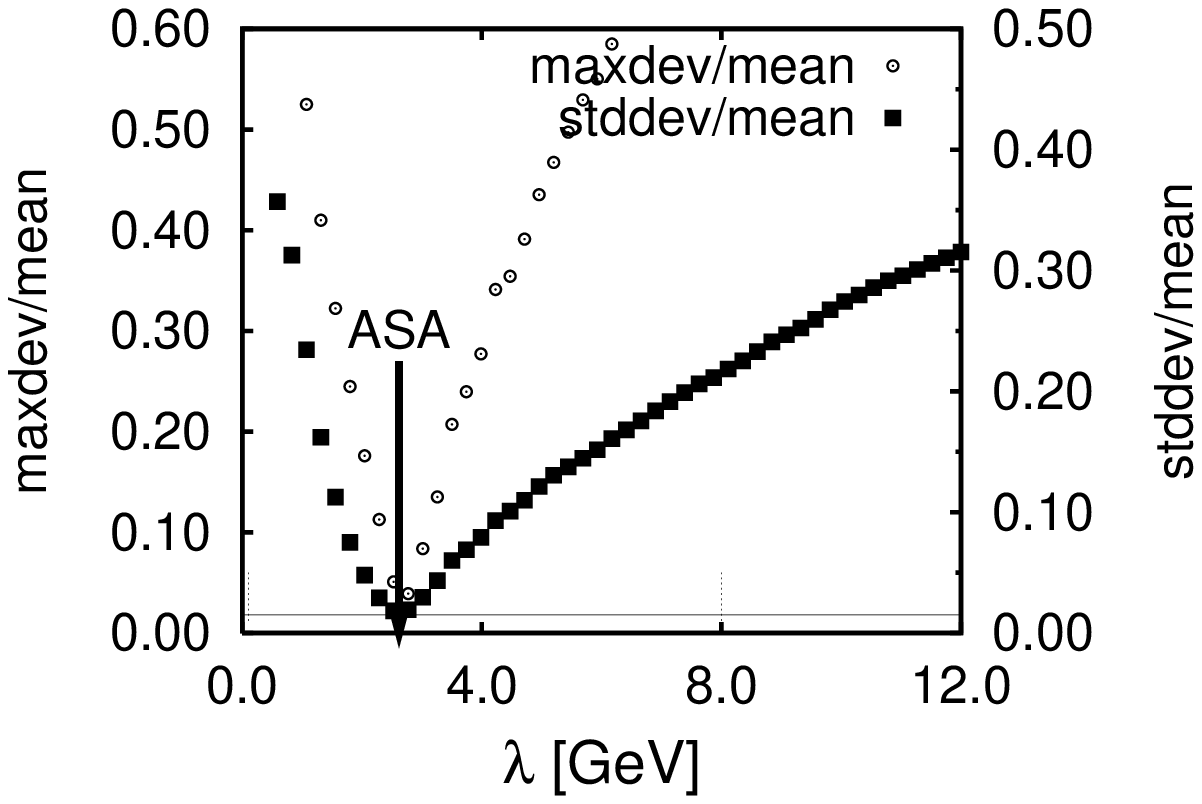}
  \includegraphics[scale=\scanscale, trim=45 0 -40 0]{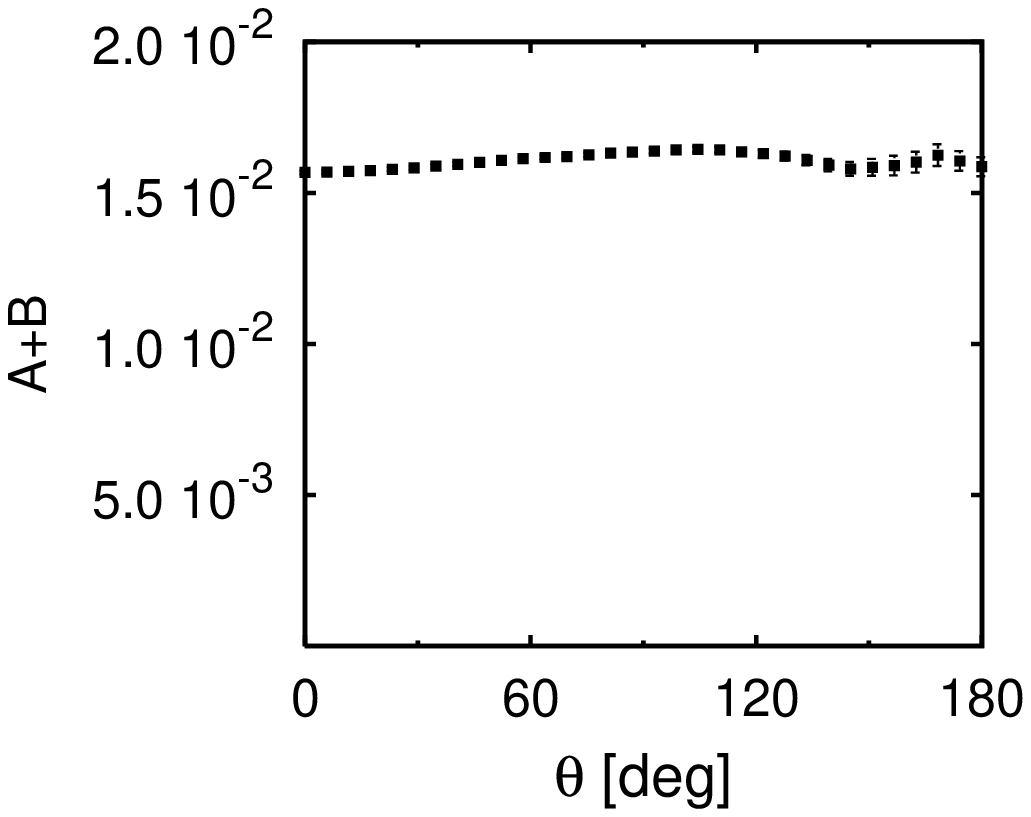}
  \caption{\label{fig:ugpv_m5-nofxx}\label{fig:ASA:uGPv_m5:f}
    Variation of rotational symmetry violation
    versus changes of parameters in the wave functions 
    in the case f) in Fig.~\ref{fig:ASA:uGPv_m5:devsqrp-1}
    $S_h = \bar{u} \tilde{G} \tilde{P} v$, with $N = 1$,
    decay into two $J^{PC} = 0^{-+}$ mesons.
    The optimal values of the parameters are given in the 
    sixth column in Table~\ref{tab:ASA:uGPv_m5:devsqrp-1}.  
    The arrow marked ``ASA'' points toward the optimal value of a 
    parameter. The last plot shows the amplitude itself for the
    optimal parameters.
  }  
\end{figure*}

\end{document}